\documentclass[letterpaper]{article}

\usepackage{arxiv}

\usepackage[T1]{fontenc}
\usepackage{float}
\usepackage{enumitem}
\usepackage[english]{babel}
\usepackage{amssymb}
\usepackage{amsmath,mathtools}
\usepackage{xcolor}
\usepackage{graphicx}
\usepackage{multirow}
\usepackage{overpic}
\usepackage{psfrag}
\usepackage{bm}
\usepackage{epsfig}
\usepackage{marvosym}
\usepackage{verbatim}
\usepackage{ulem}
\usepackage{lmodern}
\usepackage{booktabs} 
\usepackage{moresize} 
\usepackage{tikz}
\usepackage{pgfplots}
\pgfplotsset{compat=1.15}
\usetikzlibrary{shapes, arrows, arrows.meta, calc, positioning, decorations,
decorations.pathmorphing, decorations.text}
\usepackage{lineno}
\modulolinenumbers[1]
\usepackage[bookmarks=false]{hyperref}
\usepackage{cancel} 
\usepackage{rotating} 
\usepackage{array}
\usepackage{booktabs}
\usepackage{makecell}
\usepackage{threeparttable}
\usepackage{kantlipsum}
\usepackage{breakcites}

\definecolor{myblue1}	{RGB}{0,177,234}				
\definecolor{myblue2}	{RGB}{76,200,239}				
\definecolor{myblue3}	{RGB}{127,215,244}				
\definecolor{myblue4}	{RGB}{178,231,248}				
\definecolor{myblue5}	{RGB}{198,251,255}				
\definecolor{mybluegray1}{RGB}{0,127,167}				
\definecolor{mybluegray2}{RGB}{76,165,193}				
\definecolor{mybluegray3}{RGB}{127,191,211}				
\definecolor{mybluegray4}{RGB}{178,216,228}				
\definecolor{mygray1}	{RGB}{76,84,93}				
\definecolor{mygray2}	{RGB}{129,135,141}				
\definecolor{mygray3}	{RGB}{165,169,174}				
\definecolor{mygray4}	{RGB}{201,203,206}				
\definecolor{myorange1}	{RGB}{255,126,46}				
\definecolor{myorange2}	{RGB}{255,164,108}				
\definecolor{myorange3}	{RGB}{255,190,150}				
\definecolor{myorange4}	{RGB}{255,216,192}				

\newcommand\red[1]{\textcolor{black}{#1}}

\def\dt{\partial_t^{}}

\newcommand{\bsymb}[1]{\boldsymbol{#1}}
\renewcommand\vec[1]{\bsymb{#1}}
\newcommand\op[1]{{\textbf{#1}}}
\newcommand\myrefeq[1]{$($\ref{#1}$)$}

\def\bal#1\eal{\begin{linenomath}\begin{align}#1\end{align}\end{linenomath}}
\def\bov#1\eov{\begin{overpic}#1\end{overpic}}

\def\vecg{\vec{g}}
\def\nn{\vec{n}}
\def\ww{\vec{w}}

\def\uu{\vec{u}}

\def\eey{\vec{e}_y}

\def\bepsilon{\bsymb{\varepsilon}}
\def\00{\mathbf{0}}
\def\rey{\textrm{Re}}
\def\pr{\textrm{Pr}}
\def\ray{\textrm{Ra}}

\def\nus{\textrm{Nu}}
\def\vecpsi{\vec{\psi}}

\title{Deep reinforcement learning for the control of conjugate heat transfer with application to workpiece cooling}

\author{
	Elie Hachem\thanks{Corresponding author}\\
	MINES Paristech , PSL - Research University, CEMEF\\
	\texttt{elie.hachem@mines-paristech.fr}\\
\And
	H. Ghraieb\\
	MINES Paristech , PSL - Research University,  CEMEF
\And
	J. Viquerat\\
	MINES Paristech , PSL - Research University, CEMEF
\And
	A. Larcher\\
	MINES Paristech , PSL - Research University, CEMEF
\And
	P. Meliga\\
	MINES Paristech , PSL - Research University, CEMEF	
}

\begin{document} 
\newgeometry{left=2.5cm,right=2.5cm,top=3cm,bottom=3cm}
\maketitle



 

%
%
%
%
%

\begin{abstract} 

This research gauges the ability of deep reinforcement learning (DRL) techniques to assist the control of conjugate heat transfer systems
{governed by the coupled Navier--Stokes and heat equations. 
It uses  a novel, “degenerate” version of the proximal policy optimization (PPO) algorithm, intended for situations where the optimal policy to be learnt by a neural network does not depend on state, as is notably the case in optimization and open-loop control problems.
The numerical reward fed to the neural network is computed with an in-house stabilized finite elements environment combining variational multi-scale (VMS) modeling of the governing equations, immerse volume method, and multi-component anisotropic mesh adaptation.}
Several test cases of natural and forced convection in two and three dimensions are used as testbed for developing the methodology. 
{The approach successfully alleviates the natural convection induced enhancement of heat transfer in a two-dimensional, differentially heated square cavity controlled by 
piece-wise constant fluctuations of the sidewall temperature. It also proves capable of improving the homogeneity of temperature across the surface of two and three-dimensional hot workpieces under impingement cooling.  Various cases are tackled, in which the position of multiple cold air injectors is optimized relative to a fixed workpiece position. The flexibility of the numerical framework makes it tractable to solve also the inverse problem, i.e., to optimize the workpiece position relative to a fixed injector distribution. The obtained results showcase the potential of the method for black-box optimization of practically meaningful computational fluid dynamics (CFD) conjugate heat transfer systems. More significantly, they stress how DRL can reveal unanticipated solutions or parameter relations (as the optimal workpiece position under symmetrical actuation turns to be offset from the symmetry axis), in addition to being a tool for optimizing searches in large parameter spaces.}
\end{abstract}

\keywords{Deep Reinforcement Learning; Artificial Neural Networks; Conjugate heat transfer; Computational fluid dynamics; Thermal control}




\section{Introduction} 

Thermal control, defined as the ability to finesse the thermal properties of a volume of fluid (and of the solid objects inside) into a certain desired state, is a field of tremendous societal and economical importance.
For instance, heat/cool exchangers are used in a broad range of industrial applications to regulate process temperatures by heat or cool transfer between  fluid media, which in turn ensures that machinery, chemicals, water, gas, and other substances remain within safe operating conditions. Green building engineering is another field whose focus is on regulating indoor thermal conditions (temperature, humidity) under substantial variations of the ambient conditions to provide high-quality living and working environments. In many manufacturing processes, thermal conditioning is also intended to improve the final mechanical (e.g., hardness, toughness, resistance), electrical, or optical properties of the product, the general picture being that high temperature gradients are useful to speed up the process but generally harm the quality of the outcome because of heat transfer inhomogeneities caused by the increased convection by the fluid particles. All such problems fall under the purview of this line of study.

Numerous strategies have been implemented to control fluid mechanical systems (including
conjugate heat transfer systems combining thermal conduction in the solid and convective transfer in the fluid), either open-loop with passive appendices (e.g., end plate, splitter plate, small secondary cylinder, or flexible tail), or open-loop with actuating devices (e.g., plasma actuation, boundary temperatures, steady or unsteady base bleeding, rotation) or closed-loop (e.g. via transverse motion, perturbations of the thermal boundary layer, blowing/suction, rotation, all relying on an appropriate sensing of flow variables). Nonetheless, many of the proposed strategies are trial and error, and therefore require extensive and costly experimental or numerical campaigns. This has motivated the development of analytical methods and numerical algorithms for the optimal control of Navier--Stokes systems~\cite{Jameson1998,gunz02,bewl01}, and the maturing of mathematical methods in flow control and discrete concepts for PDE constrained optimization. 
Applications to the heat equation~\cite{Momose2004} and the
coupled Navier--Stokes and heat equations~\cite{Belmiloudi2002,Barwolf2006,Boldrini2007,Karkaba2020} have also been considered, including fresh developments meant to alter the linear amplification of flow disturbances~\cite{meli10tcfd}, but the general picture remains that the optimal control of conducting-convecting (possibly radiating) fluids has not been extensively studied.


The premise of this research is that the related task of selecting an optimal subset of control parameters can alternatively be assisted by machine learning algorithms. Indeed, the introduction of 
the back-propagation algorithm~\cite{Rumelhart1986} has progressively turned Artificial Neural Networks (ANN) into a family of versatile non-parametric tools that can be trained to hierarchically extract informative features from data and to provide qualitative and quantitative modeling predictions.
Together with the increased affordability of high-performance computational hardware, this has allowed leveraging the ever-increasing volume of data generated for research and engineering purposes into novel insight and actionable information, which in turn has entirely transformed scientific disciplines, such as robotics~\cite{kober2013reinforcement,Mnih2015} or image analysis~\cite{hinton2012imagenet}. 
Owing to the ability of neural networks to handle stiff, large-scale nonlinear problems~\cite{lusch2018deep}, 
machine learning algorithms have also been making rapid inroads in fluid mechanics, as a mean to solve the Navier--Stokes equations~\cite{Raissi2018} or to predict closure terms in turbulence models~\cite{Beck2018}; see also Ref.~\cite{Brunton2020} for an overview of the current developments and opportunities.

Neural networks can also be used to solve decision-making problems, which is the purpose of Deep Reinforcement Learning (DRL, where the \textit{deep} terminology generally weighs on the sizable depth of the network), an advanced branch of machine learning. Simply put, a neural network trains in finding out which actions or succession of actions maximize a numerical reward signal, with the possibility for a given action to affect not only the immediate but also the future rewards.
Successful applications of DRL range from AlphaGo, the well-known  ANN that defeated the top-level human player at the game of Go \cite{silver2017mastering} to the real-word deployment of legged robots~\cite{hwangbo2019learning}, to breakthroughs in computer vision (e.g., filtering, or extracting image features)~\cite{Bernstein2018} and optimal control problems~\cite{Lillicrap2015,Schulman2017}. 
{There is also great potential for applying DRL to fluid mechanics, for which efforts are ongoing but still at an early stage. Sustained commitment from the machine learning community has allowed expanding the scope from computationally inexpensive, low-dimensional reductions of the underlying fluid dynamics~\cite{Belus2019,Bucci2019,Novati2019} 
to complex Navier--Stokes systems~\cite{Novati2017,Verma2018}, with a 
handful of pioneering studies providing insight into the performance improvements to be delivered in shape optimization~\cite{lee2018,Yan2019,viquerat2019direct} and flow control~\cite{ma2018fluid,Biferale2019,rabault2019artificial,Ren2020,Tang2020,Paris2020,Xu2020}, \red{including recent advances assessing experimentally the effectiveness of reinforcement learning control strategies~\cite{Fan2020}}.} 
{The literature on thermal control is even more {scarce}, as our literature review
did not reveal any other study considering DRL-based control of conjugate heat transfer aside from~\cite{beintema2020controlling}, another research effort conducted in the same time frame as the present work that will be discussed further on, plus a few other publications relying on dealing with energy efficiency in civil engineering
from low-dimensional thermodynamic models basically unrelated to the equations of fluid dynamics~\cite{kazmi2018gigawatt,zhang2019flow}.}


This research assesses the feasibility of using proximal policy optimization (PPO~\cite{Schulman2017}) for control and optimization purposes of conjugate heat transfer systems, as governed by the coupled Navier--Stokes and heat equations. 
The objective here is to keep shaping the capabilities of the method (PPO is still a relatively newcomer that has quickly emerged as the go-to DRL algorithm due to its data efficiency, simplicity of implementation and reliable performance) and to narrow the gap between DRL and advanced numerical methods for multi-scale, multi-physics computational fluid dynamics (CFD). 
We investigate more specifically the ``degenerate'' single-step PPO algorithm introduced in~\cite{viquerat2019direct} {for optimization and open-loop control problems, as the optimal policy to be learnt is then state-independent, and it may be enough for the neural network to get only one attempt per episode at finding the optimal.} 
Several problems of conjugate heat transfer in two and three dimensions are used as testbed to push forward the development of this novel approach, whose potential {for reliable black-box optimization of computational fluid dynamics (CFD) systems has been recently assessed 
for open-loop drag reduction in cylinder flows at Reynolds numbers ranging from a few hundreds to a few ten thousands}~\cite{ghraieb2020}. 
To the best of the authors knowledge, this constitutes the first attempt to achieve DRL-based control of 
{conjugate \textit{forced} convection heat transfer, while~\cite{beintema2020controlling} is the first attempt to achieve DRL control of conjugate \textit{natural} convection heat transfer.}


The organization is as follows: section~\ref{section:method} outlines the main features of the finite element CFD environment used to compute the numerical reward fed to the neural network,
that combines variational multi-scale (VMS) modeling of the governing equations, immerse volume method, and multi-component anisotropic mesh adaptation. 
The baseline principles and assumptions of DRL and PPO are presented in section~\ref{section:drl}, together with the specifics of the single-step PPO algorithm. 
Section~\ref{section:natural} revisits the natural convection case of~\cite{beintema2020controlling} 
for the purpose of validation and assessment part of the method capabilities.
In section~\ref{section:forced2d}, DRL is used to control conjugate heat transfer in a  
model setup of two-dimensional workpiece cooling by impingement of a fluid. 
An extension to three-dimensional workpieces is proposed in section~\ref{section:forced3d}.

\section{Computational fluid dynamics}\label{section:method}

The focus of this research is on conjugate heat transfer and laminar, incompressible fluid flow problems in two and three-dimensions,
for which the conservation of mass, momentum and energy is described by the nonlinear, coupled 
Navier--Stokes and heat equations 
\bal
\nabla \cdot \uu&=0\,,
\label{eq:mass_cons}\\
\rho(\partial_{t}\uu + \uu\cdot\nabla \uu)&= \nabla\cdot(-p\op{I}+2\mu\bepsilon(\uu)) + \vecpsi\,,
\label{eq:momentum}\\
\rho c_{p}(\partial_{t}T + \uu\cdot\nabla T) &=\nabla\cdot (\lambda\nabla T)+\chi\,,
\label{eq:energy}
\eal
where $\uu$ is the velocity field,
$p$ is the pressure,
$T$ is the temperature, $\bepsilon(\uu)=(\nabla\uu+\nabla\uu^T)/2$ is the rate of deformation tensor, $\vecpsi$ and $\chi$ are source terms (modeling, e.g., buoyancy or radiative heat transfer), and we assume here
constant fluid density $\rho$, dynamic viscosity $\mu$, thermal conductivity $\lambda$, and specific heat $c_{p}$.

\subsection{The immersed volume method}

The numerical modeling of conjugate heat transfer mostly depends upon a heat transfer coefficient to ensure that the proper amount of heat is exchanged at the fluid/solid interface via thermal boundary conditions.
Computing said coefficient is no small task (as it requires solving an inverse problem to assimilate relevant experimental data, which in turn requires such data to be available), and is generally acknowledged to be a limiting issue for practical applications where one must vary, e.g., the shape, number and position of the solid, or the fluid and/or solid material properties. We thus rather use here the immerse volume method (IVM) to combine both the fluid and solid phases into a single fluid with variable material properties. Simply put, we solve equations formally identical to~\myrefeq{eq:mass_cons}-\myrefeq{eq:energy} on a unique computational domain $\Omega$,
but with variable density, dynamic viscosity, conductivity, and specific heat, which
removes the need for a heat transfer coefficient since the amount of heat  exchanged at the interface then proceeds solely from the individual material properties on either side of it. In order to ensure numerical accuracy, such an approach must combine three key ingredients, that are briefly reviewed in the next paragraphs: an interface capturing method, anisotropic mesh adaptation to achieve a high-fidelity description of said interface, and relevant mixing laws to describe the properties of the composite fluid. 
One point worth mentioning is that the interface here is static, although the same numerical framework can be used to dynamically track moving interfaces, and thus to encompass the effect of solid displacements. This is because the solid is fixed once an action has been taken by the PPO agent, although not fixed over the course of optimization, as the solid position can very well be the quantity subjected to optimization, as illustrated in section~\ref{section:forced2d:inverse}.\\

\paragraph*{- Level set method}The level set approach is used to localize the fluid/solid interface by the zero iso-value of a smooth function. In practice, a signed distance function $\phi$ is used to localize the interface and initialize the material properties on both either side of it, with the convention that $\phi>0$ (resp. $\phi<0$) in the fluid (resp. the solid).
\\

\paragraph*{- Anisotropic mesh adaptation}The interface may intersect arbitrarily the mesh elements
if it is not aligned with the element edges, in which case discontinuous material properties across the interface can yield oscillations of the numerical solutions. We thus use the anisotropic mesh adaptation technique presented in~\cite{gruau20053d}
 to ensure that the material properties are distributed as accurately and smoothly as possible over the smallest possible thickness around the interface. This is done computing modified distances from a symmetric positive defined tensor (the metric) whose eigenvectors define preferential directions along which mesh sizes can be prescribed from the related eigenvalues. 
The metric used here is isotropic far from the interface, with mesh size set equal to $h_\infty$ in all directions, but anisotropic near the interface, with mesh size equal to $h_\perp$ in the direction normal to the interface, and to $h_\infty$ in the other directions. This is written for an intended thickness $\delta$ as 
\bal
\bm{M}=K(\phi)\nn\otimes \nn+\frac{1}{{h_\infty^2}}\bm{I}\qquad \text{with}\qquad
K(\phi)=\left\{\begin{tabular}{ll}0&\quad if $|\phi|\geq \delta/2$\,,\\$\displaystyle\frac{1}{h_\perp^{2}}-\frac{1}{h^{2}_{\infty}}$& \quad if $|\phi|< \delta/2$\,,\end{tabular}\right.\label{eq:metric}
\eal
where $\nn=\nabla\phi/||\nabla\phi||$ is the unit normal to the fluid/solid interface computed from the level set gradient.
A posteriori anisotropic error estimator is then used to minimize the interpolation error under the constraint of a fixed number of edges in the mesh. A unique metric can be built from multi-component error vectors~\cite{gruau20053d,bernacki2008level,mesri2009advanced,coupez2011},
which is especially relevant for conjugate heat transfer optimization, as
it allows each learning episode to use an equally accurate mesh adapted from the velocity vector and magnitude, the temperature field, and the level set.\\
 
\paragraph*{- Mixing laws}The composite density, dynamic viscosity and specific heat featured in equations~\myrefeq{eq:mass_cons}-\myrefeq{eq:energy} are computed {as the arithmetic means}
of the fluid and solid values, for instance the composite density is 
\bal
\rho=\rho_{f}H_\epsilon(\phi)+\rho_{s}(1-H_\epsilon(\phi))\,,\label{eq:mixing}
\eal
where $H_\epsilon$ is the smoothed Heaviside function defined as
\bal
H_\epsilon(\phi)=\left\{\begin{tabular}{ll}0&\quad if $\phi<-\epsilon$\,,\\$\displaystyle\frac{1}{2}(1+\frac{\phi}{\epsilon}+\frac{1}{\pi}\sin(\pi\frac{\phi}{\epsilon}))$& \quad if $|\phi|\leq\epsilon$\,,\\1& \quad if $\phi>\epsilon$\,,\end{tabular}\right.\label{eq:heaviside}
\eal
and $\epsilon$ is a regularization parameter proportional to the mesh size in the normal direction to the interface, set here to $\epsilon=2h_\perp$. 
{In order to ensure continuity of the heat flux across the interface, the thermal conductivity is computed as the harmonic mean}
\bal
\frac{1}{\lambda}=\frac{1}{\lambda_f}H_\epsilon(\phi)+\frac{1}{\lambda_s}(1-H_\epsilon(\phi))\,,\label{eq:mixinglambda}
\eal
{as obtained from a steady, no source, one dimensional analysis of the heat flux when the conductivity varies stepwise from one medium to the next; see~\cite{suhas1980} for detailed derivation and analysis, and~\cite{pata78} for proof of the gain in numerical accuracy (with respect to the arithmetic mean model) by comparison with analytical solutions.}

\subsection{Variational multi-scale approach (VMS)}

In the context of finite element methods (that remain widely used to simulate engineering CFD systems due to their ability to handle complex geometries), direct numerical simulation (DNS) solves the weak form of~\myrefeq{eq:mass_cons}-\myrefeq{eq:energy}, obtained by integrating by parts the pressure, viscous and conductive terms, to give 
\bal 
&(\rho(\dt\uu+\uu\cdot\nabla\uu)\,,\,\ww)+(2\mu\bepsilon(\uu)\,,\,\bepsilon(\ww))
-(p\,,\,\nabla\cdot\ww)+(\nabla\cdot\uu\,,\,q)=(\vecpsi\,,\,\ww)\,,
\label{eq:weakdns}\\
&(\rho c_p(\dt T+\uu\cdot\nabla T)\,,\,s)+(\lambda\nabla T\,,\,\nabla s)=(\chi\,,\,s)\,,
\label{eq:weakheat}
\eal 
where $(\,,\,)$ is the $L^2$ inner product on the computational domain, 
$\ww$, $q$ and $s$ are relevant test functions for the velocity, pressure and temperature variables, and all fluid properties are those mixed with the smoothed Heaviside function~\myrefeq{eq:heaviside}.

We use here the variational multi-scale (VMS) approach~\cite{hugh98,codi00,bazi07} to solve a stabilized formulation of~\myrefeq{eq:weakdns}-\myrefeq{eq:weakheat}, which allows circumventing the Babuska---Brezzi condition (that otherwise imposes that different interpolation orders be used to discretize the velocity and pressure variables, while we use here simple continuous piecewise linear P$_1$ elements for all variables) and prevents numerical instabilities in convection regimes at high Reynolds numbers. We shall not go into the extensive details about the derivation of the stabilized formulations, for which the reader is referred to~\cite{hach10,hach12}. Suffice it to say here that the flow quantities are split into coarse and fine scale components, that correspond to different levels of resolution. The fine scales are solved in an approximate manner to allow modeling their effect into the large-scale equations. This gives rise to additional terms in the right-hand side of~\myrefeq{eq:weakdns}-\myrefeq{eq:weakheat}, and yields the following weak forms for the large scale
\bal 
\nonumber
& (\rho(\dt\uu+\uu\cdot\nabla\uu)\,,\,\ww)+(2\mu\bepsilon(\uu)\,,\,\bepsilon(\ww))
-(p\,,\,\nabla\cdot\ww)+(\nabla\cdot\uu\,,\,q)=(\vecpsi\,,\,\ww)\\
&\qquad\qquad+\sum_{K\in\mathcal{T}_h}[(\tau_1\mathcal{R}_M\,,\,\uu\cdot\nabla\ww)_K		
+(\tau_1\mathcal{R}_M\,,\,\nabla q)_K+(\tau_2\mathcal{R}_C\,,\,\nabla\cdot \ww)_K]\,,			
\label{eq:weakvms}\\
\nonumber
&(\rho c_p(\dt T+\uu\cdot\nabla T)\,,\,s)+(\lambda\nabla T\,,\,\nabla s)=(\chi\,,\,s)\\
&\qquad\qquad+\sum_{K\in\mathcal{T}_h}[(\tau_{3}\mathcal{R}_T\,,\,\uu\cdot\nabla s)_K
+{(\tau_{4}\mathcal{R}_T\,,\,\zeta\nabla T\cdot\nabla s)_K}]\,,\label{eq:weakheatvms}
\eal
where $(\,,\,)_K$ is the inner product on element $K$, {we denote by $\zeta=\uu\cdot\nabla T/||\nabla T||^2$ the (normalized) velocity projected along the direction of the temperature gradient}, and the $\mathcal{R}$ terms are the governing equations residuals
\bal
-\mathcal{R}_C=\nabla\cdot\uu\,,\quad
-\mathcal{R}_M=\rho(\dt\uu+\uu\cdot\nabla\uu)+\nabla p-\vecpsi\quad
-\mathcal{R}_T=\rho c_p(\dt T+\uu\cdot\nabla T)-\chi\,,
\eal
{whose second derivatives vanish since we use linear interpolation functions.} 
In~\myrefeq{eq:weakvms}, $\tau_{1,2}$ are ad-hoc mesh-dependent stabilization parameters defined in~\cite{codi02,hach13}. {Conversely, in}~\myrefeq{eq:weakheatvms}, 
$\tau_{3,4}$ are mesh-independent stabilization parameters acting both in the direction of the solution and of its gradient, that proceed from the stabilization of the ubiquitous convection-diffusion-reaction equation~\cite{codina1998comparison,badia2006analysis}, whose definition is given in~\cite{brooks1982streamline,galeao1988consistent}.

The governing equations are solved sequentially, i.e.,
we solve first~\myrefeq{eq:weakvms}, then use the resulting fluid velocity to solve~\myrefeq{eq:weakheatvms}.
All linear systems are preconditioned with a block Jacobi method supplemented by an incomplete LU factorization, and solved with the GMRES algorithm, with tolerance threshold set to $10^{-6}$ for the Navier--Stokes equations, and  $10^{-5}$ for the heat equation.
{The time derivatives and convection terms of the Navier--Stokes equations and related VMS source terms are integrated semi-implicitly using the first-order backward 
differentiation formula and Newton--Gregory backward polynomial.
The viscous, pressure and divergence terms are treated implicitly with the backward Euler scheme. Finally, the VMS stabilization terms $\tau_{1,2}$ are treated explicitly with the forward Euler scheme, which yields }
\bal
&\nonumber 
(\rho(\frac{\uu^{i+1}-\uu^{i}}{\Delta t}+\uu^{i}\cdot\nabla\uu^{i+1})\,,\,\ww)+(2\mu\bepsilon(\uu^{i+1})\,,\,\bepsilon(\ww))
-(p^{i+1}\,,\,\nabla\cdot\ww)+(\nabla\cdot\uu^{i+1}\,,\,q)=(\vecpsi^i\,,\,\ww)\\
&\qquad\qquad+\sum_{K\in\mathcal{T}_h}[(\tau_1^i\mathcal{R}_M^{i+1}\,,\,\uu^i\cdot\nabla\ww)_K		
+(\tau_1^i\mathcal{R}_M^{i+1}\,,\,\nabla q)_K+(\tau_2^i\mathcal{R}_C^{i+1}\,,\,\nabla\cdot \ww)_K]\,,			
\label{eq:weakvms2}
\eal
with residuals 
\bal
-\mathcal{R}_C^{i+1}=\nabla\cdot\uu^{i+1}\,,\quad
-\mathcal{R}_M^{i+1}=\rho(\frac{\uu^{i+1}-\uu^{i}}{\Delta t}+\uu^i\cdot\nabla\uu^{i+1})+\nabla p^{i+1}-\vecpsi^i\,,
\eal
where the superscript $i$ refers to the solution at time $t_i = i\Delta t$. 
{The time derivatives, convection and conduction terms of the heat equation and related VMS source terms are integrated implicitly with the backward Euler scheme (modeling the velocity after $\uu^{i+1}$ wherever needed on behalf of the sequential resolution process, although we drop the dependence in the notation to ease the reading). The VMS stabilization terms $\tau_{3,4}$ are treated explicitly with the forward Euler scheme\footnote{That is, with respect to $T$, but the velocity is modeled after its latest computed approximation $\uu^{i+1}$ wherever needed.}, to give}
\bal 
\nonumber
&(\rho c_p(\frac{T^{i+1}-T^{i}}{\Delta t}+\uu^{i+1}\cdot\nabla T^{i+1})\,,\,s)+(\lambda\nabla T^{i+1}\,,\,\nabla s)=(\chi^i\,,\,s)\\
&\qquad\qquad+\sum_{K\in\mathcal{T}_h}[(\tau_{3}^i\mathcal{R}_T^{i+1}\,,\,\uu^{i+1}\cdot\nabla s)_K
+(\tau_{4}^i\mathcal{R}_T^{i+1}\,,\,{\zeta^i\nabla T^{i}\cdot\nabla s})_K]\,,\label{eq:weakheatvms2}
\eal
with residual 
\bal
-\mathcal{R}_T^{i+1}=\rho c_p(\frac{T^{i+1}-T^{i}}{\Delta t}+\uu^{i+1}\cdot\nabla T^{i+1})-\chi^i\,.
\eal
We solve equations~\myrefeq{eq:weakvms2}-\myrefeq{eq:weakheatvms2} with an in-house VMS solver whose flexibility, accuracy and reliability is assessed in a series of previous papers {to which the reader is referred for further information, see in particular~\cite{hachem2012immersed,hach13} for the detailed mathematical formulation of the IVM in the context of finite element VMS methods. The ability of the IVM to handle the abrupt conductivity change across the fluid/solid interface is documented in~\cite{hach12,hach12b,hach13b}.
Excellent agreement with reference solutions available from the literature and in-house data obtained enforcing proper thermal conditions at the boundary of body-fitted meshes is reported for several 
time-dependent conjugate heat transfer test cases (e.g., mixed convection in a plane channel flow, combined convection in square enclosures and conduction/radiation heat transfer, all in two dimensions).
Ref.~\cite{hach12b} also reports favorable agreement between the IVM and in-house experimental data pertaining to a three-dimensional test case representative of an industrial cooling system, which provides strong evidence of relevance for the intended application.}




\section{Deep reinforcement learning and proximal policy optimization}\label{section:drl} 

\subsection{Neural networks}

\begin{figure}[t!]
\centering
\begin{tikzpicture}[	arrow/.style=		{thick,color=mybluegray1,rounded corners},
				netnode/.style=		{circle, inner sep=0pt, text width=22pt, align=center, very thick},
				inputnode/.style=	{netnode, fill=red, draw=black},
				hiddennode/.style=	{netnode, fill=orange, draw=black},
				outputnode/.style=	{netnode, fill=yellow, 	draw=black},
				signal/.style=		{arrows={-stealth},draw=black}]
	\def\nodedist{30pt}
	\def\layerdist{60pt}
    
	\foreach \y in {1,...,3}
		\node[inputnode] (I\y) at (0,-\y*\nodedist) {$x_\y$};  
	\foreach \y in {1,...,4}
		\node[hiddennode] (H1\y) at ($(\layerdist,-\y*\nodedist) +(0, 0.5*\nodedist)$) {};
	\foreach \y in {1,...,4}
		\node[hiddennode] (H2\y) at ($(2*\layerdist,-\y*\nodedist) +(0, 0.5*\nodedist)$) {};
	\foreach \y in {1,...,2}
		\node[outputnode] (O\y) at ($(H21) + (\layerdist, -\y*\nodedist)$) {$y_\y$};

	\foreach \dest in {1,...,4}
		\foreach \source in {1,...,3}
			\draw[signal] (I\source) -- (H1\dest);
	\foreach \dest in {1,...,4}
		\foreach \source in {1,...,4}
			\draw[signal] (H1\source) -- (H2\dest);
	\foreach \dest in {1,...,2}
		\foreach \source in {1,...,4}
			\draw[signal] (H2\source) edge (O\dest);
\end{tikzpicture}
\caption{Fully connected neural network with two hidden layers, modeling a mapping from $\mathbb{R}^3$ to $\mathbb{R}^2$.}
\label{fig:simple_network}
\end{figure}
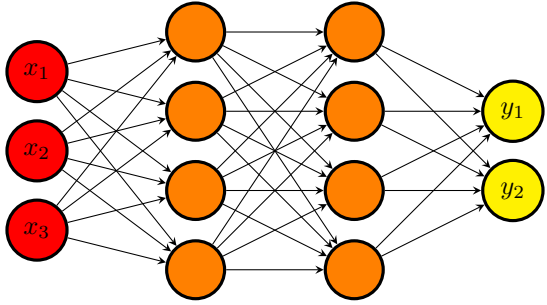

A neural network (NN) is a collection of artificial neurons, i.e., connected computational units
that can be trained to arbitrarily well approximate the mapping function between input and output spaces. Each connection provides the output of a neuron as an input to another neuron.
Each neuron performs a weighted sum of its inputs, to assign significance to the inputs with regard to the task the algorithm is trying to learn. It then adds a bias to
better represent the part of the output that is actually independent of the input. Finally, it feeds an activation function that determines whether and to what extent the computed value should affect the outcome.
As sketched in figure~\ref{fig:simple_network}, a
fully connected network is generally organized into layers, with the neurons of one layer being connected solely to those of the immediately preceding and following layers. The layer that receives the external data is the input layer, the layer that produces the outcome is the output layer, and in between them are zero or more hidden layers.

The design of an efficient neural network requires a proper optimization of the weights and biases, together with a relevant nonlinear activation function. The abundant literature available on this topic points to a relevant network architecture (e.g., type of network, depth, width of each layer), finely tuned hyper parameters (i.e.,  parameters whose value cannot be estimated from data, e.g., optimizer, learning rate, batch size) and a sufficiently large amount of data to learn from as being the key ingredients for success; see, e.g., Ref.~\cite{IanGoodfellow2017} and the references therein.

\subsection{Deep reinforcement learning} 

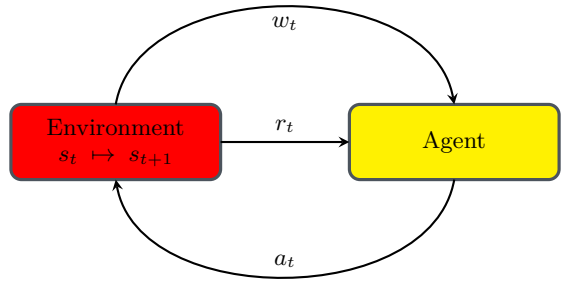
\begin{figure}[!t]
\centering
\begin{tikzpicture}[	fontsize/.style={		font=\footnotesize},
				intnode/.style={		black, fontsize, pos=0.5},
				arrow/.style={		thick,color=black, rounded corners,thick,-stealth},
				box/.style={		rectangle,rounded corners, draw=mygray1, very thick, 
								text width=2.7cm,minimum height=1cm,text centered,
								inner sep=1.25pt, outer sep=0pt, fontsize}]

	\node[box,fill=red] 	(env) 	at (0,0) {Environment\\ $s_t \mapsto s_{t+1}$};
	\node[box,fill=yellow] (agent) 	at (4.5,0) {Agent};
	
	\draw[arrow] 	(env.east) 		to [out=0,in=180] 	node[intnode, above] {$r_t$} 	(agent.west);
	\draw[arrow] 	(agent.south) 	to [out=-100,in=-80] 	node[intnode, above] {$a_t$} 	(env.south);
	\draw[arrow] 	(env.north) 	to [out=80,in=100] 	node[intnode, below] {$w_{t}$} 	(agent.north);
\end{tikzpicture}

\caption{RL agent and its interactions with its environment.}
\label{fig:RL_agent}
\end{figure}

Deep reinforcement learning (DRL) is an advanced branch of machine learning in which deep neural networks train in solving sequential decision-making problems.
It is a natural extension of reinforcement learning (RL), in which an agent (the neural network) 
is taught how to behave in an environment by taking actions and by receiving feedback from it under the form of a reward (to measure how good or bad the action was) and information (to gauge how the action has affected the environment). This can be formulated as a Markov Decision Process, for which a typical execution goes as follows (see also figure~\ref{fig:RL_agent}):
\begin{itemize}
	\item assume the environment is in state $s_{t} \in \mathcal{S}$ at iteration $t$, where $\mathcal{S}$ is a set of states,
	\item the agent uses $w_{t}$, an observation of the current environment state (and possibly a partial subset of $s_t$) to take action $a_{t} \in \mathcal{A}$, where $\mathcal{A}$ is a set of actions,
	\item the environment reacts to the action and transitions from $s_t$ to state $s_{t+1}\in \mathcal{S}$,
	\item the agent is fed with a reward $r_{t} \in \mathcal{R}$, where $\mathcal{R}$ is a set of rewards, and a new observation $w_{t+1}$,
\end{itemize}
This repeats until some termination state is reached, the succession of states and actions defining a trajectory $\tau = \big( s_0, a_0, s_1, a_1, ... \big)$. In any given state, the objective of the agent is to determine the action maximizing its cumulative reward over an episode, i.e., over one instance of the scenario in which the agent takes actions. 
Most often, the quantity of interest is the discounted cumulative reward along a trajectory defined as
\bal
	R(\tau) = \displaystyle \sum_{t=0}^{T} \gamma^t r_{t}\,,
\eal
where $T$ is the horizon of the trajectory, and $\gamma\in[0,1]$ is a discount factor that weighs the relative importance of present and future rewards (the agent being short-sighted in the limit where $\gamma\rightarrow0$, since it then cares solely about the first reward, and far-sighted in the limit where $\gamma\rightarrow1$, since it then cares equally about all rewards).

There exist two main types of RL algorithms, namely model-based methods, in which the agent tries to build a model of how the environment works 
to make predictions about what the next state and reward will be before taking any action, and 
model-free methods, in which the agent conversely interacts with the environment without trying to understand it, and are prominent in the DRL community.
Another important distinction to be made within model-free algorithms is that between value-based methods, in which the agent learns to predict the future reward of taking an action when provided a given state, then selects the maximum action based on these estimates, and policy-based methods, in which it optimizes the expected reward of a decision policy mapping states to actions.
Many of the most successful algorithms in DRL 
(including proximal policy optimization, whose assessment for flow control and optimization purposes is the primary motivation for this research) 
proceed from policy gradient methods, in which gradient ascent is used to optimize a parameterized policy with respect to the expected return, as further explained in the next section. 
The reader interested in a more thorough introduction to the zoology of RL methods (together with their respective pros and cons) is referred to Ref.~\cite{sutton2018reinforcement}.

\subsection{From policy methods to Proximal policy optimization}\label{subsection:toppo}

This section intended for the non-specialist reader briefly reviews the basic principles and assumptions of policy gradient methods, together with the various steps taken for improvement.\\

\paragraph*{- Policy methods}A policy method maximizes the expected 
discounted cumulative reward of a decision policy mapping states to actions. It resorts not to a value function, but to a probability distribution over actions given states, that fully defines the behavior of the agent. Since policies are most often stochastic, the following notations are introduced:
\begin{itemize} 
\item $\pi(s,a)$ is the probability of taking action $a$ in state $s$ under policy $\pi$,

\item $Q^\pi(s,a)$ is the expected value of the return of the policy after taking action $a$ in state $s$ (also termed 
state-action value function or Q-function)
\bal
Q^\pi (s,a) = \mathbb{E}_{{\pi}} \big[ R(\tau) \vert s, a \big]\,,
\eal
{where we use $\mathbb{E}_{\pi}$ for the expected value $\mathbb{E}$ under policy $\pi$.}
\item $V^\pi(s)$ is the expected value of the return of the policy in state $s$ (also termed value function or V-function)
\bal
V^\pi (s) =  \mathbb{E}_{{\pi}} \big[ R(\tau) \vert s \big]\,.
\eal
%
The V and Q functions are therefore such that 
\bal
V^\pi (s) = \sum_a\pi(s,a)Q^\pi(s,a)\,,
\eal
so $V^\pi (s)$ can also be understood as the probability-weighted average of discounted cumulated rewards over all possible actions in state $s$. 
\end{itemize}

\paragraph*{- Policy gradient methods}A policy gradient method aims at optimizing 
a parametrized policy $\pi_\theta$, where $\theta$ denotes the free parameters whose value can be learnt from data (as opposed to the hyper parameters). In practice, one   
defines an objective function based on the expected discounted cumulative reward
\bal
J(\theta) =\mathbb{E}_{{\pi_\theta}} \big[ R(\tau) \big]\,,
\eal
and seeks the parameterization $\theta^*$ maximizing $J(\theta)$, hence such that
\bal
\theta^* = \arg \max_\theta\mathbb{E}_{{\pi_\theta}} \big[ R(\tau) \big]\,,    
\eal
which can be done on paper by plugging an estimator of the policy gradient $\nabla_\theta J(\theta)$ into a gradient ascent algorithm.
This is no small task as one is looking for the gradient with respect to the policy parameters, in a context where the effects of policy changes on the state distribution are unknown (since modifying the policy will most likely modify the set of visited states, which will in turn affect performance in some indefinite manner).
One commonly used estimator, derived in~\cite{sutton2018reinforcement} using the log-probability trick, reads
\bal
\label{eq:policy_gradient}
\nabla_\theta J(\theta) &=  \mathbb{E}_{{\pi_\theta}}\left[ \sum_{t=0}^T \nabla_\theta \log \left( \pi_\theta (s_t, a_t) \right) R(\tau) \right]\sim \mathbb{E}_{{\pi_\theta}}\left[ \sum_{t=0}^T \nabla_\theta \log \left( \pi_\theta (s_t, a_t) \right) {\widehat{A}^{\pi}}(s_t,a_t) \right]\,,
\eal
where {$\widehat{A}^{\pi}$ is some biased estimator (here its normalization to zero mean and unit variance) of the advantage function}
\bal
A^{\pi}(s,a) = Q^{\pi} (s,a) - V^{\pi} (s)\,,
\eal
that measures the improvement (if $A{^{\pi}}>0$, otherwise the lack thereof) associated with taking action $a$ in state $s$ compared to taking the average over all possible actions. This is because the value function does not depend on $\theta$, so taking it off changes neither the expected value, nor the gradient, but it does reduce the variance, and speeds up the training. Furthermore, when the policy $\pi_\theta$ is represented by a neural network (in which case $\theta$ simply denotes the network weights and biases to be optimized), the focus is rather on the policy loss defined as
\bal
\label{eq:gradient_loss}
L(\theta) = \mathbb{E}_{{\pi_\theta}}\left[ \sum_{t=0}^T \log \left( \pi_\theta (a_t \vert s_t) \right) 
{\widehat{A}}(s_t,a_t)\right]\,,
\eal 
whose gradient is equal to the (approximated) policy gradient \myrefeq{eq:policy_gradient} (since the gradient operator acts only on the log-policy term, not on the advantage) and is computed with respect to each weight and bias by the chain rule, one layer at the time, using the back-propagation algorithm~\cite{Rumelhart1986}.\\

\paragraph*{- Trust regions}The performance of policy gradient methods is hurt by the high 
sensitivity to the learning rate, i.e., the size of the step to be taken in the gradient direction. Indeed, small learning rates are detrimental to learning, but large learning rates can lead to a performance collapse if the agent falls off the cliff and restarts from a poorly performing state with a locally bad policy. This is all the more harmful as the learning rate cannot be tuned locally, meaning that an above average learning rate will speed up learning in some regions of the parameter space where the policy loss is relatively flat, but will possibly trigger an exploding policy update in other regions exhibiting sharper variations. One way to ensure continuous improvement is by imposing a trust region constraint to limit the difference between the current and updated policies, which can be done by determining first a maximum step size relevant for exploration, then by locating the optimal point within this trust region. We will not dwell on the intricate details of the many algorithms developed to solve such trust region optimization problems, e.g., natural policy gradient (NPG~\cite{Kakade01}), or trust region policy optimization (TRPO~\cite{Schulman2015}).
Suffice it to say that they use the minorize-maximization algorithm to maximize iteratively a surrogate policy loss (i.e. a lower bound approximating locally the actual loss at the current policy), but are difficult to implement and can be computationally expensive, as they rely on an estimate of the second-order gradient of the policy log probability.\\

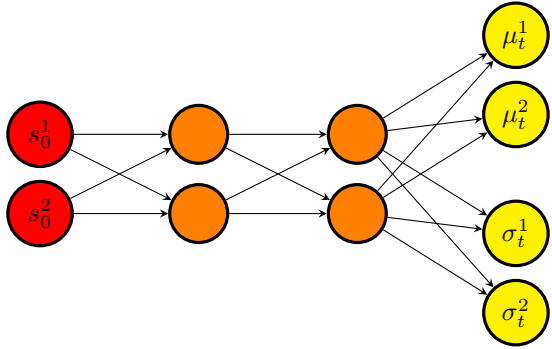
\begin{figure}[!t]
\centering

\begin{tikzpicture}[	arrow/.style=		{thick,color=mybluegray1,rounded corners},
				netnode/.style=		{circle, inner sep=0pt, text width=22pt, align=center, very thick},
				inputnode/.style=	{netnode, fill=red, draw=black},
				hiddennode/.style=	{netnode, fill=orange, draw=black},
				outputnode/.style=	{netnode, fill=yellow, 	draw=black},
				signal/.style=		{arrows={-stealth},draw=black}]
	\def\nodedist{30pt}
	\def\layerdist{60pt}
    
	\foreach \y in {1,...,2}
		\node[inputnode] (I\y) at (0,-\y*\nodedist) {$s^\y_0$};  
	\foreach \y in {1,...,2}
		\node[hiddennode] (H1\y) at (\layerdist,-\y*\nodedist) {};
	\foreach \y in {1,...,2}
		\node[hiddennode] (H2\y) at (2*\layerdist,-\y*\nodedist) {};
	\foreach \y in {1,...,2}
		\node[outputnode] (O\y) at ($(H21) + (\layerdist, -\y*\nodedist) +(0, 2.25*\nodedist)$) {$\mu^\y_t$};
	\foreach \y in {3,...,4}
		\pgfmathsetmacro\idx{int(\y-2)}
		\node[outputnode] (O\y) at ($(H21) + (\layerdist, -\y*\nodedist) +(0, 1.75*\nodedist)$) {$\sigma^{\idx}_t$};

	\foreach \dest in {1,...,2}
		\foreach \source in {1,...,2}
			\draw[signal] (I\source) -- (H1\dest);
	\foreach \dest in {1,...,2}
		\foreach \source in {1,...,2}
			\draw[signal] (H1\source) -- (H2\dest);
	\foreach \dest in {1,...,4}
		\foreach \source in {1,...,2}
			\draw[signal] (H2\source) edge (O\dest);
\end{tikzpicture}

\caption{Agent network example used to map states to policy. The input state $\bm{s_0}$, here of size 2, is mapped to a mean $\bm{\mu}$ and a standard deviation $\bm{\sigma}$ vectors, each of size 2. All activation functions are ReLu, except for that of the last layer, which are linear for the $\mu$ output, and softplus for the $\sigma$ output. Orthogonal weights initialization is used throughout the network.}
\label{fig:network_PPO}
\end{figure}

\paragraph*{- Proximal policy optimization}Proximal policy optimization (PPO) is another 
approach with simple and effective heuristics, that uses a probability ratio between the two policies
to maximize improvement without the risk of performance collapse~\cite{Schulman2017}. The focus here is on the PPO-clip algorithm\footnote{There is also a PPO-Penalty variant  which uses a penalization on the average Kullback--Leibler divergence between the current and new policies, but PPO-clip performs better in practice.}, that optimizes the surrogate loss 
\bal
\label{eq:ppo_loss}
L(\theta) = \mathbb{E}_{{\pi_\theta}} \left[ \min \left( \frac{\pi_\theta (a \vert s)}{\pi_{\theta_{old}} (a \vert s)} , g( \epsilon, \widehat{A}^{\pi} (s,a)) \right)\widehat{A}^{\pi} (s,a) \right]\,,
\eal
where 
\bal
g(\epsilon, A) = 
\begin{cases}
1+\epsilon 	& A \geq 0\,,\\
1-\epsilon 	& A < 0\,,
\end{cases}
\eal
and $\epsilon\in\left[0.1, 0.3\right]$ is the clipping range, a small hyper parameter defining how far away the new policy is allowed to go from the old.
{The general picture is that a positive (resp. negative) advantage increases (resp. decreases) the probability of taking action $a$ in state $s$, but always by a proportion smaller than $\epsilon$, otherwise the min kicks in~\myrefeq{eq:ppo_loss} and its argument hits a ceiling of $1+\epsilon$ (resp. a floor of $1-\epsilon $).  This prevents stepping too far away from the current policy, and ensures that the new policy will
behave similarly. }

{There exist more sophisticated PPO algorithms (e.g., Trust region PPO~\cite{Wang2019}, that determines first a maximum step size relevant for exploration, then adaptively adjusts the clipping range to
find the optimal within this trust region), but standard PPO has simple and effective heuristics. Namely, it is computationally inexpensive, easy to implement (as only the first-order gradient of the policy log probability is needed to calculate the clipped surrogate), and remains regarded as one of the most successful RL algorithms, achieving} state-of-the-art performance across a wide range of challenging tasks.

\subsection{Single-step PPO} 

\begin{figure}[t!]
\centering
\begin{tikzpicture}[	every loop/.style={	min distance=20mm,looseness=20},
				fontsize/.style={		font=\footnotesize},
				intnode/.style={		black, fontsize, pos=0.5},
				arrow/.style={		thick,color=black, rounded corners,thick,-stealth},
				box/.style={		rectangle,rounded corners, draw=mygray1, very thick, 
								text width=1.5cm,minimum height=0.75cm,text centered,
								inner sep=2pt, outer sep=0pt, fontsize},
				backbox/.style={	box, opacity=0.5,text width=8cm,minimum height=4cm}]

	\node[backbox]			(bb)		at (6,0) {};
	\node[box,fill=red] 	(s0) 		at (0,0) {$\bm{s}_0$};
	\node[box,fill=orange] (agent) 	at (4,0) {Agent};
	\node[box,fill=yellow] 	(env) 	at (8,0) {Parallel envs.};
	
	\draw[arrow] 	(s0) 			to 	[out=0,in=180] 						 								           (agent.west);
	\draw[arrow] 	(agent) 		to 	[out=0,in=180] 				node[intnode, above] {$\bm{a}_t$} 				           (env.west);
	\draw[arrow] 	(env.south) 	to 	[out=-110,in=-80] 			node[intnode, below] {$\bm{r}_t$} 						   (agent.south);
	\draw[] 		(agent.north) 	edge	[out=150,in=30, loop,arrow] 	node[intnode, above] {$\,\,\bm{\theta}_t \to \bm{\theta}_{t+1}$} (agent.north);
\end{tikzpicture}
\caption{Action loop for single-step PPO. At each episode, the input state $\bm{s}_0$ is provided to the agent, which in turn provides $n$ actions to $n$ parallel environments. The latter return $n$ rewards, that evaluate the quality of each action taken. Once all the rewards are collected, an update of the agent parameters is made using the PPO loss (\ref{eq:ppo_loss}).}
\label{fig:single_step_PPO}
\end{figure}
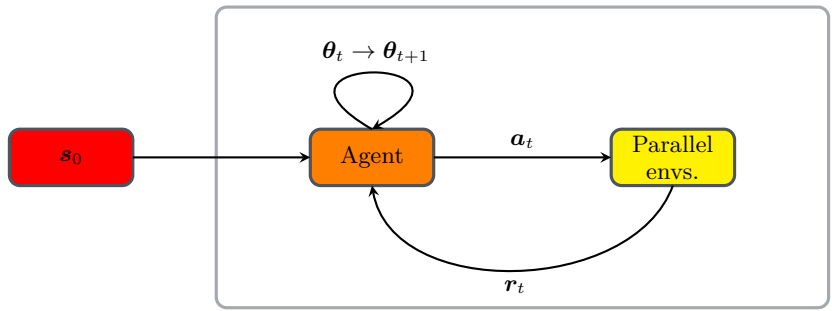

{We now come to single-step PPO, a ``degenerate'' version of PPO {introduced in~\cite{viquerat2019direct} and} intended for situations where the optimal policy to be learnt by the neural network is state-independent, as is notably the case in optimization and open-loop control problems (closed-loop control problems conversely require state-dependent policies for which standard PPO is best suited). The main difference between {standard and single-step} PPO can be summed up as follows: where standard PPO seeks the optimal set of actions $a^\star$ yielding the largest possible reward, single-step PPO seeks the optimal mapping $f_{\theta^\star}$ such that $a^\star = f_{\theta^\star} (s_0)$, where $\theta$ denotes the network free parameters and $s_0$ is some input state (usually a vector of zeros) consistently fed to the agent for the optimal policy to eventually embody the transformation from $s_0$ to $a^\star$. The agent initially implements a random state-action mapping $f_{\theta_0}$ from $s_0$ to an initial policy determined by the free parameters initialization $\theta_0$, after which it gets only one attempt per learning episode at finding the optimal (i.e., it  interacts with the environment only once per episode). This is illustrated in figure~\ref{fig:single_step_PPO} showing the agent draw a population of actions $a_t= f_{\theta_t}(s_0)$ from the current policy, and being returned incentives from the associated rewards to update the free parameters for the next population of actions $a_{t+1} = f_{\theta_{t+1}}(s_0)$ to yield larger rewards.}

{In practice, the agent outputs a policy parameterized by the mean and variance of the probability density function of a $d$-dimensional multivariate normal distribution, with $d$ the dimension of the action required by the environment. Actions drawn in $[-1,1]^d$ are then mapped into relevant physical ranges, a step deferred to the environment as being problem-specific. The resolution essentially follows the process described in section~\ref{subsection:toppo}, only a normalized averaged reward substitutes for the advantage function. This is because classical PPO is actor-critic, i.e., it improves the learning performance by updating two different networks, a first one called actor that controls the actions taken by the agent, and a second one called critic, that learns to estimate the advantage from the value function as\bal
\label{eq:advdef}
A(s_t,a_t)=r_t + \gamma V(s_{t+1})-V(s_t)\,.
\eal
In single-step PPO, the trajectory consists of a single state-action pair, so the discount factor can be set to $\gamma=1$ with no loss of generality. In return, the advantage reduces to the whitened reward since the two rightmost terms cancel each other out in~\myrefeq{eq:advdef}. This means that the approach can do without the value-function evaluations of the critic network, i.e., it is not actually actor-critic.}

\subsection{{Numerical implementation}} 

The present workflow relies on the online PPO implementation of Stable Baselines, a toolset of reinforcement learning algorithms dedicated to the research community and industry~\cite{stable-baselines}, for which a custom OpenAI environment has been designed using the Gym library~\cite{1606.01540}. 
{Hyperbolic tangent is used as default activation function.}
The instant reward $r_t$ used to train the neural network is simply the quantity subjected to optimization (modulo a plus or minus sign to tackle both maximization and minimization problems). A moving average reward is also computed on the fly as the sliding average over the $100$ latest values of $r_t$ (or the whole sample if {it has insufficient size)}.
All other relevant hyper parameters are documented in the next sections, with the exception of the discount factor {(set to $\gamma=1$).}

{In practice, actions are distributed to multiple environments running in parallel, each of which executes a self-contained MPI-parallel CFD simulation and feeds data to the DRL algorithm  (hence, two levels of parallelism related to the environment and the computing architecture). The algorithm waits for the simulations running in all parallel environments to be completed, then shuffles and splits the rewards data set collected from all environments into several buffers (or mini-batches) used sequentially to compute the loss and perform a network update. 
The process repeats for several epochs, i.e., several full passes of the training algorithm over the entire data set (so the policy network ends up being trained on samples generated by older policies, which is customary in standard PPO operation).
This simple parallelization technique is key to use DRL in the context of CFD applications, as a sufficient number of actions drawn from the current policy must be evaluated to accurately estimate the policy gradient.
This comes at the expense of computing the same amount of reward evaluations, and yields a substantial computational cost for high-dimensional fluid dynamics problems (typically from 
a few tens to several thousand hours for the steady-state optimization
problems considered herein). In the same vein, it should be noted that the common practice in DRL studies to gain insight into the performances of the selected algorithm by averaging results over multiple independent training runs with different random seeds is not tractable, as it would trigger a prohibitively large computational burden. The same random seeds have thus been deliberately used over the whole course of study to ensure a minimal level of performance comparison between cases.}

\section{Control of natural convection in 2-D closed cavity}\label{section:natural} 

\subsection{Case description}

\begin{figure}[t!]
\setlength{\unitlength}{1cm}
\begin{picture}(20,4.5)
\put(4.3,-0.5){\includegraphics[trim=175 110 300 340pt,clip,height=5cm]{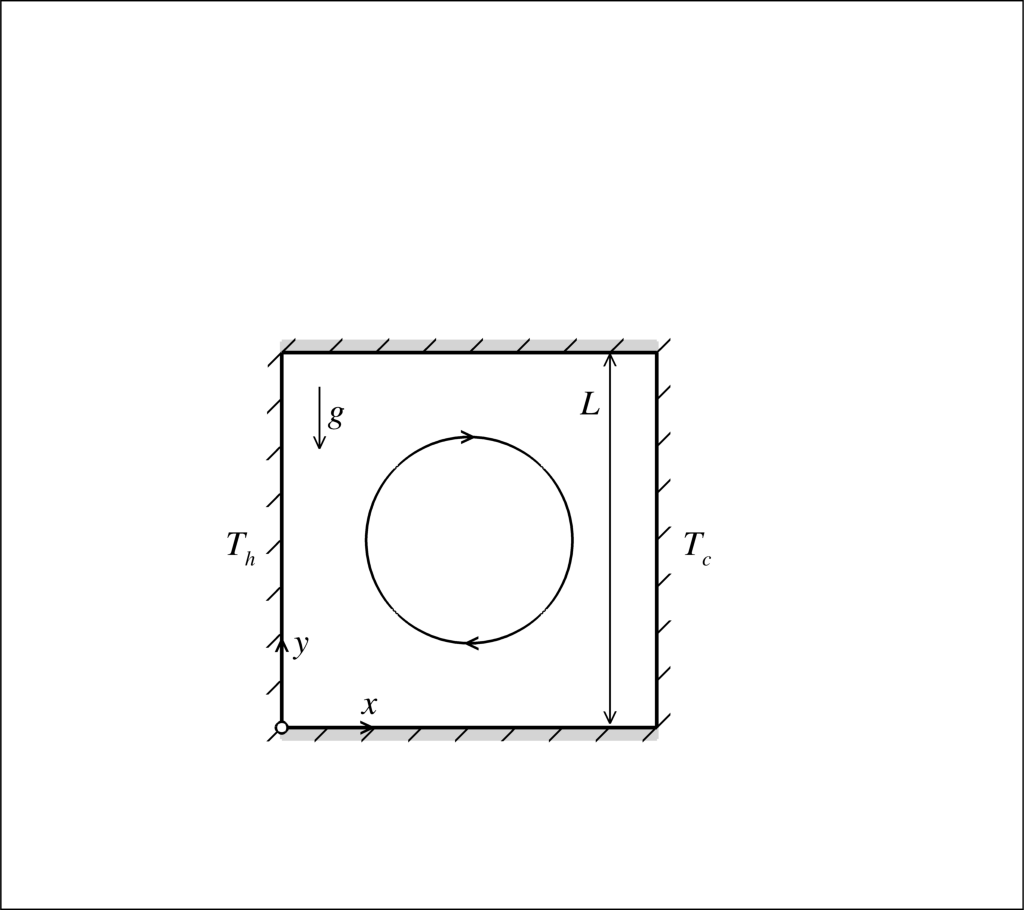}}
\end{picture}
\caption{Schematic of the two-dimensional Rayleigh--B\'enard set-up. }
\label{fig:natural_conf}
\end{figure}

We address first the control of natural convection in the two-dimensional differentially heated square cavity schematically illustrated in figure~\ref{fig:natural_conf}(a). This is a widely studied benchmark system for thermally-driven flows, relevant in nature and technical applications (e.g., ocean and atmospheric convection, materials processing, metallurgy), that is thus suitable to validate and compare numerical solution algorithms while enriching the knowledge base for future projects in this field. 
A Cartesian coordinate system is used with origin at the lower-left edge, horizontal $x$-axis, and vertical $y$-axis. The cavity has side $L$, its top and bottom horizontal walls are perfectly insulated from the outside, and the vertical sidewalls are isothermal. Namely, the right sidewall is kept at a constant, homogeneous ``cold'' temperature $T_c$, and the left sidewall is entirely controllable via a constant in time, varying in space ``hot'' distribution $T_h(y)$ such that 
\bal
\langle T_h\rangle>T_c\,,
\eal
where the brackets denote the average over space (here over the vertical position along the sidewall). 

In the following, we neglect radiative heat transfer ($\chi=0$) and consider a Boussinesq system driven by buoyancy, hence
\bal
\vecpsi=\rho_0\beta(T-T_c)g\eey\,,
\eal 
where $\vecg$ is the gravitational acceleration parallel to the sidewalls, $\beta$ is the 
thermal expansion coefficient, and we use the cold sidewall temperature as Boussinesq reference temperature.
By doing so, the pressure featured in the momentum equation~\myrefeq{eq:momentum} and related weak forms must be understood as the pressure correction representing the deviation from hydrostatic equilibrium.
The governing equations are solved with no-slip conditions $\uu=\00$ on $\partial\Omega$ and temperature boundary conditions
\bal
\partial_y T(x,0,t)=\partial_y T(x,L,t)=0\,,\qquad
T(0,y,t)=\langle T_h\rangle+\tilde{T}_h(y)\,,\qquad T(L,y,t)=T_c\,,
\eal
where $\tilde{T}_h$ is a zero-mean (in the sense of the average over space) distribution of hot temperature fluctuations subjected to optimization, whose magnitude is bounded by some constant $\Delta T_{max}$ according to
\bal
|\tilde{T}_h(y)|&\leq \Delta T_{max}\,,\label{eq:constraintrbc2}
\eal
to avoid extreme and nonphysical temperature gradients.
All results are made non-dimensional using the 
cavity side, the heat conductivity time, and the well-defined, constant in time difference between the averaged sidewall temperatures.
The retained fluid properties yield values of the Rayleigh and Prandtl numbers
\bal
\ray=\frac{g\beta (\langle T_h\rangle-T_c)L^3}{\nu \alpha}=10^4\,,
\qquad\qquad\pr=\frac{\nu}{\alpha}=0.71\,,
\eal
where $\alpha=\lambda/(\rho c_p)$ is the thermal diffusivity. 

\begin{figure}[t!]
\setlength{\unitlength}{1cm}
\begin{picture}(20,5.1)
\put(9.8,0){\includegraphics[trim=175 100 360 325pt,clip,height=5cm]{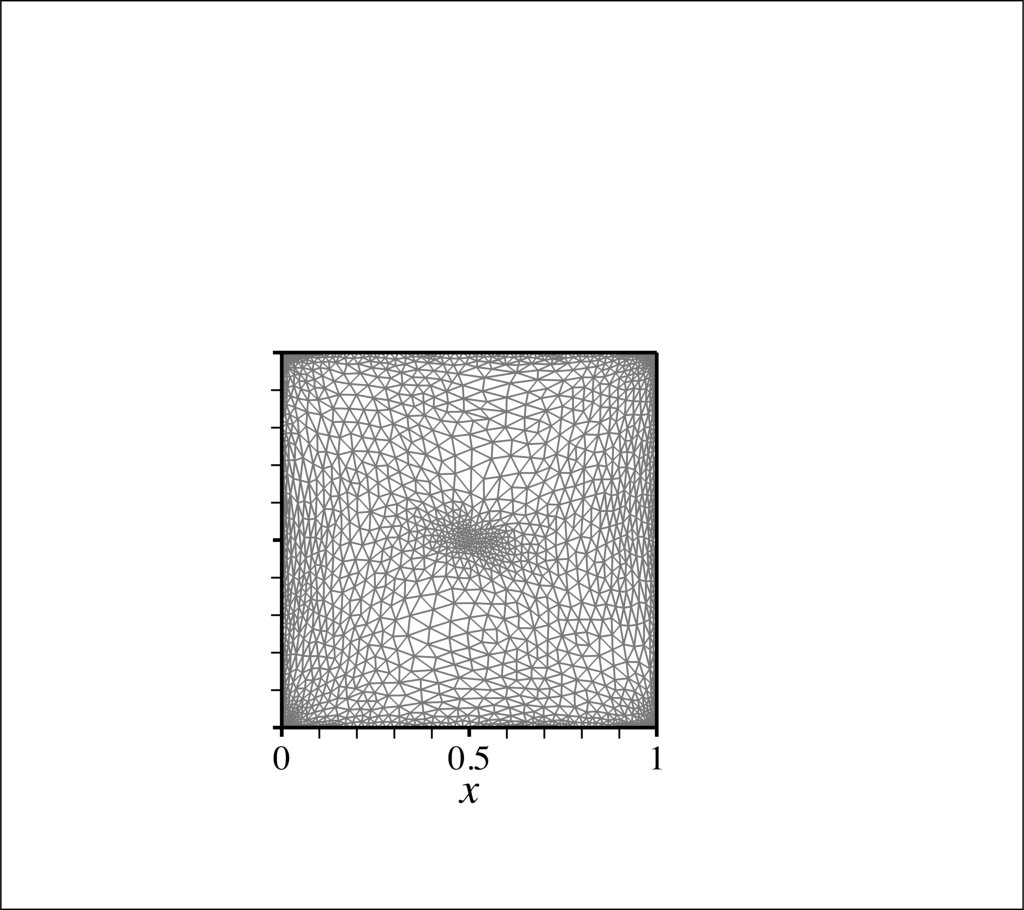}}
\put(4.9,0){\includegraphics[trim=175 100 360 325pt,clip,height=5cm]{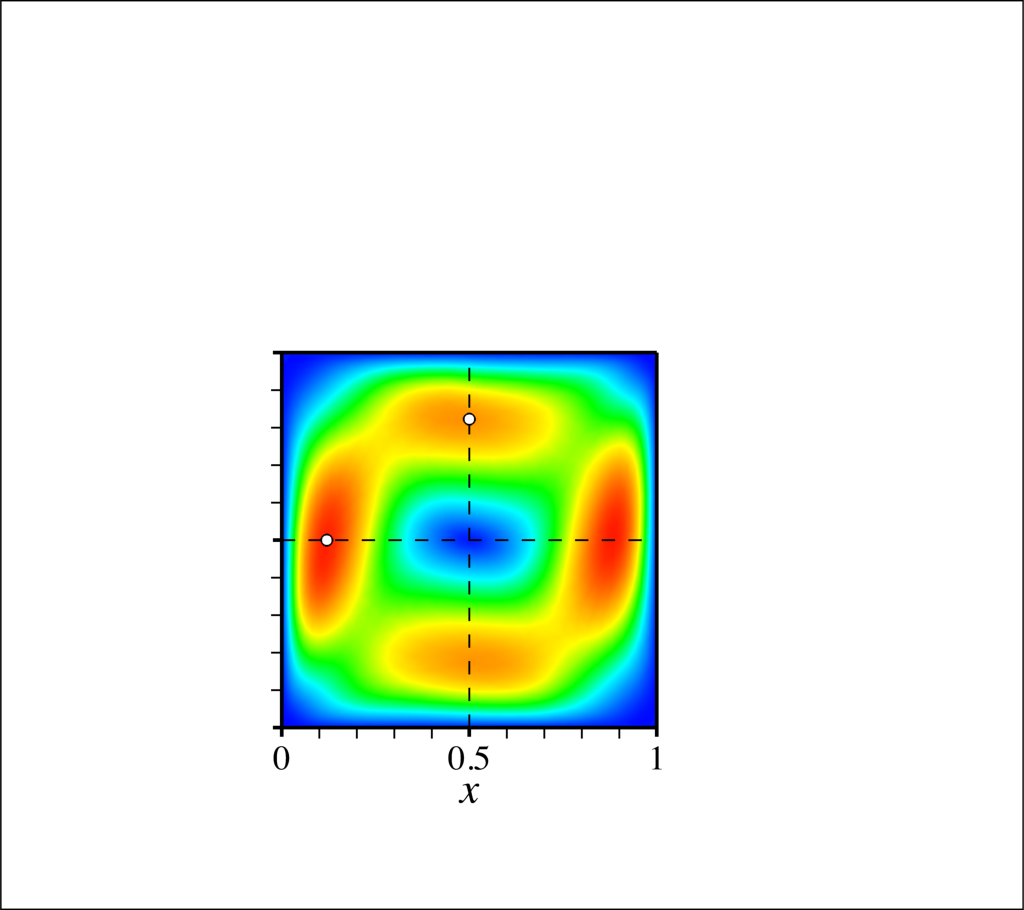}}
\put(0,0){\includegraphics[trim=175 100 360 325pt,clip,height=5cm]{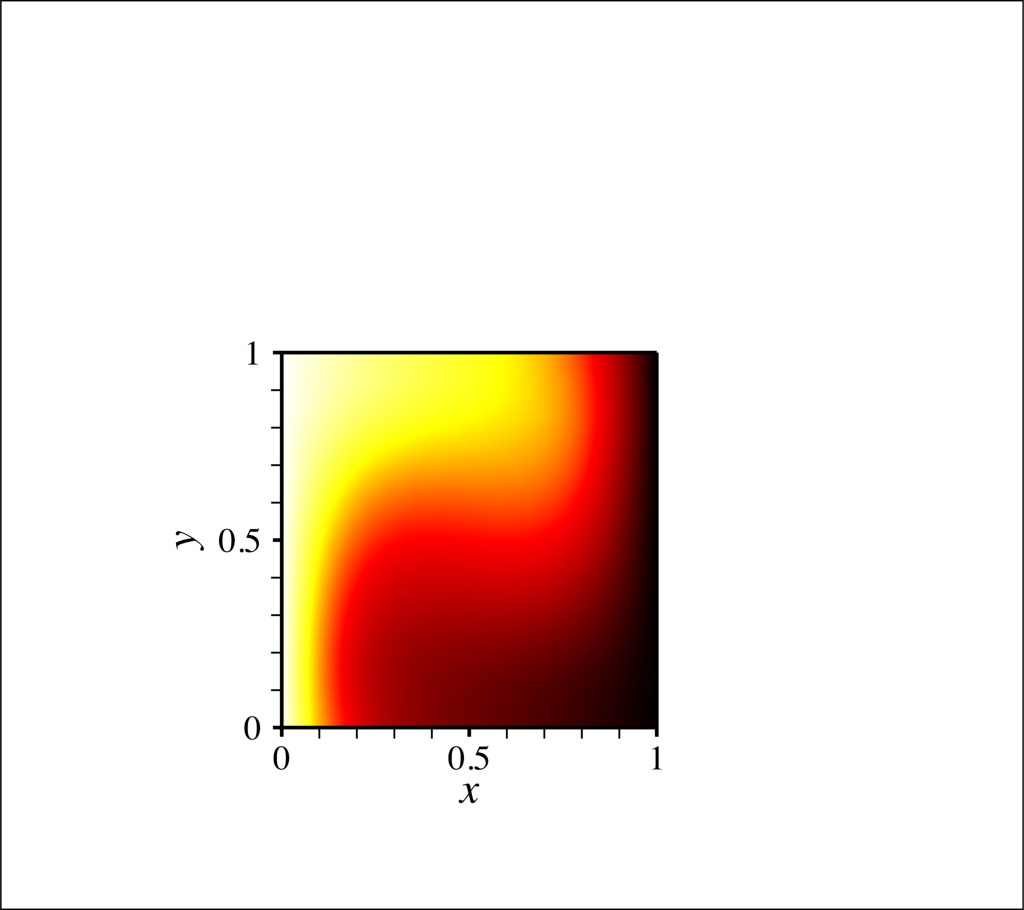}}
\put(3.25,4.8){\includegraphics[trim=420 635 440 245pt,clip,height=0.45cm]{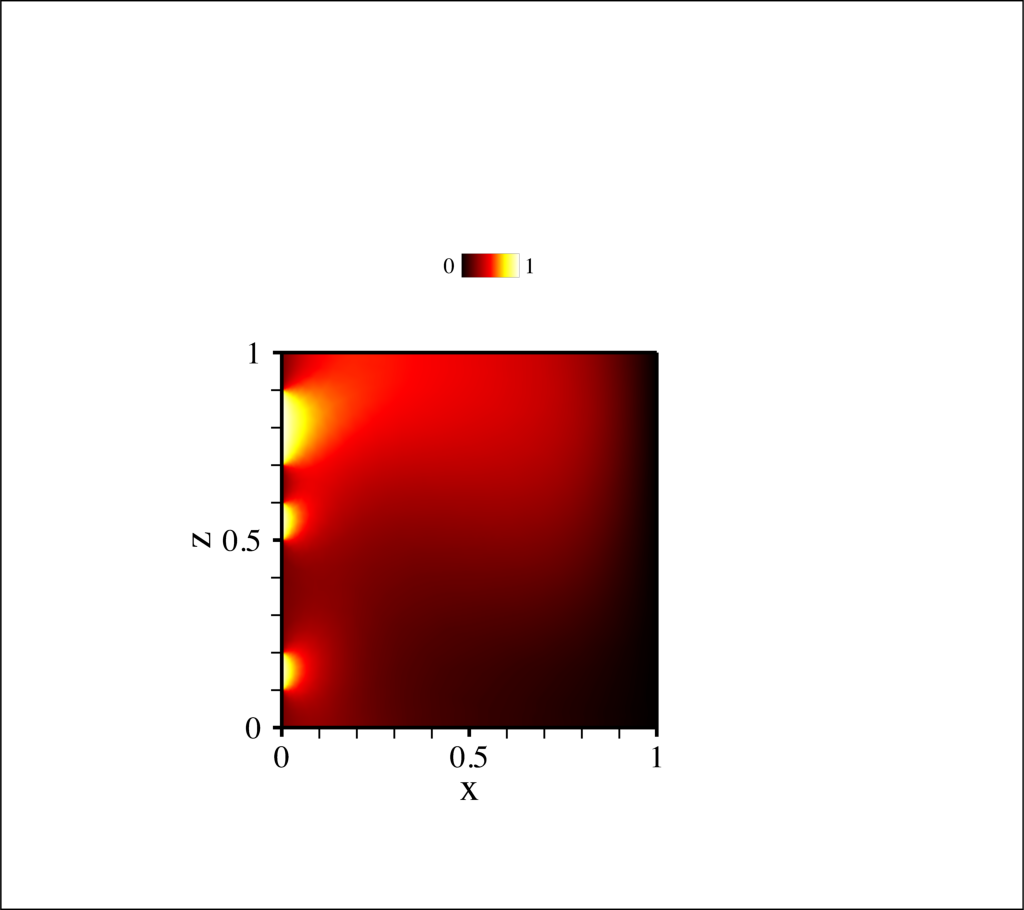}}
\put(8,4.8){\includegraphics[trim=420 635 440 245pt,clip,height=0.45cm]{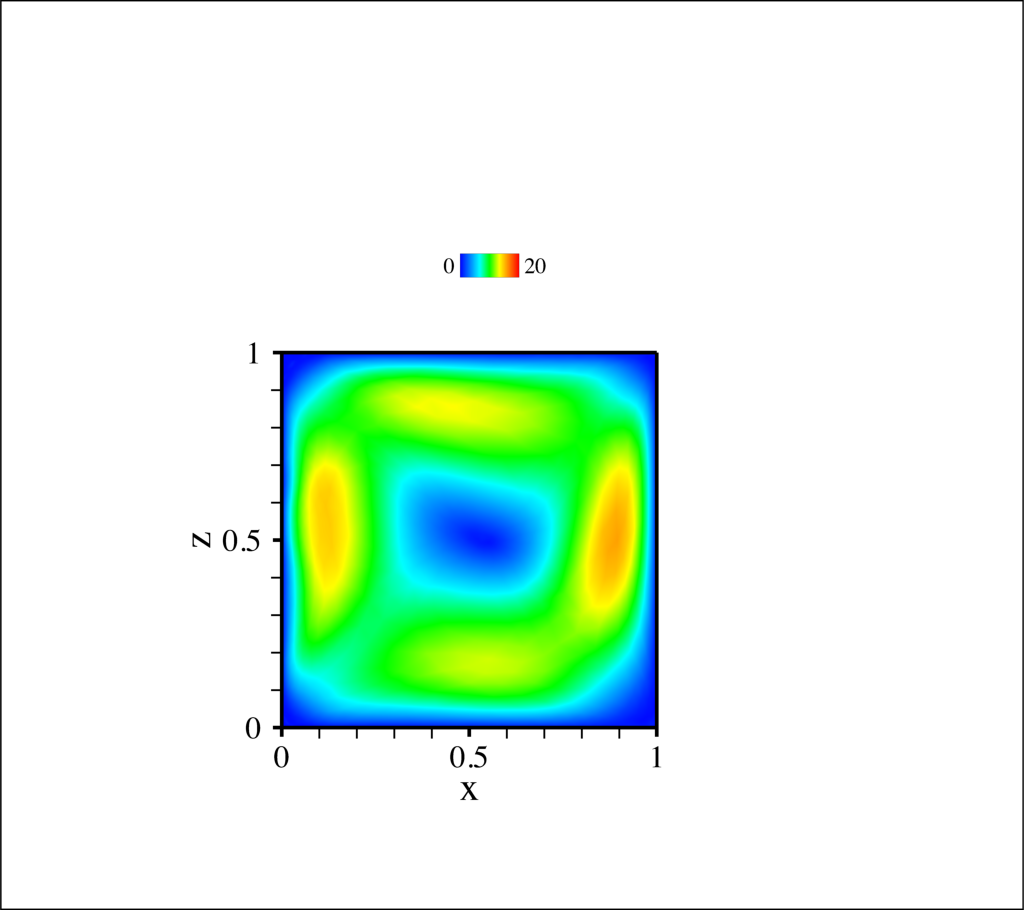}}
\put(0.35,5.1){(a)}
\put(5.3,5.1){(b)}
\put(10.2,5.1){(c)}
\end{picture}
\caption{Iso-contours of the uncontrolled steady state (a) temperature and (b) velocity magnitude. (c) Adapted mesh.  The circle symbols in (b) mark the positions of the maximum horizontal and vertical velocity along the centerlines reported in table~\ref{table:natural_valid}.}
\label{fig:natural_nocontrol_vtu}
\end{figure}
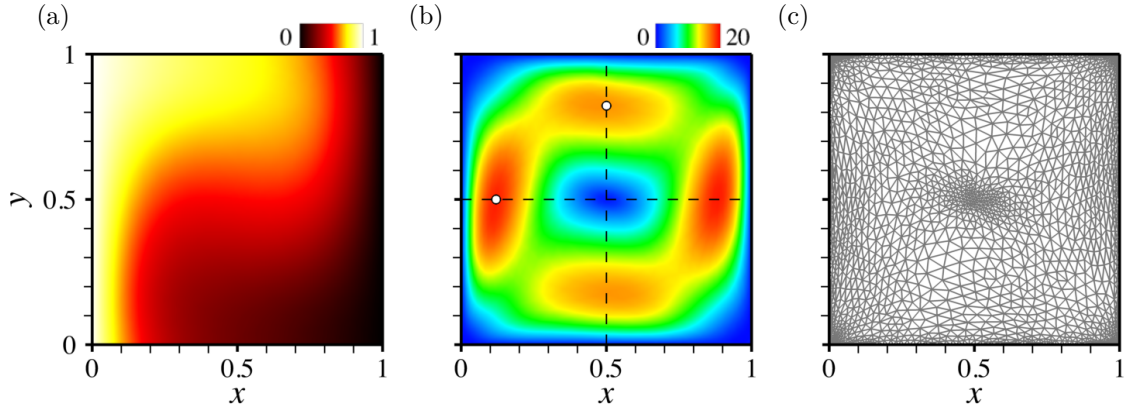

In order to assess the accuracy of the numerical framework, the uncontrolled solution has been computed by performing $60$ iterations with time step $\Delta t = 0.5$ to march the initial solution (consisting of zero velocity and uniform temperature, except at the hot sidewall) to steady state. At each time step, an initially isotropic mesh is adapted under the constraint of a fixed number of elements $n_{el}=4000$ using
a multiple-component criterion featuring velocity and temperature, but no level-set. 
This is because the case is heat transfer but not conjugate heat transfer, as the solid is solely at the boundary $\partial\Omega$ of the computational domain, where either the temperature is known, or the heat flux is zero. It is thus implemented without the IVM and without a level set (although accurate IVM numerical solutions have been obtained in~\cite{hach12} using thick sidewalls with high thermal conductivity). 
The solution shown in figure~\ref{fig:natural_nocontrol_vtu}(a,b) features a centered roll confined by the cavity walls, consistently with the fact that $\ray$ exceeds the critical value $\ray_c\sim 920$ for the onset of convection (as extrapolated from the near-critical benchmark data in~\cite{de1983natural})
by one order of magnitude, and heat transfer is thus driven by both conduction and convection.
This shows in the Nusselt number, i.e., the non-dimensional temperature gradient averaged over the hot sidewall
\bal
\nus = - \langle\partial_x T\rangle\,,
\eal
whose present value $\nus=2.27$ (as computed from $68$ points uniformly distributed along the sidewall) exceeds that $\nus=1$ of the purely conductive solution, and exhibits excellent agreement with benchmark results from the literature. This is 
evidenced in table~\ref{table:natural_valid} where we also report the magnitude and position of the maximum horizontal velocity $u$ (resp. the vertical velocity $v$) along the vertical centerline (resp. the horizontal centerline).
The corresponding adapted mesh shown in figure~\ref{fig:natural_nocontrol_vtu}(c) stresses that all boundary layers are sharply captured via extremely stretched elements, and that the adaptation strategy yields refined meshes near high temperature gradients and close to the side walls. Note however, the mesh refinement is not only along the boundary layers but also close to the recirculation regions near the cavity center, while the elements in-between are coarse and essentially isotropic.

\begin{table}[!t]
\begin{center}
\begin{tabular}{clcccccc}
\toprule
\multicolumn{3}{r}{Present} 
& {\centering \makecell{Ref.~\cite{de1983natural}}} 
& {\centering \makecell{Ref.~\cite{dixit2006simulation}}} 
& {\centering \makecell{Ref.~\cite{markatos1984}}} 
& {\centering \makecell{Ref.~\cite{barakos1994}}} 
& {\centering \makecell{Ref.~\cite{khanafer2003}}} 
\\
\cmidrule(lr){1-8}
\multicolumn{1}{l}{\multirow{5}{*}{$\ray=10^4$}}&
\multicolumn{1}{r}{$\nus\quad$}& \multicolumn{1}{r}{2.267} & \multicolumn{1}{r}{2.238} & \multicolumn{1}{r}{2.245} &\multicolumn{1}{r}{2.201} & \multicolumn{1}{r}{2.245} & \multicolumn{1}{r}{2.245}\\
&\multicolumn{1}{r}{$\max{u(0.5,y)}\quad$}& \multicolumn{1}{r}{16.048} & \multicolumn{1}{r}{16.178} & \multicolumn{1}{r}{16.179} &\multicolumn{1}{r}{--} & \multicolumn{1}{r}{16.262} & \multicolumn{1}{r}{16.178}\\
&\multicolumn{1}{r}{$y_{max}\quad$}& \multicolumn{1}{r}{0.823} & \multicolumn{1}{r}{0.823} & \multicolumn{1}{r}{0.824} &\multicolumn{1}{r}{0.832} & \multicolumn{1}{r}{0.818} & \multicolumn{1}{r}{0.827}\\
&\multicolumn{1}{r}{$\max{v(x,0.5)}\quad$}& \multicolumn{1}{r}{19.067} & \multicolumn{1}{r}{19.617} & \multicolumn{1}{r}{19.619} &\multicolumn{1}{r}{--} & \multicolumn{1}{r}{19.717} & \multicolumn{1}{r}{19.633}\\
&\multicolumn{1}{r}{$x_{max}\quad$}& \multicolumn{1}{r}{0.120} & \multicolumn{1}{r}{0.119} & \multicolumn{1}{r}{0.121} &\multicolumn{1}{r}{0.113} & \multicolumn{1}{r}{0.119} & \multicolumn{1}{r}{0.123}\\
\bottomrule
\end{tabular}
\caption{Comparison of the present numerical results in the absence of control with reference benchmark solutions from the literature.}
\label{table:natural_valid}
\end{center}	
\end{table}

\subsection{Control}

The question now being raised is whether DRL can be used to find a distribution of temperature fluctuations $\tilde{T}_h$ capable of alleviating convective heat transfer. 
To do so, we follow~\cite{beintema2020controlling} and train a DRL agent in selecting piece-wise constant temperature distributions over $n_s$ identical segments,
each of which allows only two pre-determined states referred to as hot or cold. This is intended to reduce the complexity and the computational resources, as large/continuous action spaces are  known to be challenging for the convergence of RL methods~\cite{lazaric2008,lee2018}.
Simply put, the network action output consists of $n_s$ values $\hat{T}_{hk\in\{1\dots n_s\}}=\pm \Delta T_{max}$,
mapped into the actual fluctuations according to
\bal
\tilde{T}_{hk}=\frac{\hat{T}_{hk}-\langle\hat{T}_{hk}\rangle}{\max_l\{1,\displaystyle\frac{|\hat{T}_{hl}-\langle\hat{T}_{hl}\rangle|}{\Delta T_{max}}\}}\,,
\label{eq:normal_natural_conv}
\eal
to fulfill the zero-mean and upper bound constraints.\footnote{Another possible approach would have been to penalize the reward passed to the DRL for those temperature distributions deemed non-admissible (either because the average temperature is non-zero or the temperature magnitude is beyond the threshold). However, this would have made returning admissible solutions part of the tasks the network is trained on (not to mention that non-zero average temperatures amount to a change in the Rayleigh number), which would likely have slowed down learning substantially.}
Ultimately, the agent receives the reward $r_t=-\nus$ to minimize the space averaged heat flux at the hot sidewall.

\begin{figure}[t!]
\setlength{\unitlength}{1cm}
\begin{picture}(20,18.3)
\put(9.8,13.2){\includegraphics[trim=175 100 360 325pt,clip,height=5cm]{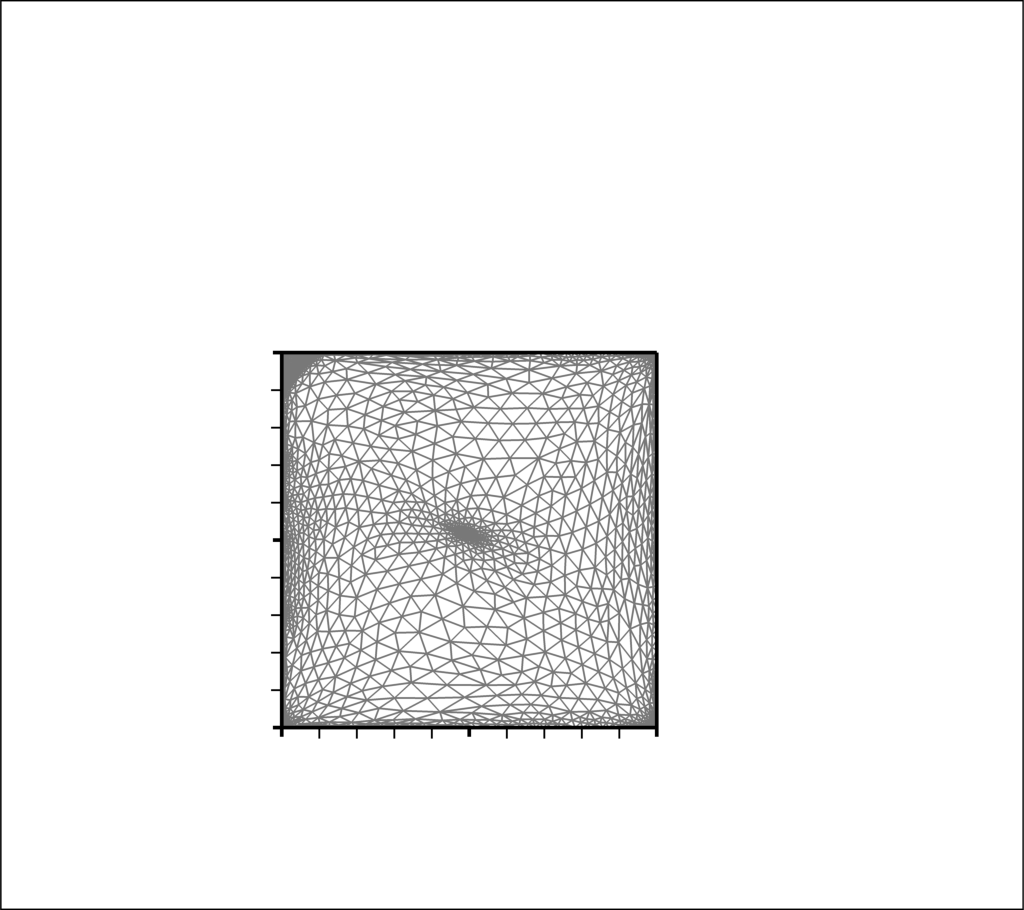}}
\put(4.9,13.2){\includegraphics[trim=175 100 360 325pt,clip,height=5cm]{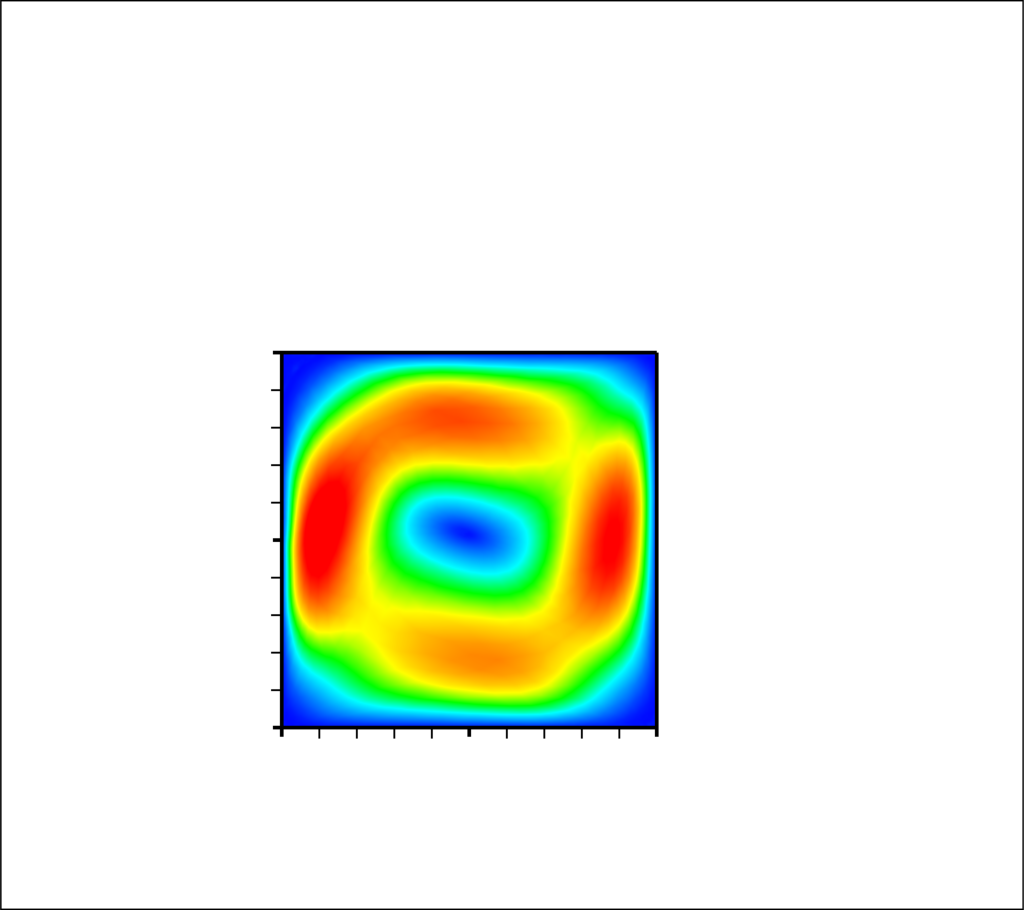}}
\put(0,13.2){\includegraphics[trim=175 100 360 325pt,clip,height=5cm]{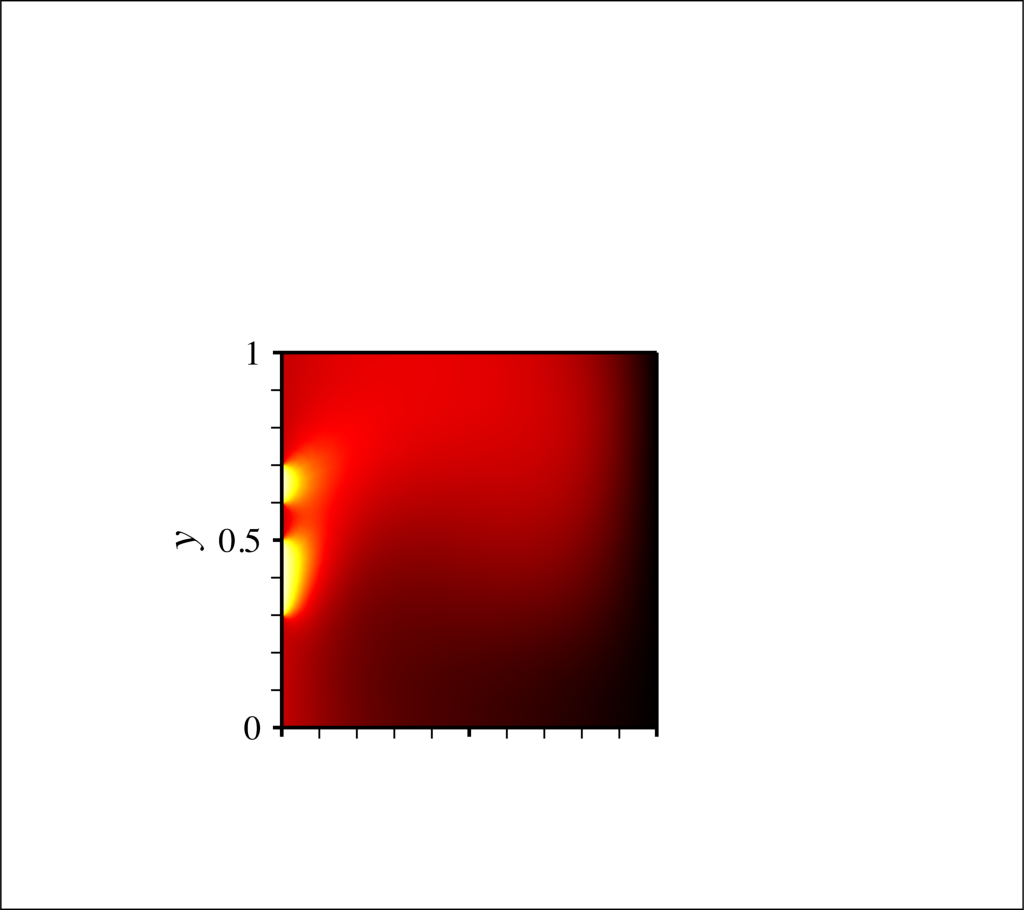}}
\put(9.8,8.8){\includegraphics[trim=175 100 360 325pt,clip,height=5cm]{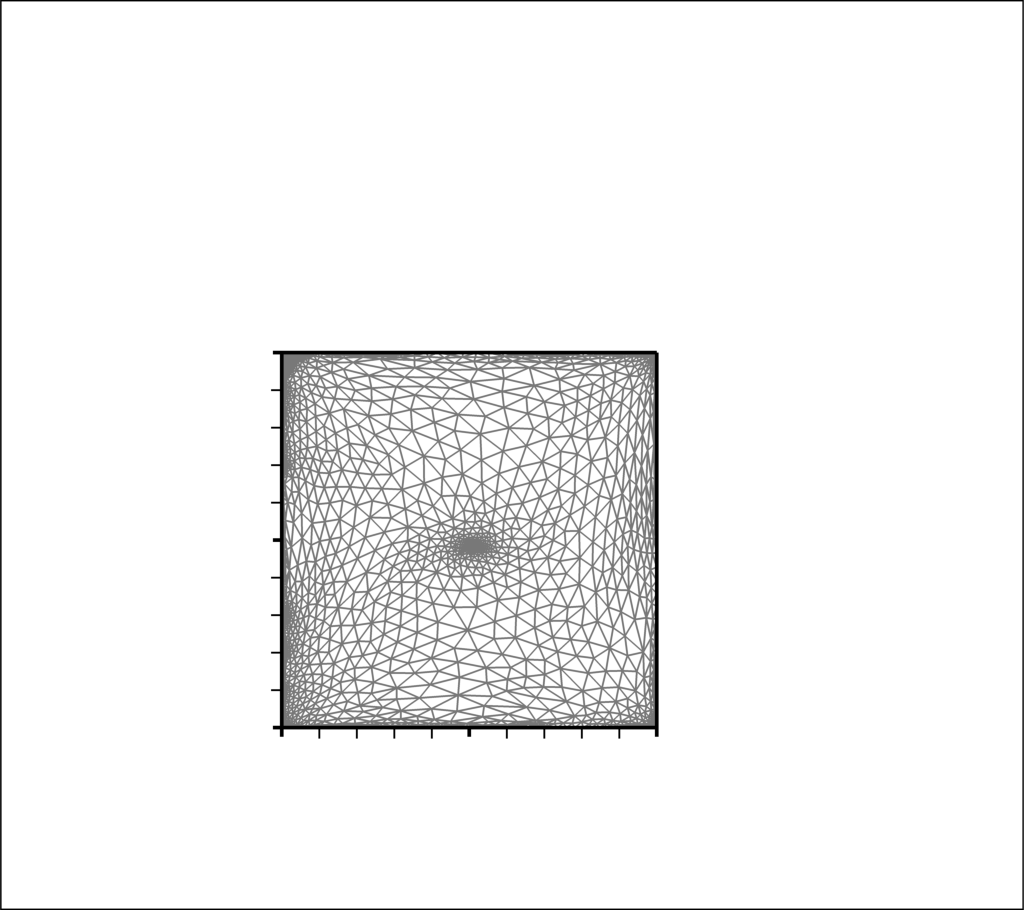}}
\put(4.9,8.8){\includegraphics[trim=175 100 360 325pt,clip,height=5cm]{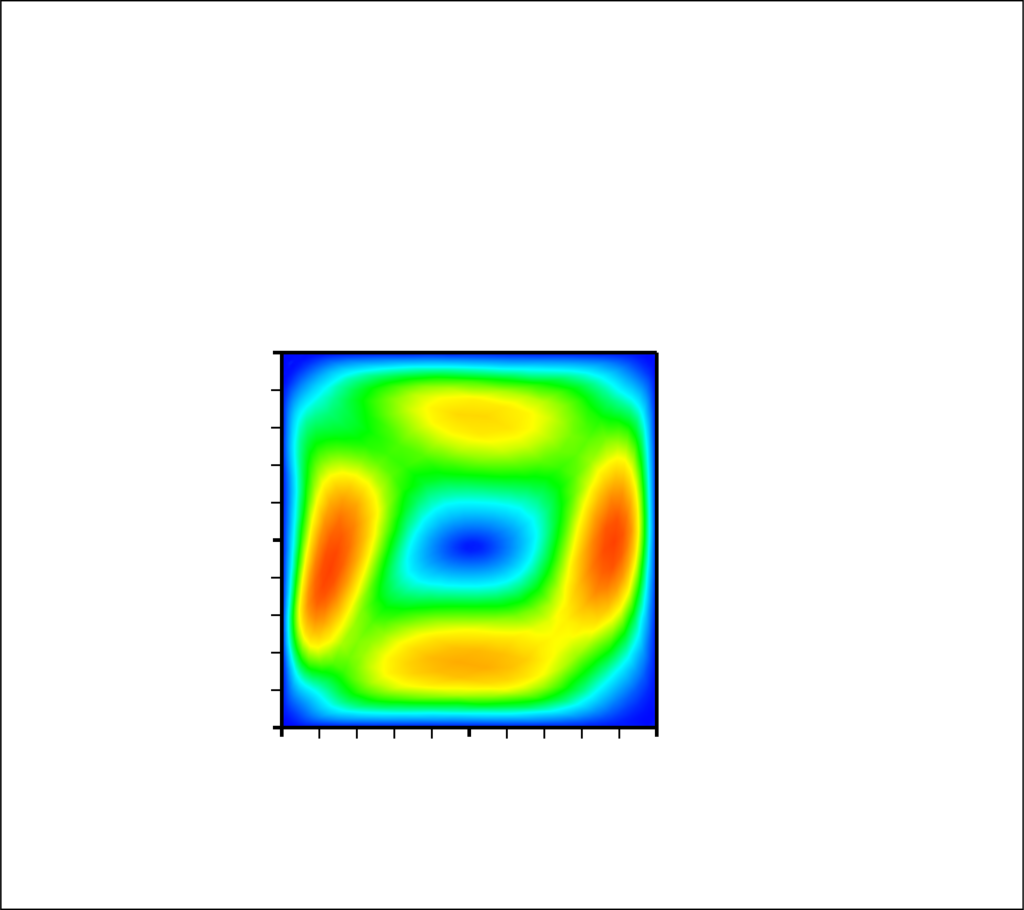}}
\put(0,8.8){\includegraphics[trim=175 100 360 325pt,clip,height=5cm]{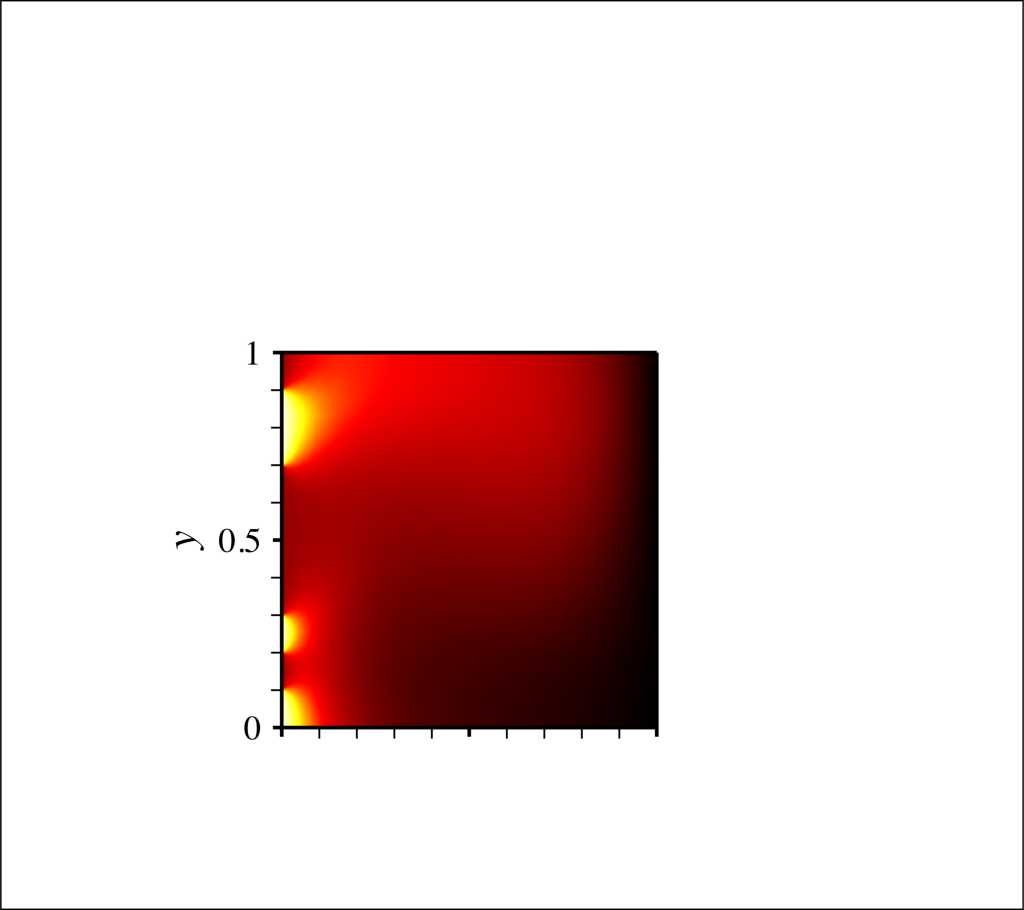}}
\put(9.8,4.4){\includegraphics[trim=175 100 360 325pt,clip,height=5cm]{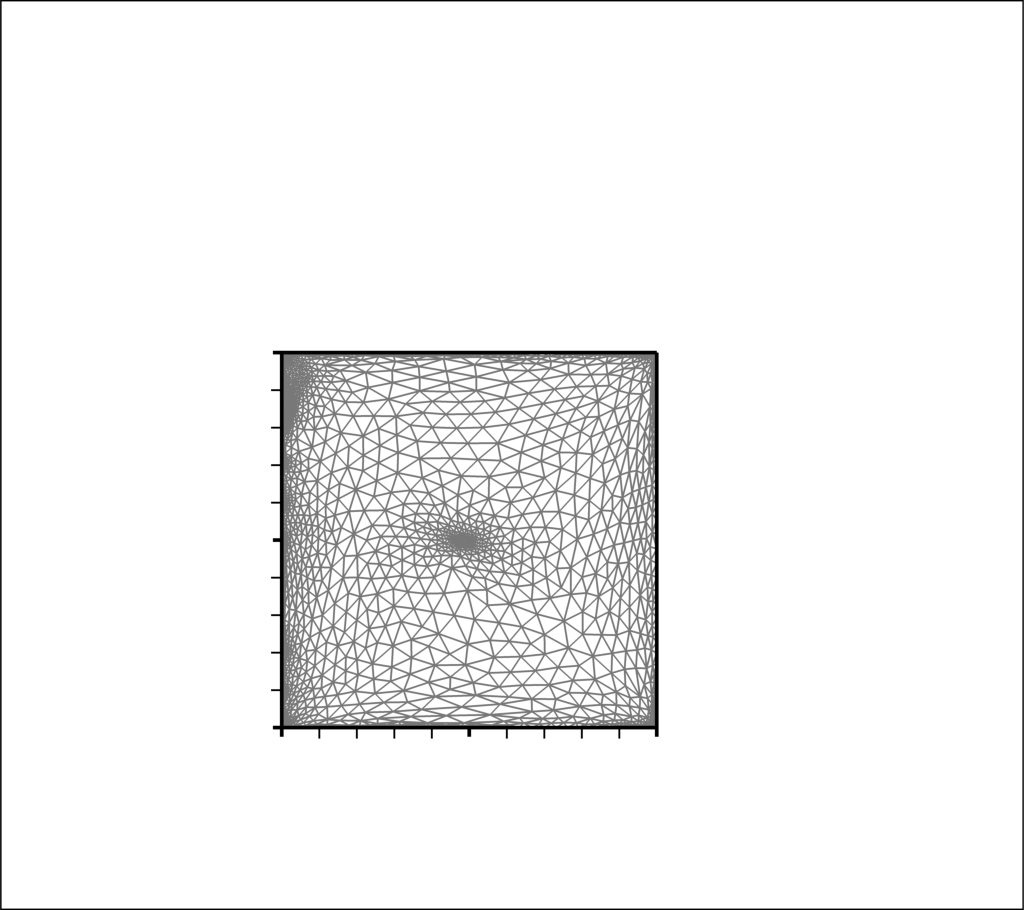}}
\put(4.9,4.4){\includegraphics[trim=175 100 360 325pt,clip,height=5cm]{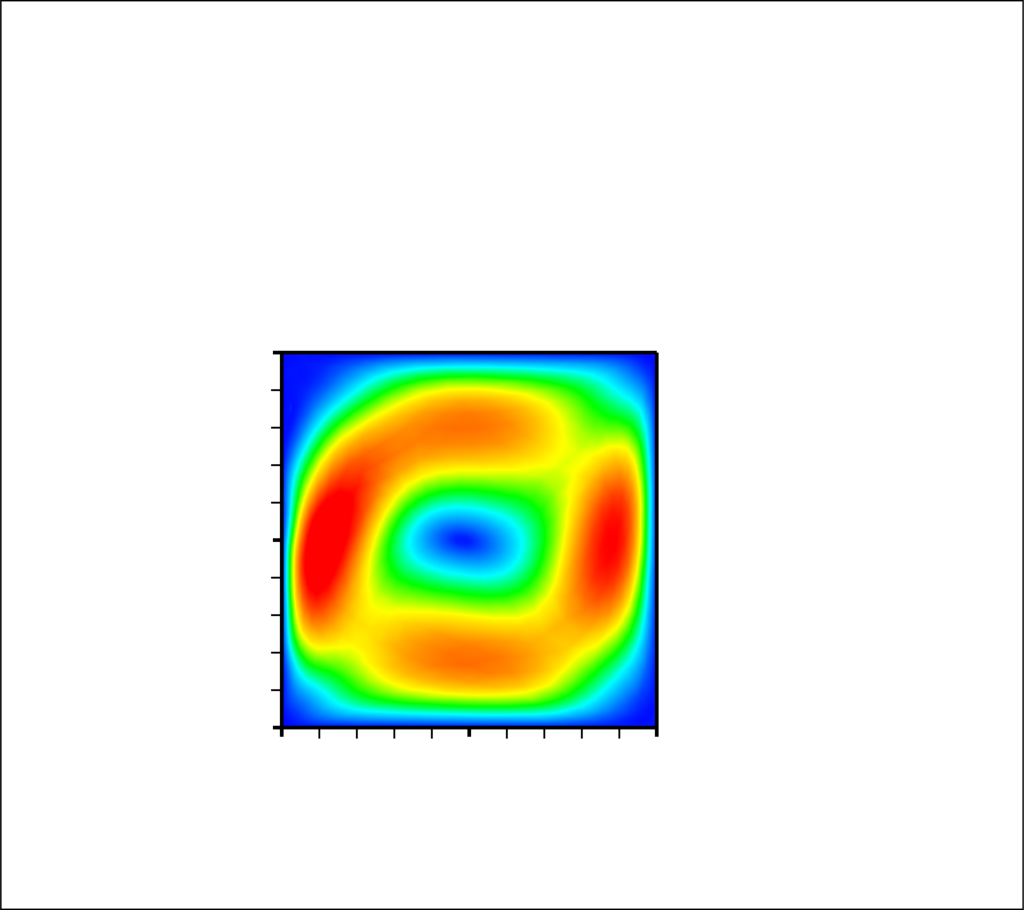}}
\put(0,4.4){\includegraphics[trim=175 100 360 325pt,clip,height=5cm]{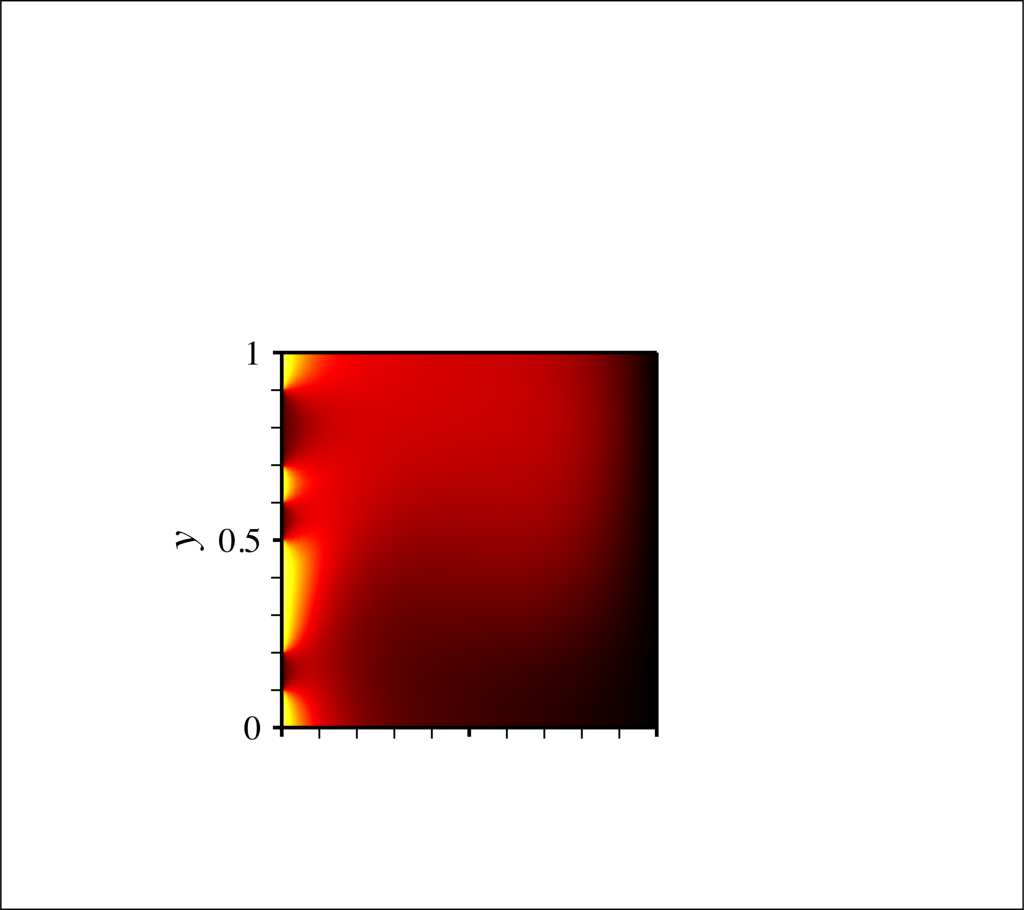}}
\put(9.8,0){\includegraphics[trim=175 100 360 325pt,clip,height=5cm]{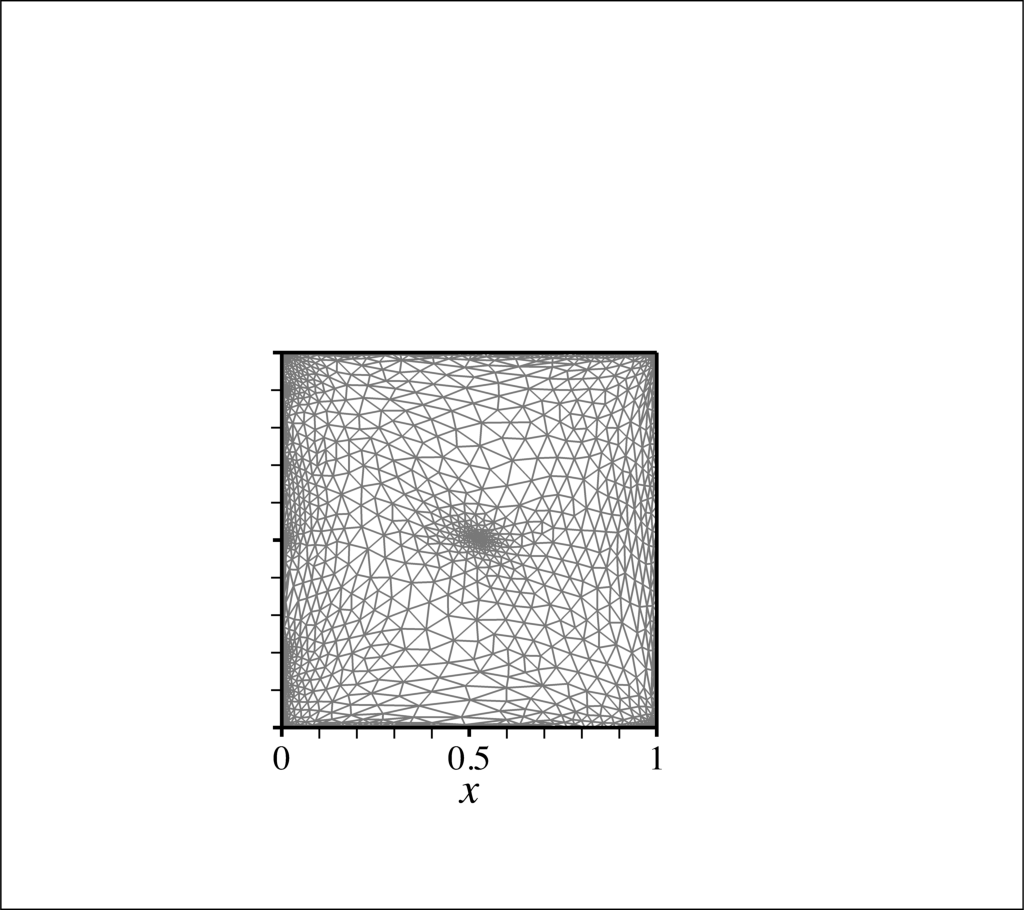}}
\put(4.9,0){\includegraphics[trim=175 100 360 325pt,clip,height=5cm]{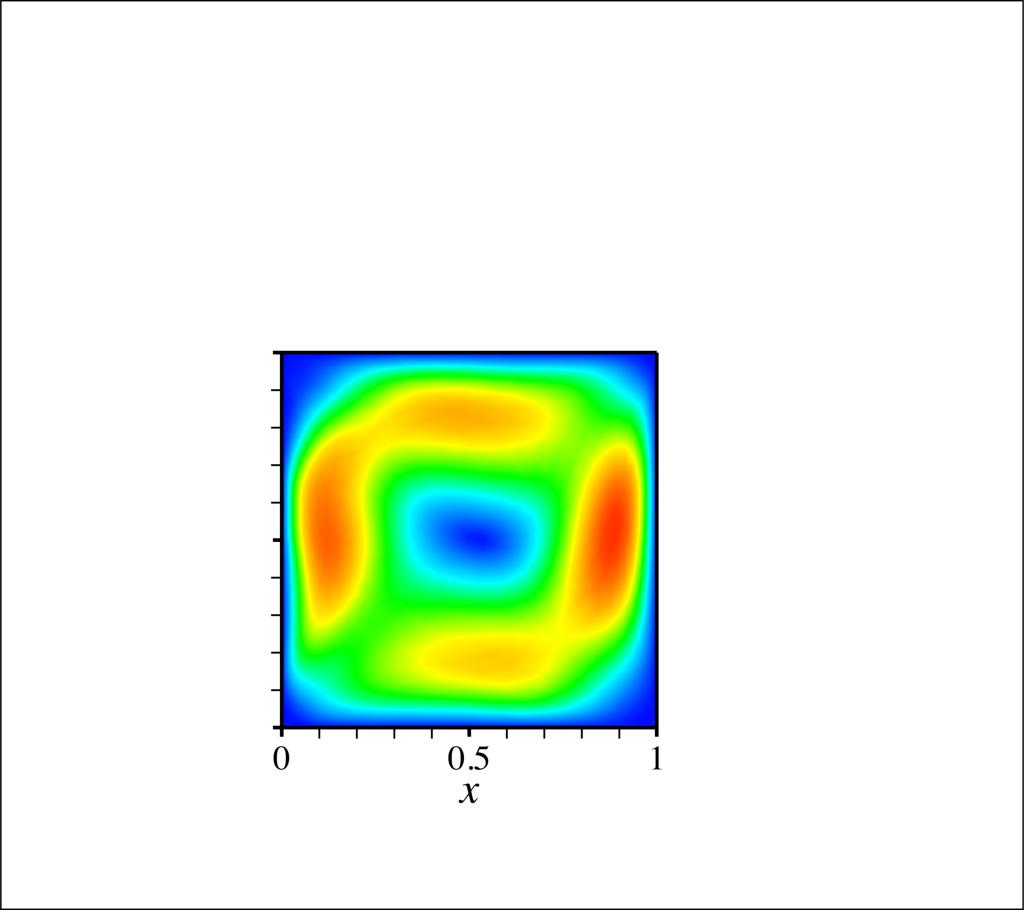}}
\put(0,0){\includegraphics[trim=175 100 360 325pt,clip,height=5cm]{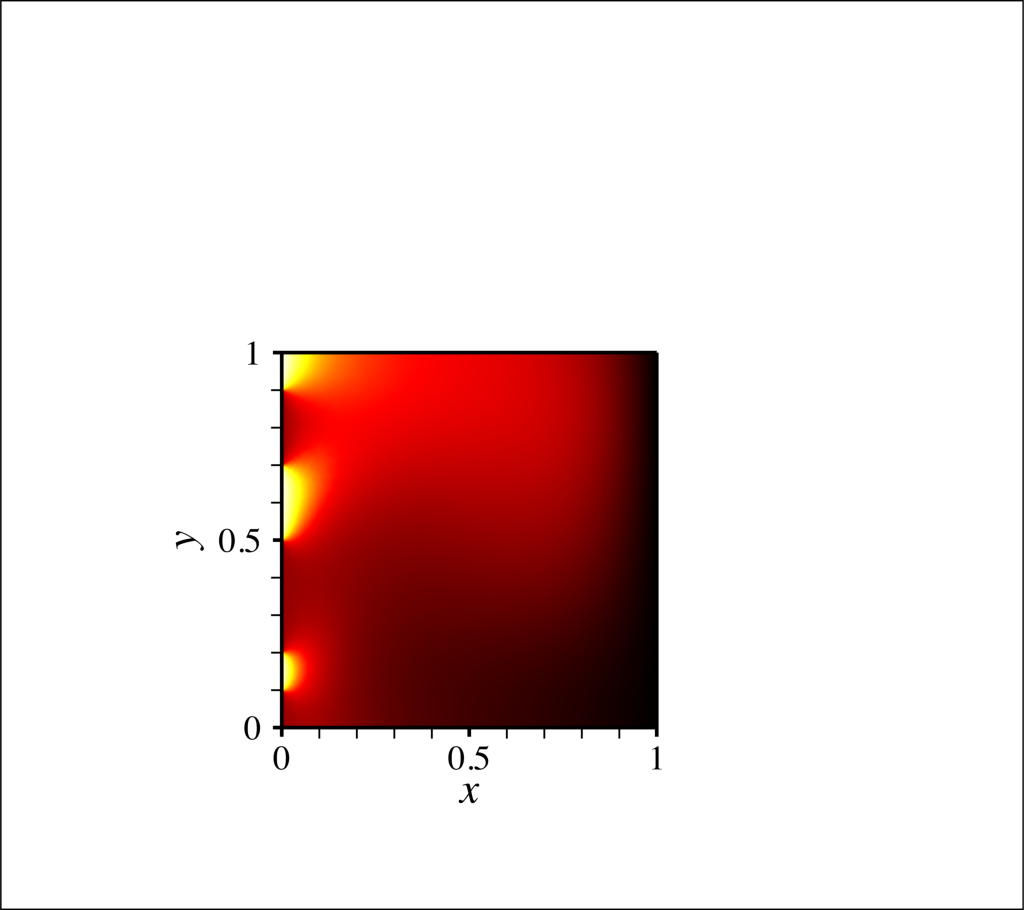}}
\put(2.825,18){\includegraphics[trim=420 635 440 245pt,clip,height=0.45cm]{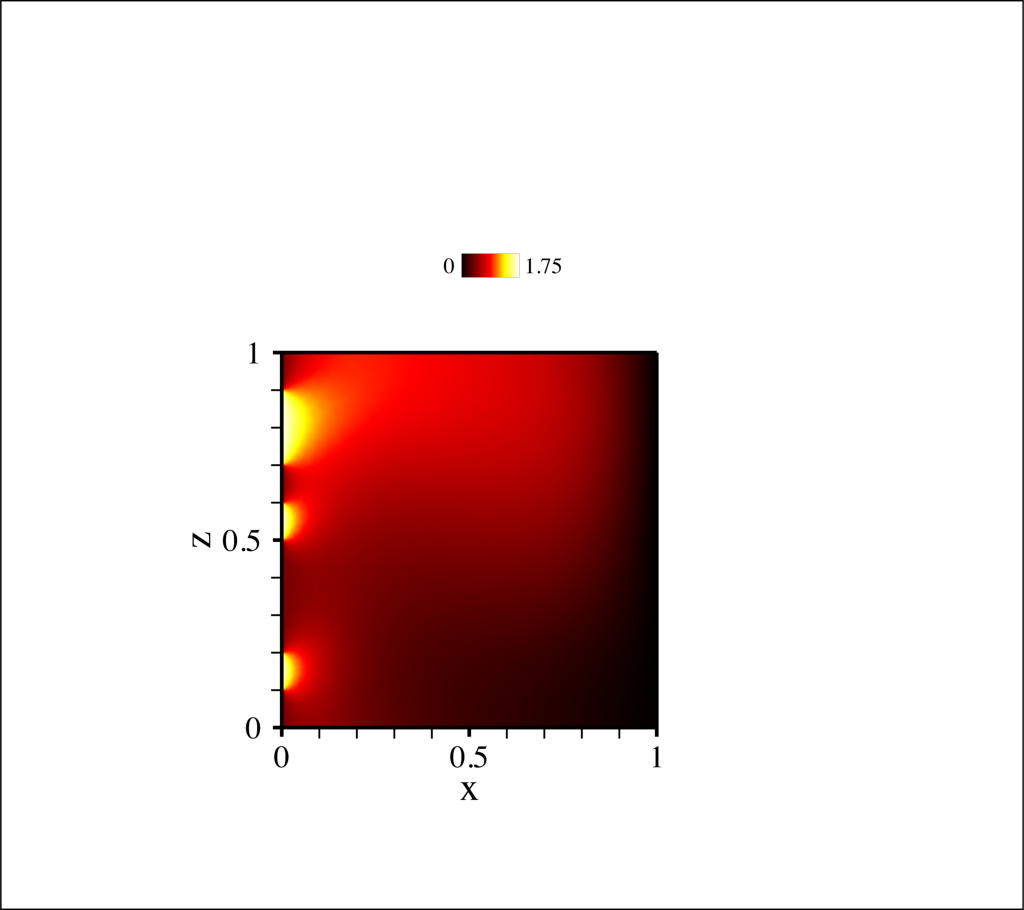}}
\put(8,18){\includegraphics[trim=420 635 440 245pt,clip,height=0.45cm]{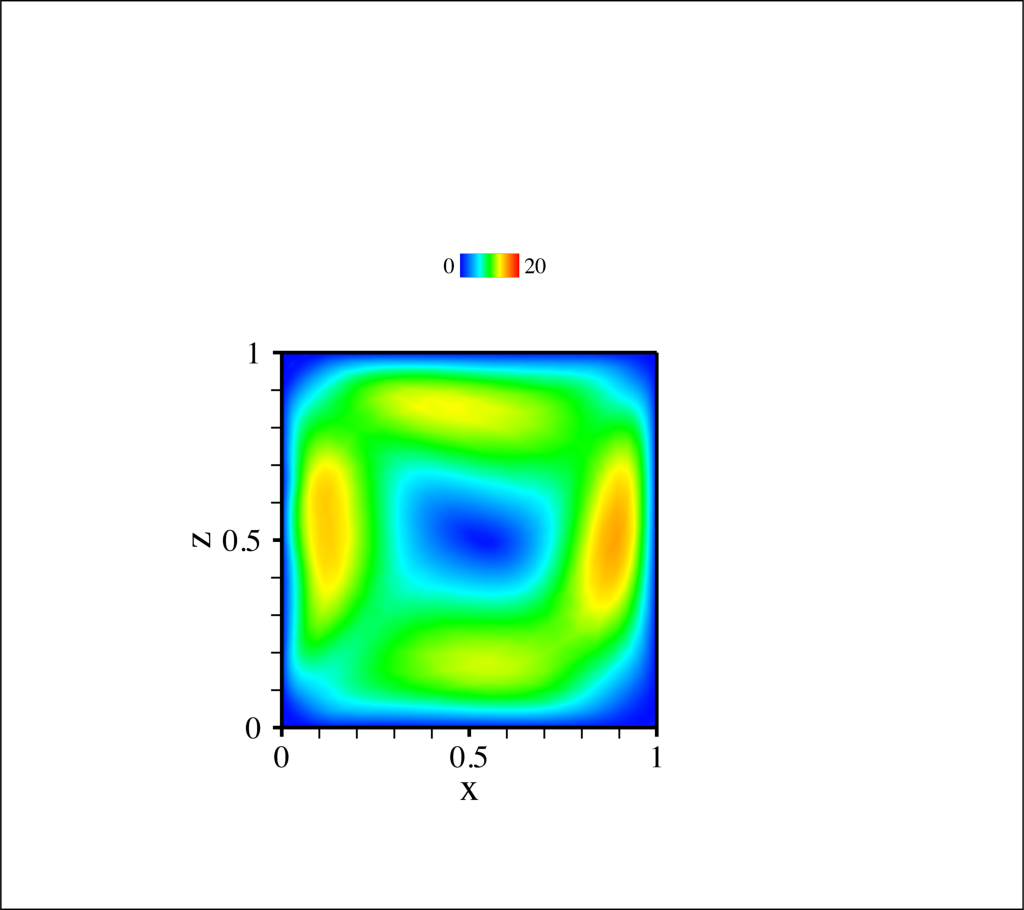}}
\put(0.35,18.3){(a)}
\put(5.3,18.3){(b)}
\put(10.2,18.3){(c)}
\end{picture}
\caption{(a,b) Steady-state (a) temperature and (b) velocity magnitude against zero-mean temperature distributed at the left sidewall. (c) Adapted meshes. }
\label{fig:natural_forced_vtu}
\end{figure}
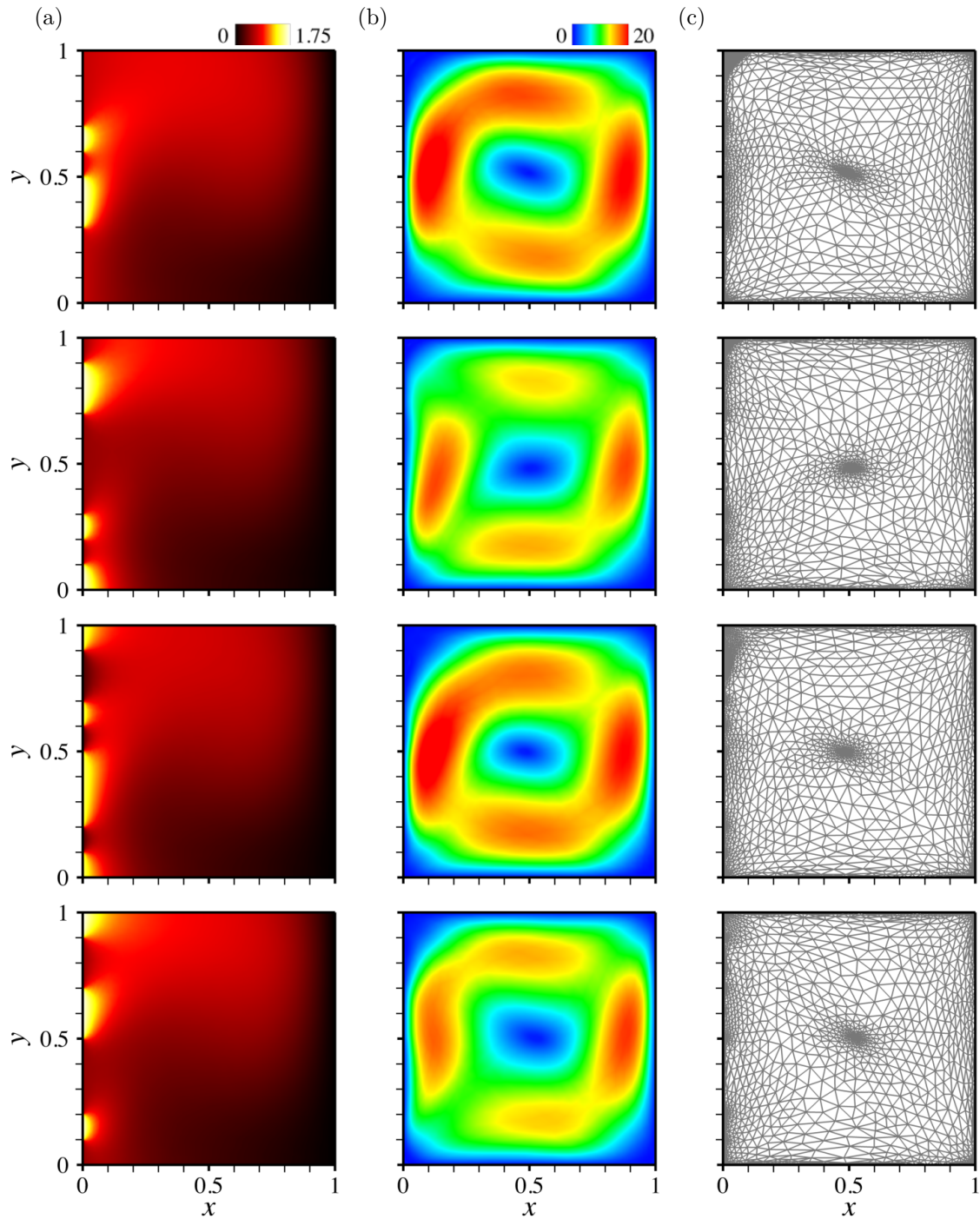


All results reported herein are for $\Delta T_{max}=0.75$ (so the hot temperature varies in the range $[0.25;1.75]$) and $n_s=10$ segments, as \cite{beintema2020controlling} report that $n_s=20$ was computationally too demanding for their case, and that $n_s=5$ yielded poor control efficiency.
The agent is a fully-connected network with two hidden layers, each holding 2 neurons. The resolution process uses 8 environments and 2 steps mini-batches to update the network for 32 epochs, with
learning rate $5\times 10^{-3}$, and PPO loss clipping range $\epsilon= 0.2$. 
 

\subsection{Results}

For this case, 120 episodes have been run, each of which follows the exact same procedure as above and performs 60 iterations with time step $\Delta t = 0.5$ to march the zero-initial condition to steady state.
This represents 960 simulations, each of which is performed on 4 cores and lasts 20s, hence 5h of total CPU cost.
We present in figure~\ref{fig:natural_forced_vtu} representative iso-contours of the steady-state temperature and velocity magnitude computed over the course of the optimization. The latter exhibit strong temperature gradients at the hot sidewall, together with a robust steady roll-shaped pattern accompanied by a small corner eddy at the upper-left edge of the cavity, whose size and position depends on the specifics of the temperature distribution. The corresponding meshes are displayed in figure~\ref{fig:natural_forced_vtu}(c) to stress the ability of the adaptation procedure to handle well the anisotropy of the solution caused by the intrinsic flow dynamics and the discontinuous boundary conditions. 

\begin{figure}[t!]
\setlength{\unitlength}{1cm}
\begin{picture}(20,5)
\put(4,0){\includegraphics[trim=175 87.5 140 340pt,clip,height=5cm]{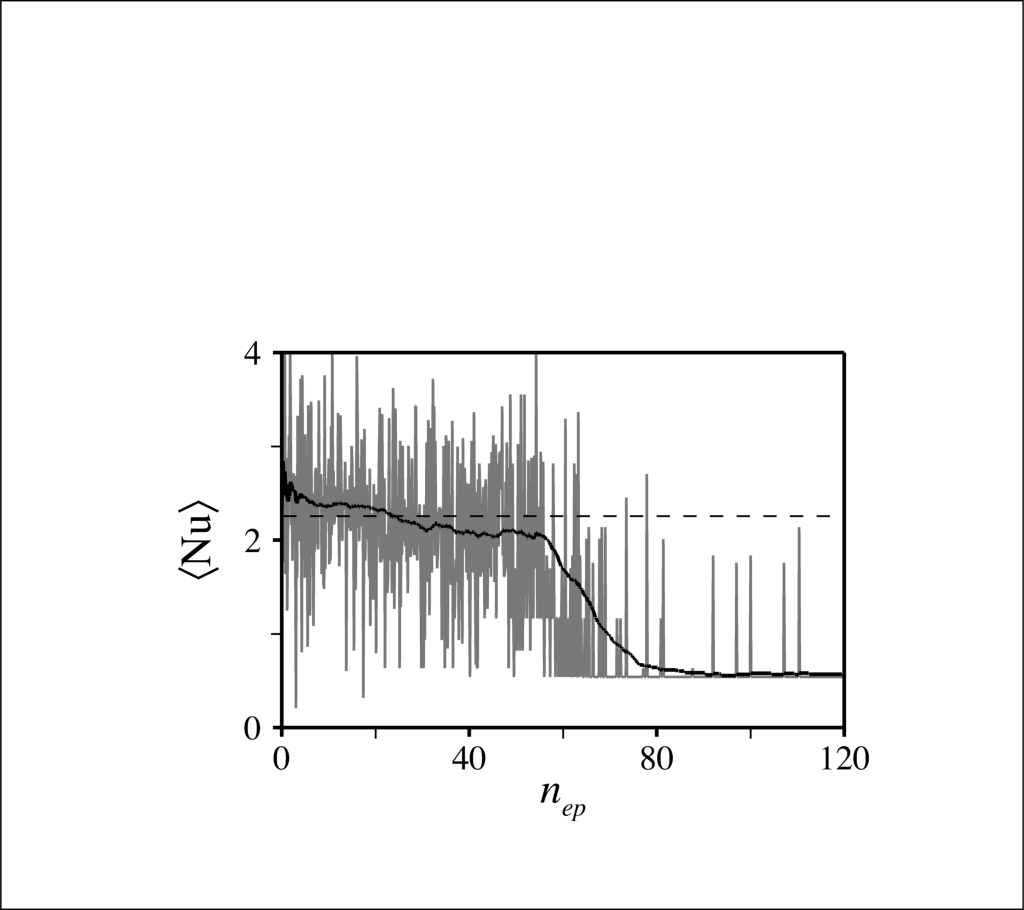}}
\end{picture}
\caption{Evolution per learning episode of the instant (in grey) and moving average (in black) Nusselt number. The horizontal dashed line marks the uncontrolled value.}
\label{fig:natural_reward}
\end{figure}

\begin{figure}[t!]
\setlength{\unitlength}{1cm}
\begin{picture}(20,5.1)
\put(9.8,0){\includegraphics[trim=175 100 360 325pt,clip,height=5cm]{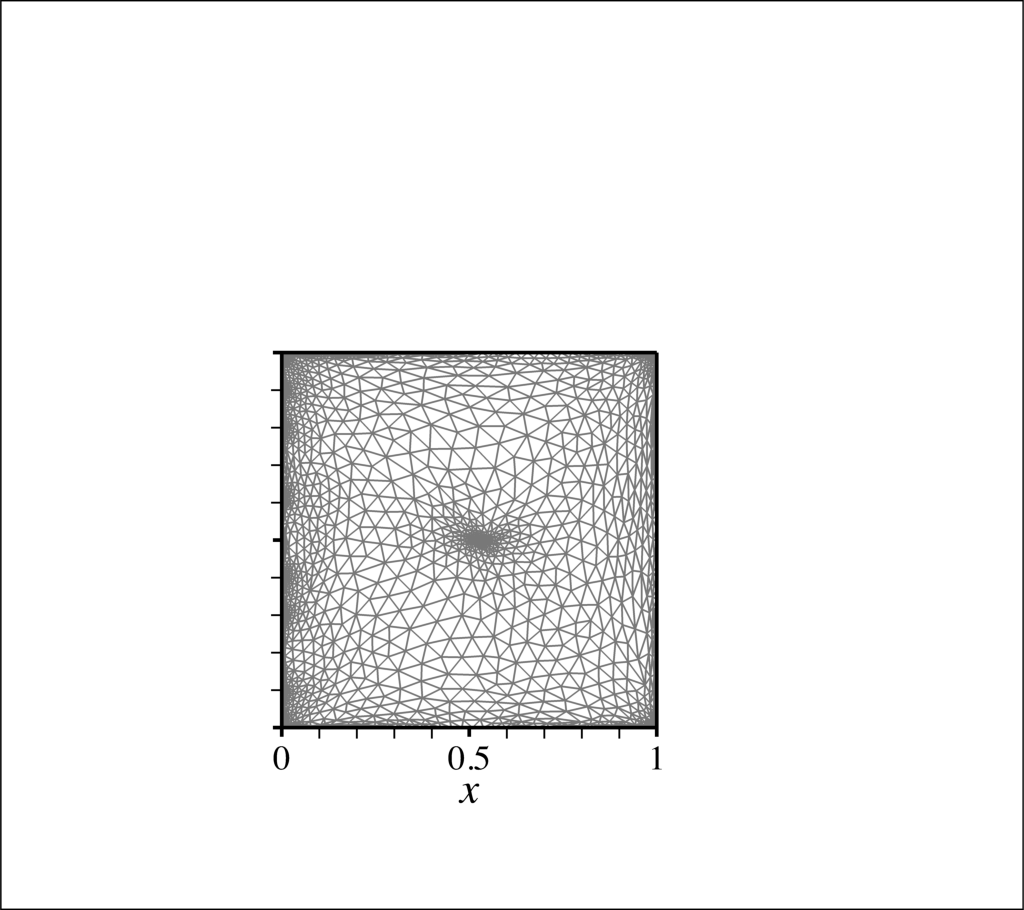}}
\put(4.9,0){\includegraphics[trim=175 100 360 325pt,clip,height=5cm]{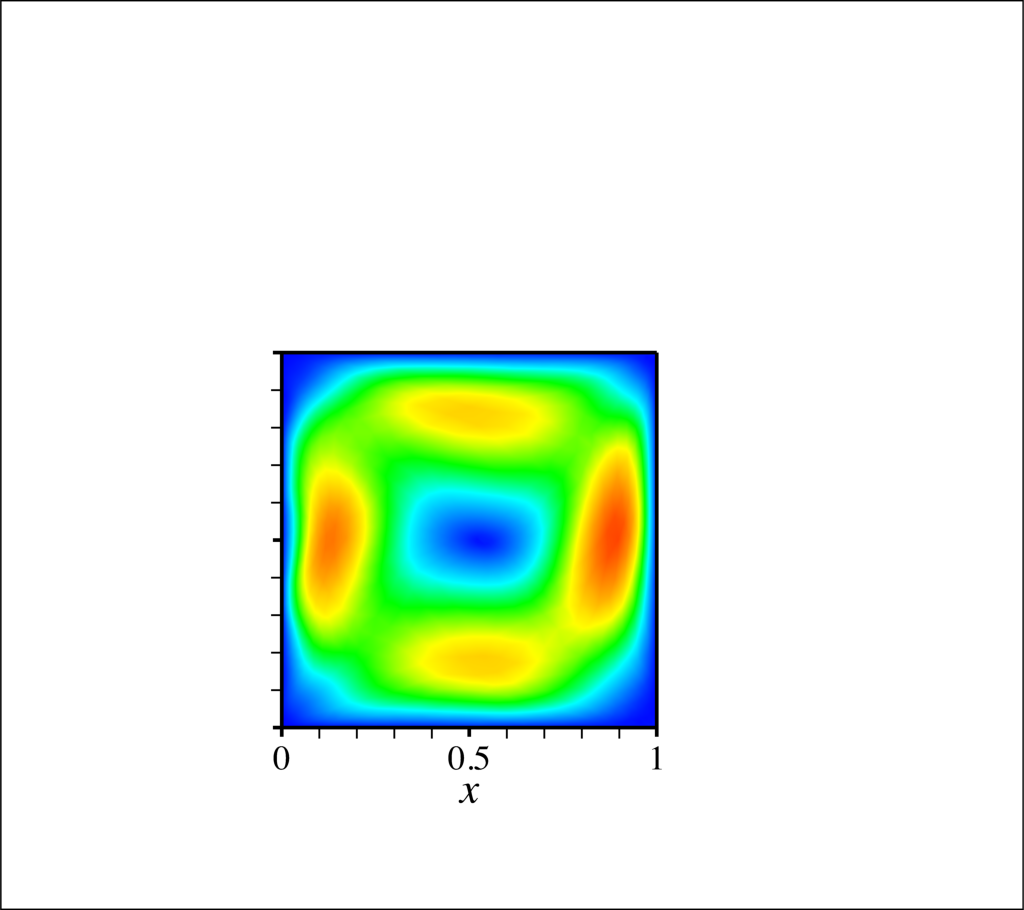}}
\put(0,0){\includegraphics[trim=175 100 360 325pt,clip,height=5cm]{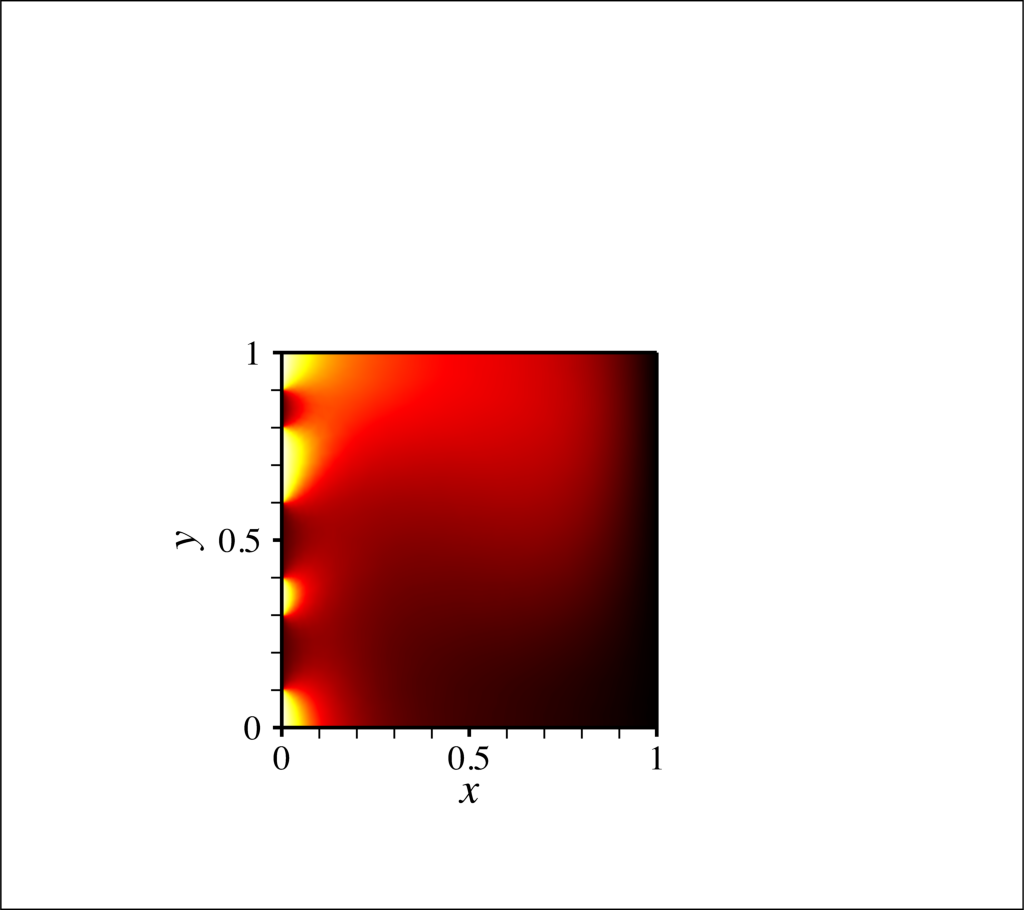}}
\put(2.825,4.8){\includegraphics[trim=420 635 440 245pt,clip,height=0.45cm]{fig7a_bar}}
\put(8,4.8){\includegraphics[trim=420 635 440 245pt,clip,height=0.45cm]{fig7b_bar}}
\put(0.35,5.1){(a)}
\put(5.3,5.1){(b)}
\put(10.2,5.1){(c)}
\end{picture}
\caption{(a,b) Steady-state (a) temperature and (b) velocity magnitude for the optimal zero-mean temperature distribution. (c) Adapted mesh. }
\label{fig:natural_opt_vtu}
\end{figure}

We show in figure~\ref{fig:natural_reward} the evolution of the controlled averaged Nusselt number, whose moving average decreases monotonically and reaches a plateau after about 90 episodes, although we notice that sub-optimal distributions keep being explored occasionally. The optimal computed by averaging over the $10$ latest episodes (hence the $800$ latest instant values) is $\langle \nus\rangle{^\star}\sim 0.57$, with variations $\pm0.01$ computed from the root-mean-square of the moving average over the same interval (which is a simple yet robust criterion to assess qualitatively convergence a posteriori). Interestingly, the optimized Nusselt number is almost twice as small as the purely conductive value ($\nus=1$), meaning that the approach successfully alleviates the heat transfer enhancement generated by the onset of convection, although it does not alleviate convection itself, as evidenced by the consistent roll-shaped pattern in figure~\ref{fig:natural_opt_vtu}. Similar results are reported in~\cite{beintema2020controlling}, for a different set-up in which the horizontal cavity walls are isothermal and control is applied at the bottom at the cavity (hence a different physics because of buoyancy), albeit with lower numerical and control efficiency since  the authors report an optimal Nusselt number $\nus\sim 1$ using up to 512 DRL environments with learning rate of $2.5\times10^{-4}$. 
The reason for such discrepancies probably lies in different ways of achieving and assessing control, as we use 
single-step PPO to optimize the steady-state Nusselt number via a time-independent control, which requires choosing a sidewall temperature, marching the controlled solution to steady state, then computing the reward. The problem considered in~\cite{beintema2020controlling} is more intricate, as classical PPO is used to optimize the reward accumulated over time via a time-dependent control temperature updated with a certain period scaling with the convection time in the cavity (the so-determined optimal control being ultimately time-independent for the considered value of $\ray$, but truly time-dependent for Rayleigh numbers above $\sim 10^5$).



\section{Control of forced convection in 2-D open cavity}\label{section:forced2d} 

\subsection{Case description}\label{section:forced2d:description}

\begin{figure}[t!]
\setlength{\unitlength}{1cm}
\begin{picture}(20,2.7)
\put(7.65,-1.1){\includegraphics[trim=175 110 155 515pt,clip,height=3.265cm]{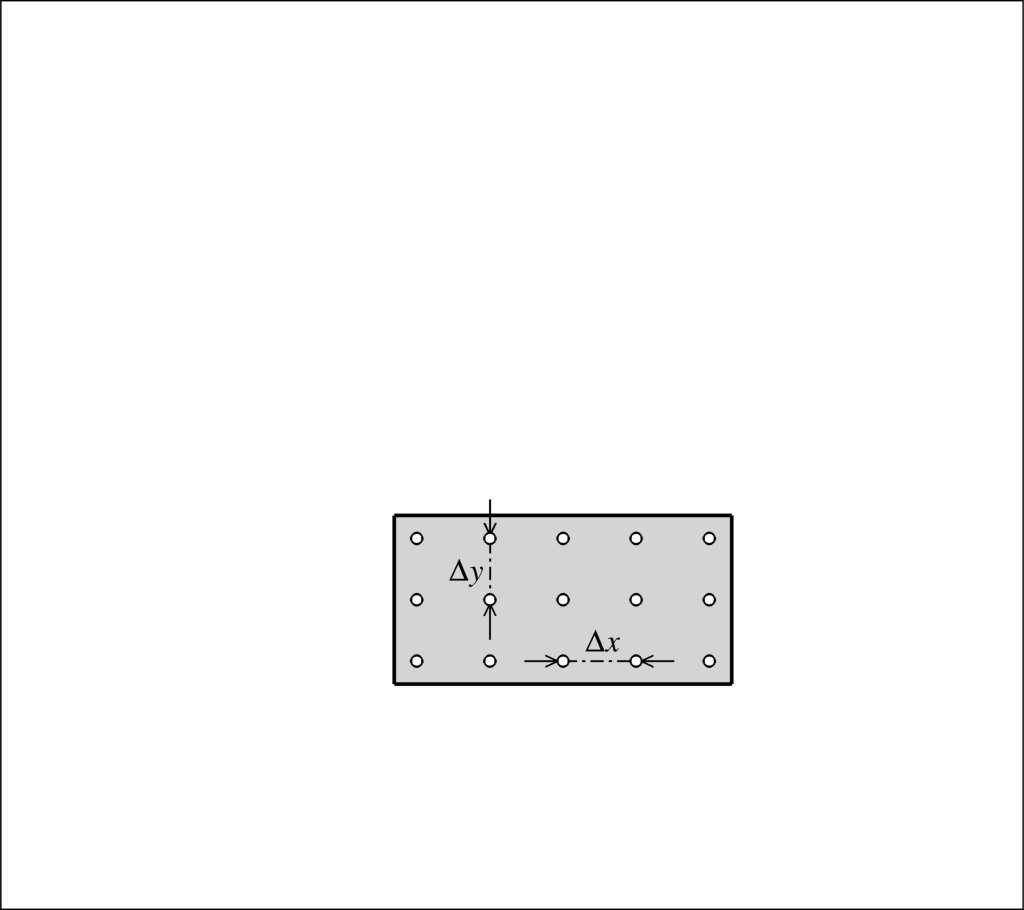}}
\put(0.1,-0.5){\includegraphics[trim=175 110 140 575pt,clip,height=2.575cm]{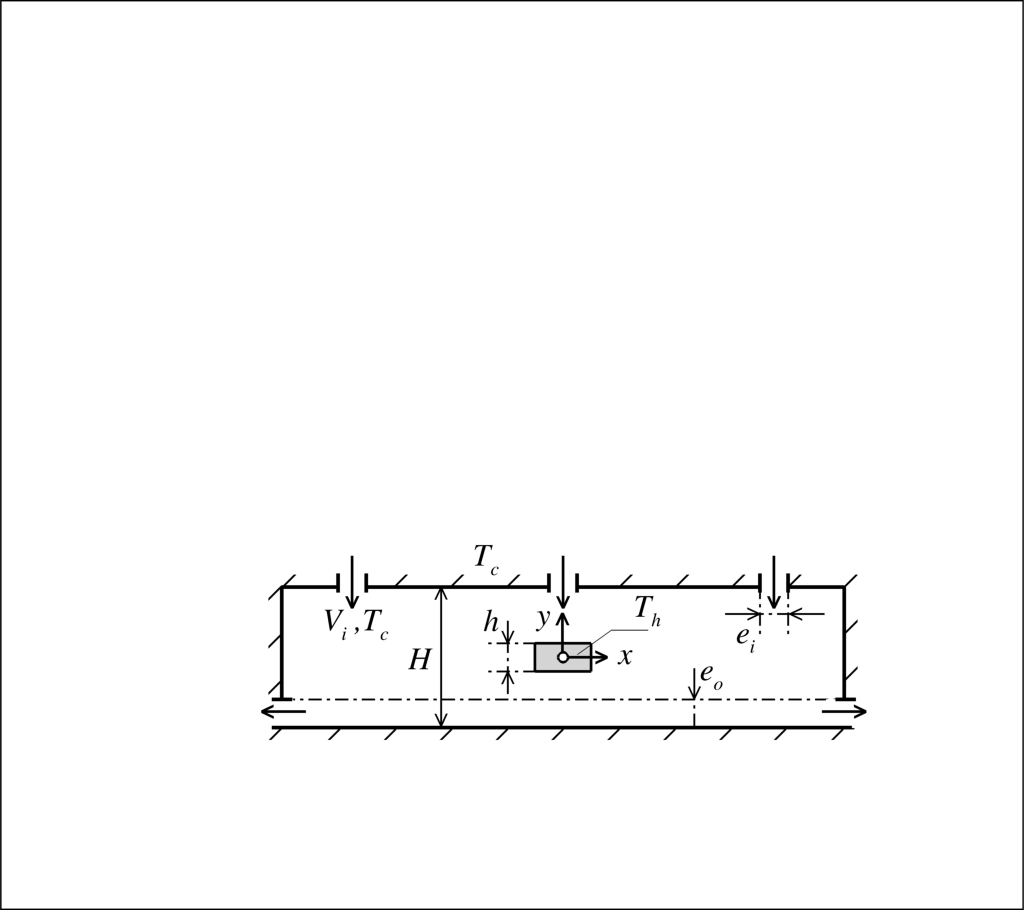}}
\put(0.55,2.7){(a)}
\put(7.85,2.7){(b)}
\end{picture}
\caption{(a) Schematic of the 2-D forced convection set-up. (b) Sensors positions in the solid domain. }
\label{fig:forced2d_conf}
\end{figure}

This second test case addresses the control of actual conjugate heat transfer in a model setup for the cooling of a hot solid by impingement of a fluid; see figure~\ref{fig:forced2d_conf}(a). A Cartesian coordinate system is used with origin at the center of mass of the solid, horizontal $x$-axis, and vertical $y$-axis. The solid has rectangular shape with height $h$ and aspect ratio 2:1, and is initially at the hot temperature $T_h$. It is fixed at the center of a rectangular cavity with height $H$ and aspect ratio 4:1, whose walls are isothermal and kept at temperature $T_w$. 
The top cavity side is flush with $n_j$ identical holes of width $e_i$ whose distribution is subjected to optimization, each of which models the exit plane of an injector blowing cold air at velocity $V_i$ and temperature $T_c$, and is identified by the horizontal position of its center $x_{k\in\{1\dots n_j\}}$.
Hot air is released through the cavity sidewalls, blown with two identical exhaust areas of height $e_o$, and identified by the vertical position of their center $(e_0-H)/2$.

For this case, both buoyancy and radiative heat transfer are neglected ($\vecpsi=\00$ and $\chi=0$), meaning that temperature evolves as a passive scalar, similar to the mass fraction of a reactant in a chemical reaction. All relevant parameters are provided in Table~\ref{table:forced2d}, including the material properties used to
model the composite fluid, that yield fluid values of the Reynolds and Prandtl numbers
\bal
\rey=\frac{\rho V_i e}{\mu}=200
\,,\qquad\qquad \pr=2\,.
\eal
Note the very high value of the solid to fluid viscosity ratio, that ensures that the velocity is zero in the solid domain and that the no-slip interface condition is satisfied. By doing so, the convective terms drop out in the energy equation, that reduces to the pure conduction equation for the solid. 
The governing equations are solved with no-slip isothermal conditions $\uu=\00$ and $T=T_w$ on $\partial\Omega$, except at the injection exit planes ($\uu=-V_i\eey$, $T=T_c$), and at the exhaust areas, where a zero-pressure outflow condition is imposed ($p=\partial_x u =\partial_x T=0$). No thermal condition is imposed at the interface, where heat exchange is implicitly driven by the difference in the individual material properties.

\begin{table}[t!]
\begin{center}
\begin{tabular}{ccccccccccccc}
\toprule
\multicolumn{1}{p{0.65cm}}{\makecell[r]{$H$}} & \multicolumn{1}{p{0.65cm}}{\makecell[r]{$h$}} & \multicolumn{1}{p{0.65cm}}{\makecell[r]{$e_i$}} & \multicolumn{1}{p{0.65cm}}{\makecell[r]{$e_0$}} & \multicolumn{1}{p{0.65cm}}{\makecell[r]{$V_i$}} & \multicolumn{1}{p{0.65cm}}{\makecell[r]{$T_w$}} & \multicolumn{1}{p{0.65cm}}{\makecell[r]{$T_c$}} & \multicolumn{1}{p{0.65cm}}{\makecell[r]{$T_h$}} & \multicolumn{1}{p{0.9cm}}{\makecell[r]{$\mu$}} & \multicolumn{1}{p{0.65cm}}{\makecell[r]{$\rho$}} & \multicolumn{1}{p{0.65cm}}{\makecell[r]{$\lambda$}} & \multicolumn{1}{p{0.9cm}}{\makecell[r]{$c_p$}}\\
\cmidrule(lr){1-13}
\multicolumn{1}{r}{\multirow{2}{*}{$1$}} & \multicolumn{1}{r}{\multirow{2}{*}{$0.2$}} & \multicolumn{1}{r}{\multirow{2}{*}{$0.2$}} & \multicolumn{1}{r}{\multirow{2}{*}{$0.2$}} & \multicolumn{1}{r}{\multirow{2}{*}{$1$}} & \multicolumn{1}{r}{\multirow{2}{*}{$10$}} & \multicolumn{1}{r}{\multirow{2}{*}{$10$}} & \multicolumn{1}{r}{\multirow{2}{*}{$150$}} & \multicolumn{1}{r}{\makecell[r]{$0.001$}} & \multicolumn{1}{r}{\makecell[r]{$1$}} & \multicolumn{1}{r}{\makecell[r]{$0.5$}} & \multicolumn{1}{r}{\makecell[r]{$1000$}} & \multicolumn{1}{r}{\makecell[r]{Fluid}}\\
\cmidrule(lr){9-13}
\multicolumn{8}{r}{\makecell[r]{}} & \multicolumn{1}{r}{\makecell[r]{$1000$}} & \multicolumn{1}{r}{\makecell[r]{$100$}} & \multicolumn{1}{r}{\makecell[r]{$15$}} & \multicolumn{1}{r}{\makecell[r]{$300$}} & \multicolumn{1}{r}{\makecell[r]{Solid}}\\
\bottomrule
\end{tabular}
\caption{Numerical parameters used in the 2-D forced convection problem. All values in SI units, with the exception of temperatures given in Celsius.}
\label{table:forced2d}
\end{center}	
\end{table}

\subsection{Control}

The quantity being optimized is the distribution of the injectors center positions $x_{k\in\{1\dots n_j\}}$.
{Several control strategies are assessed in the following, whose ability to manage increasing design complexity translates into less constrained operation when it comes to optimizing a practically meaningful device. In practice, each injector} is forced to sit in an interval $[x_{k}^-;x_{k}^+]$
whose edge values are
 determined beforehand of recomputed on the fly (depending on the control strategy), and 
bounded according to 
\bal
|x_k^\pm|\leq x_m\,,
\eal
where we set $x_m=2H-0.75e_i$
to avoid numerical issues at the upper cavity edges.
The network action output therefore consists of $n_j$ values $\hat{x}\in[-1;1]$, 
mapped
 into the actual positions according to
\bal
x_k=\frac{x_{k}^+(\hat{x}_k+1)-x_{k}^-(\hat{x}_k-1)}{2}\,.\label{eq:mapping}
\eal

In order to compute the reward passed to the DRL, we distribute uniformly $15$ probes in the solid domain, into $n_x=5$ columns and $n_y=3$ rows with resolutions $\Delta x=0.09$ and $\Delta y=0.075$, respectively; see figure \ref{fig:forced2d_conf}(b). Selected tests have been carried out to check that the outcome of the learning process does not change using $n_y=5$ rows of $n_x=5$ probes (not shown here).
The magnitude of the tangential heat flux is estimated by averaging the 
norm of the temperature gradient over all columns and rows, i.e., $i$-th column (resp. the $j$-th row) as
\bal
\langle ||\nabla_{\|}T||\rangle_{i}=&\frac{2}{n_y-1}|\sum_{j\neq 0}\text{sgn}(j)||\nabla T||_{ij}|\,,\quad\quad\langle ||\nabla_{\|}T||\rangle_{j}=&\frac{2}{n_x-1}|\sum_{i\neq 0}\text{sgn}(i)||\nabla T||_{ij}|\,,\label{eq:forced2dreward}
\eal
where subscripts $i$, $j$ and $ij$ denote quantities evaluated at $x=i\Delta x$, $y=j\Delta y$ and $(x,y)=(i\Delta x,j\Delta y)$, respectively, and symmetrical numbering is used for the center probe to sit at the intersection of the zero-th column and row. The numerical reward $r_t=-\langle ||\nabla_{\|}T||\rangle$ fed to the DRL agent deduces ultimately by averaging over all rows and columns, to give
\bal
\langle ||\nabla_{\|}T||\rangle=\frac{1}{n_x+n_y}\sum_{i,j}\langle ||\nabla_{\|}T||\rangle_{i}+\langle ||\nabla_{\|}T||\rangle_{j}\,,\label{eq:forced2dreward2}
\eal
which especially yields $r_t=0$ for a perfectly homogeneous cooling.

All results reported in the following are for $n_j=3$ injectors. The agent is a fully-connected network with two hidden layers, each holding 2 neurons. The resolution process uses 8 environments and 2 steps mini-batches to update the network for 32 epochs, with learning rate set to $5\times 10^{-3}$, and PPO loss clipping range to  $\epsilon= 0.3$.



\begin{figure}[t!]
\setlength{\unitlength}{1cm}
\begin{picture}(20,7.3)
%
\put(7.2,4.4){\includegraphics[trim=175 110 140 575pt,clip,height=2.575cm]{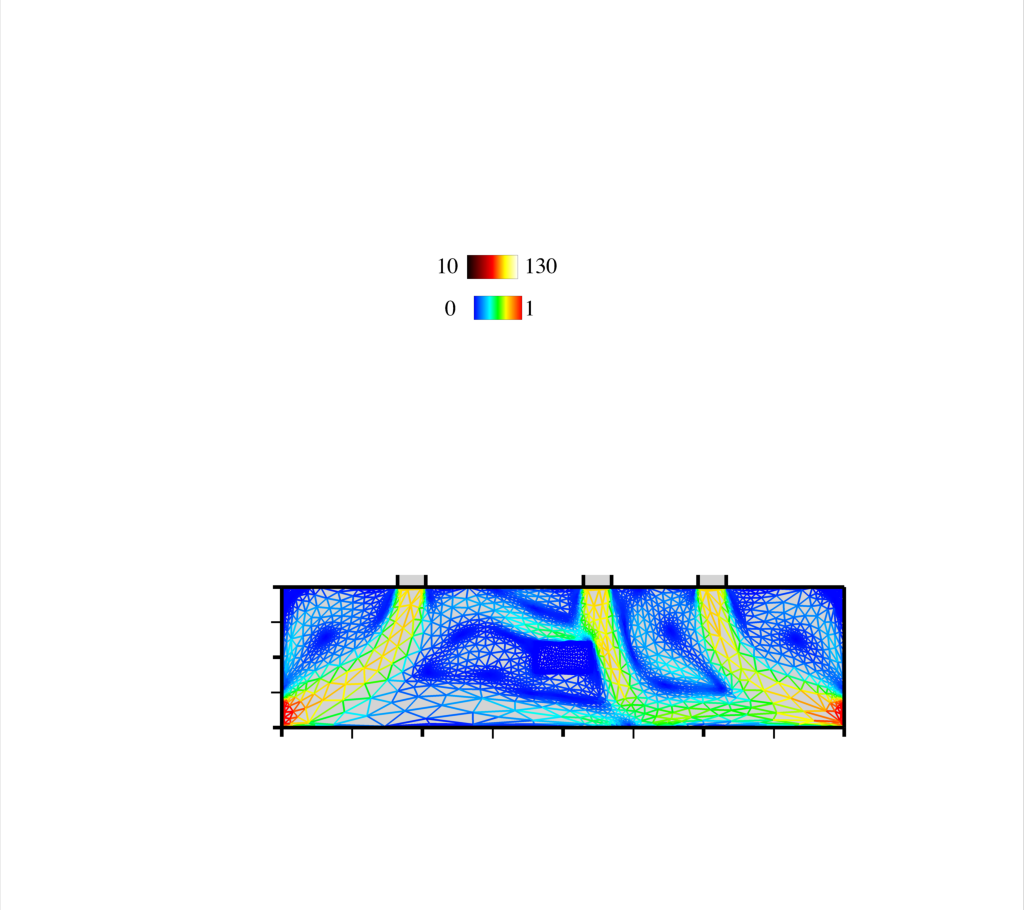}}
\put(0,4.4){\includegraphics[trim=175 110 140 575pt,clip,height=2.575cm]{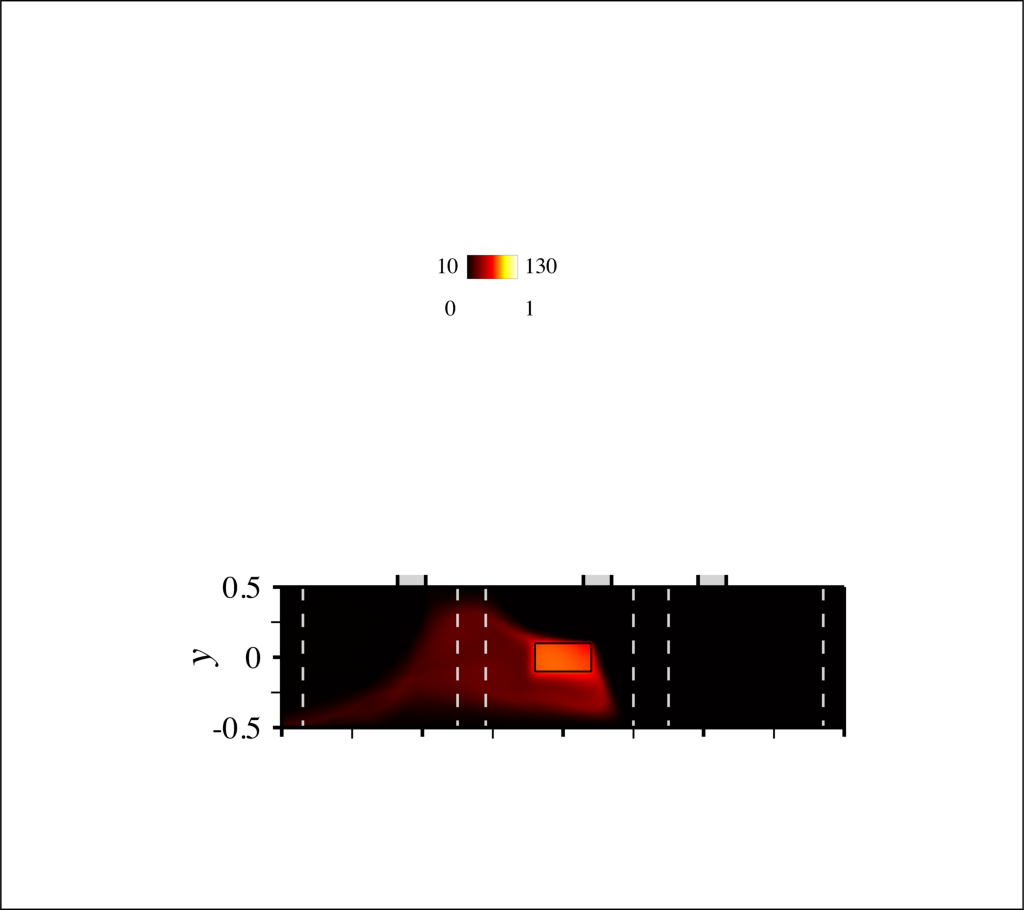}}
\put(7.2,2.2){\includegraphics[trim=175 110 140 575pt,clip,height=2.575cm]{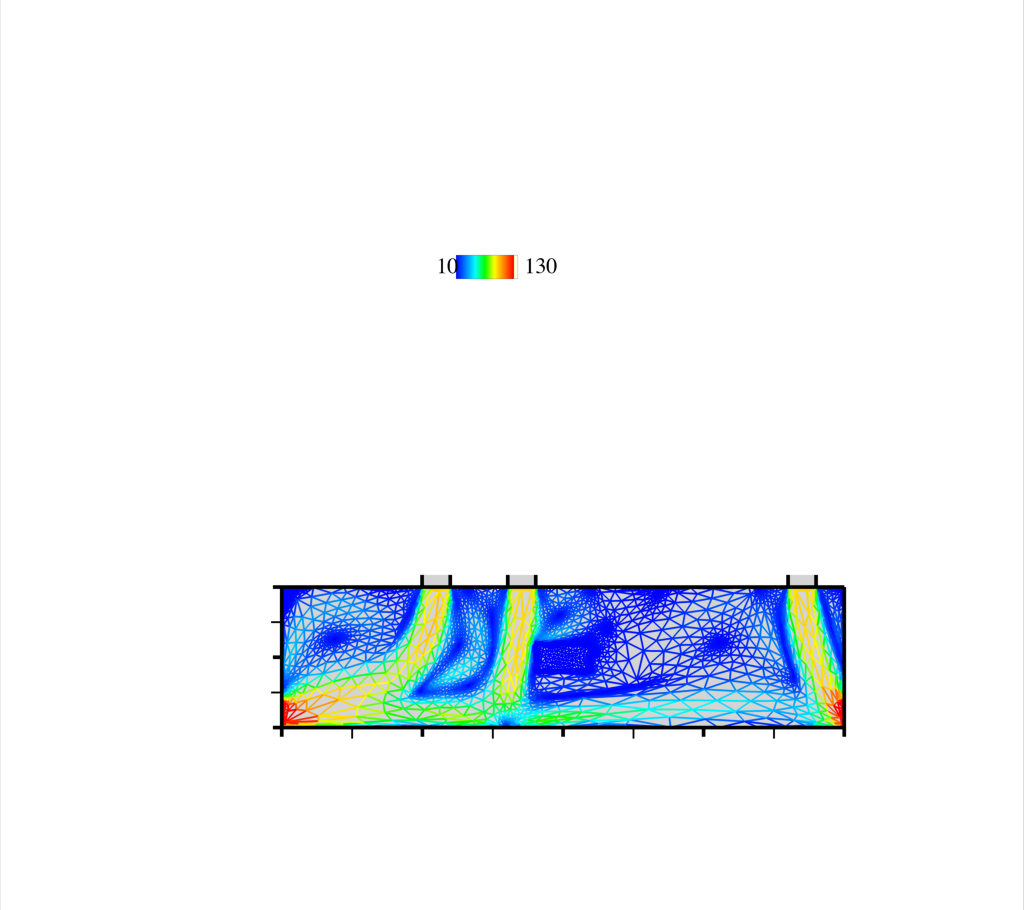}}
\put(0,2.2){\includegraphics[trim=175 110 140 575pt,clip,height=2.575cm]{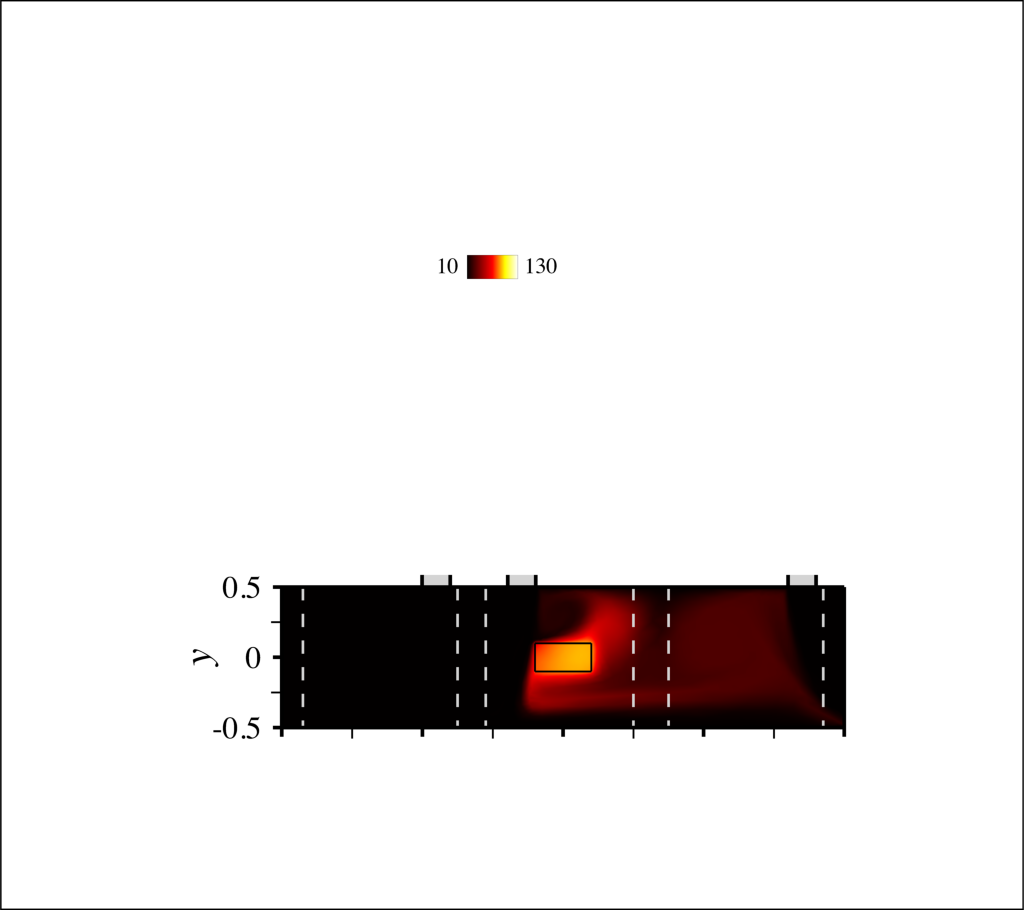}}
\put(7.2,0){\includegraphics[trim=175 110 140 575pt,clip,height=2.575cm]{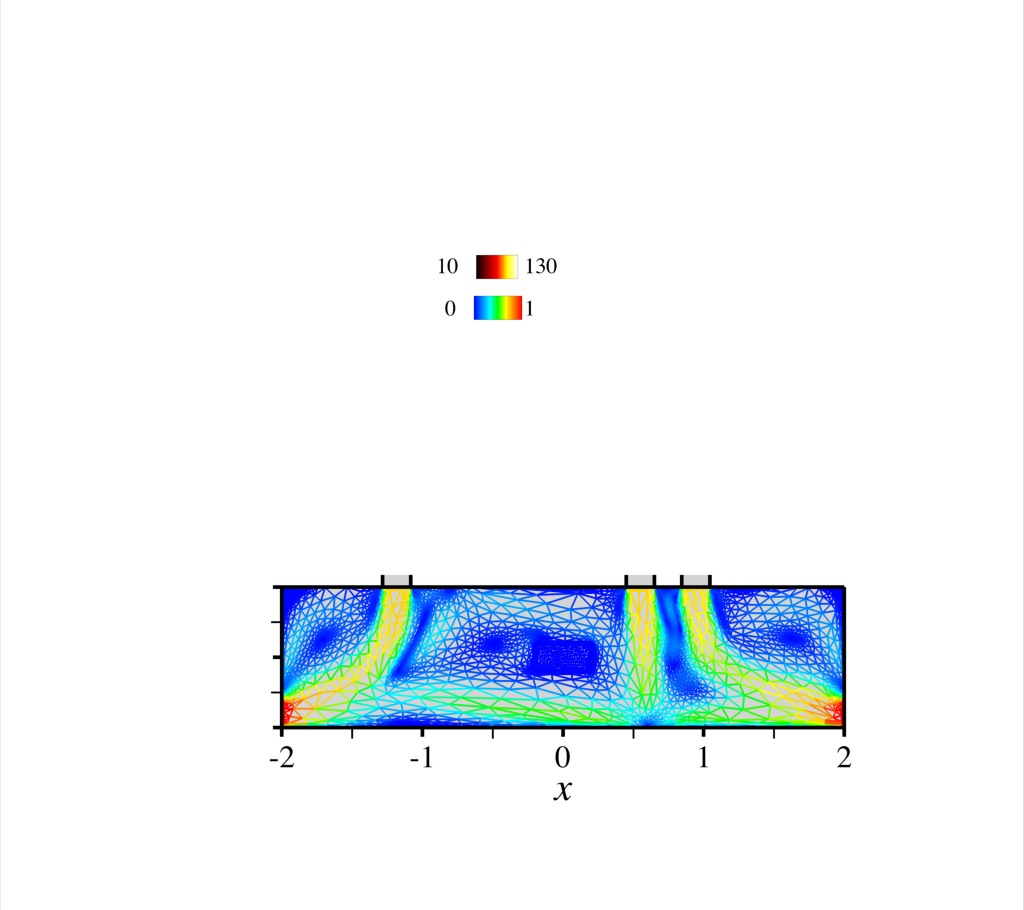}}
\put(0,0){\includegraphics[trim=175 110 140 575pt,clip,height=2.575cm]{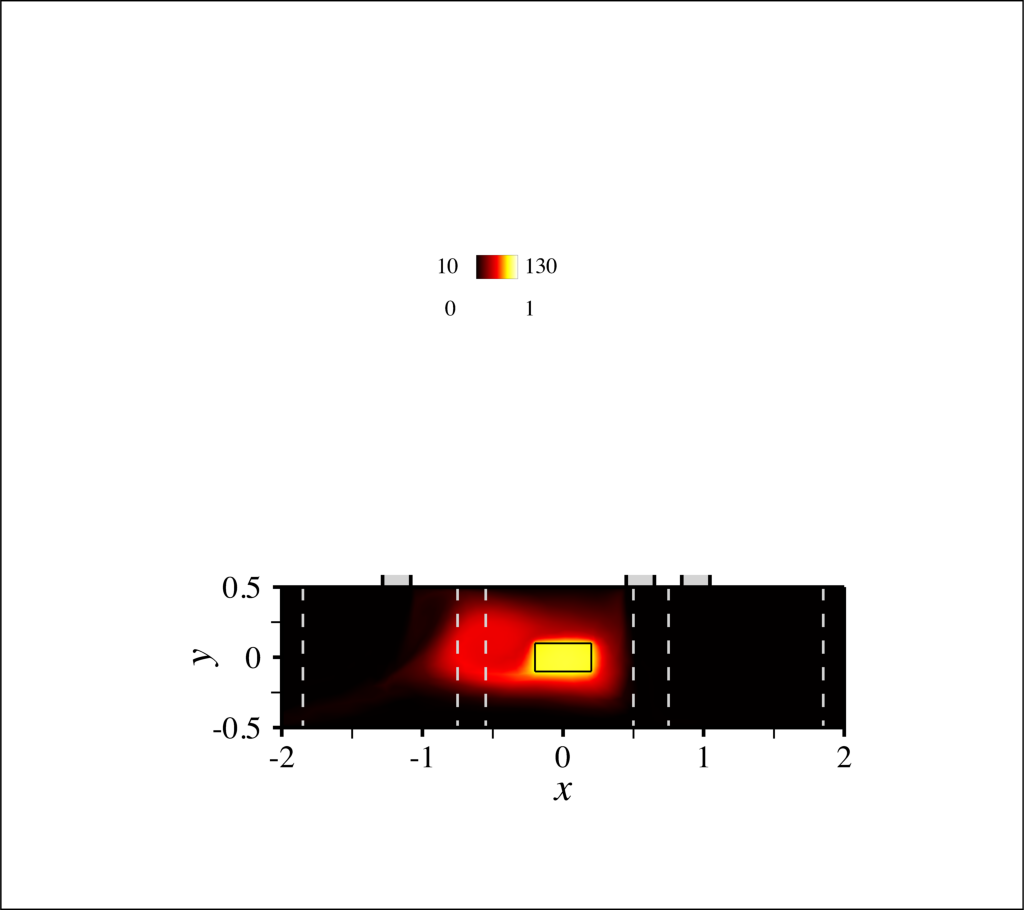}}
\put(5.6,7.1){\includegraphics[trim=420 635 440 245pt,clip,height=0.45cm]{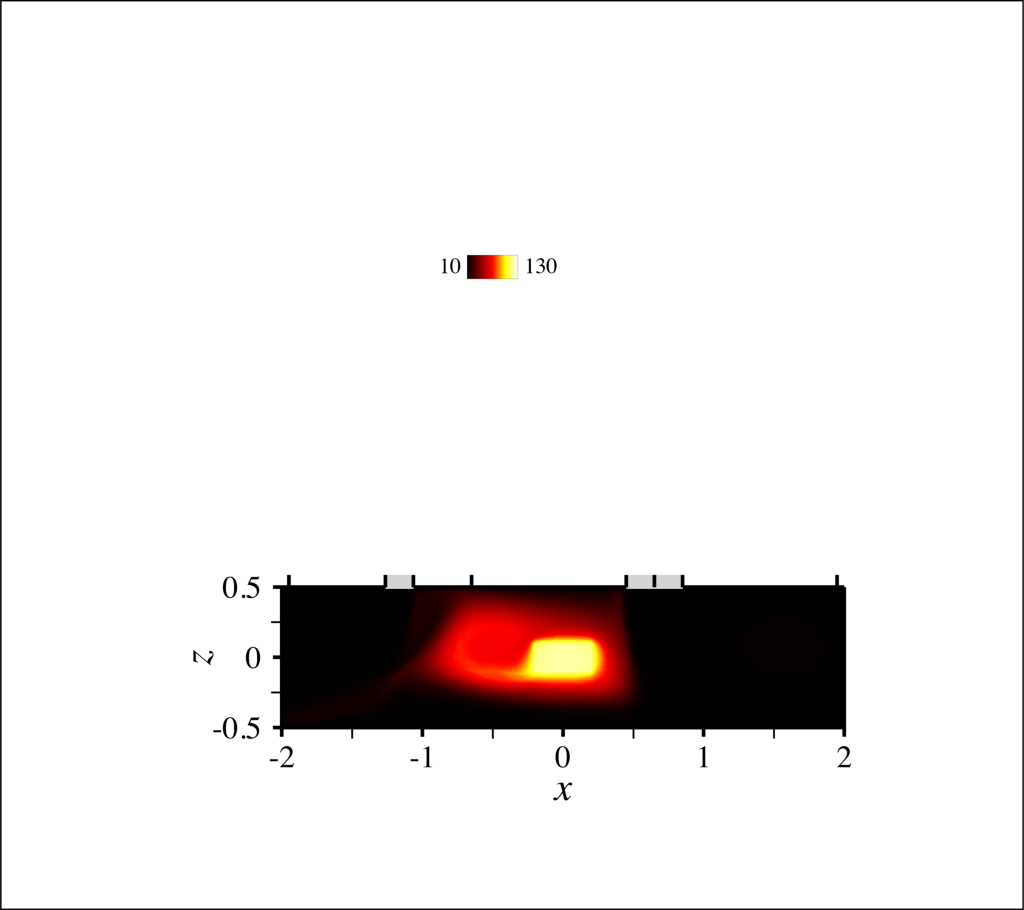}}
\put(12.9,7.1){\includegraphics[trim=420 635 440 245pt,clip,height=0.45cm]{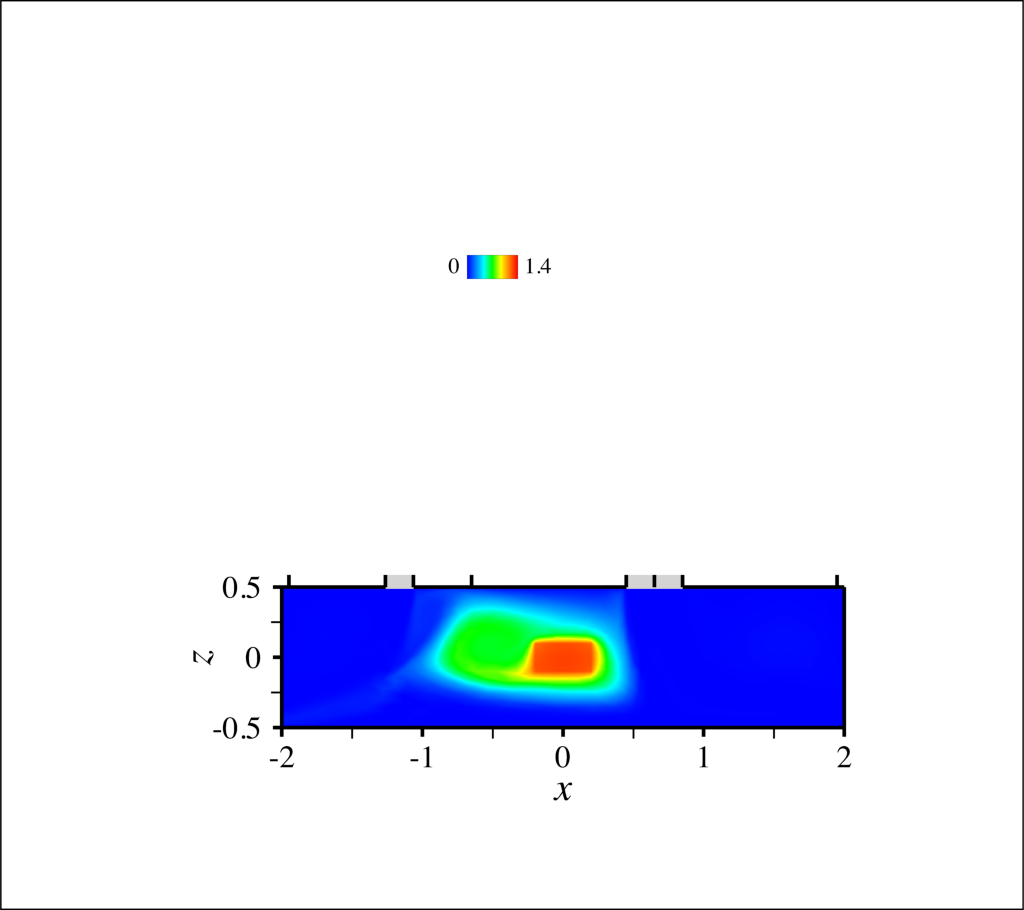}}
\put(7.85,7.3){(b)}
\put(0.55,7.3){(a)}
\end{picture}
\caption{(a) Steady-state temperature against arrangements of 3 injectors, with admissible values under the fixed domain decomposition strategy $S_1$ delimited by the dashed lines. (b) Adapted meshes colored by the magnitude of velocity.}
\label{fig:forced2d_S1_vtu}
\end{figure}

\subsection{Results}

\subsubsection{Fixed domain decomposition strategy}

We consider first the so-called fixed domain decomposition strategy $S_1$ in which the top cavity wall is split into $n_j$ equal subdomains, and each injector is forced to sit in a different subdomain {(a somehow heavily constrained optimization problem if $n_j$ is not to small, relevant for cases where the design is rigid and the
practitioner has limited freedom to act).} The 
edge values for the 
position $x_{k}$ of the $k$-th injector read 
\bal
x_{k}^-=-x_m+(k-1)\frac{2x_m +e_i}{n_j}\,,\qquad \qquad x_{k}^+=x_{k}^-+\frac{2x_m-(n_j-1)e_i}{n_j}\,.\label{eq:edgeS1}
\eal
It can be checked that $x_{k}^-=x_{k-1}^++e_i$, so, it is possible to end up with two side-by-side injectors, which is numerically equivalent to having $n_j-1$ injectors, $n_j-2$ of width $e_i$ plus one of width $2e_i$.
For this case, 60 episodes have been run, each of which performs 1500 iterations with time step $\Delta t = 0.1$ to march the same initial condition (consisting of zero velocity and uniform temperature, except in the solid domain) to steady state, using the level set, velocity and temperature as multiple-component criterion to adapt the mesh (initially pre-adapted using the sole level set) every $5$ time steps under the constraint of a fixed number of elements $n_{el} = 15000$. This represents 480 simulations, each of which is performed on 8 cores and lasts 10mn, hence 80h of total CPU cost. 

It is out of the scope of this work to analyze in details the many flow patterns that develop 
when the blown fluid travels through the cavity. Suffice it to say that the outcome depends dramatically on the injectors arrangement, and features complex rebound phenomena (either fluid/solid, when a jet impinges on the cavity walls or on the workpiece itself, or fluid/fluid, when a deflected jet meets the crossflow of another jet), leading to the formation of multiple recirculation varying in number, position and size.
Several such cases are illustrated in figure~\ref{fig:forced2d_S1_vtu} via iso-contours of the steady-state temperature distributions, together with the corresponding adapted meshes colored by the magnitude of velocity to illustrate the ability of the numerical framework 
to capture accurately all boundary layers and shear regions via extremely stretched elements. 

\begin{figure}[t!]
\setlength{\unitlength}{1cm}
\begin{picture}(20,5.2)
\put(0.4,-0.1){\includegraphics[trim=175 87.5 140 340pt,clip,height=5cm]{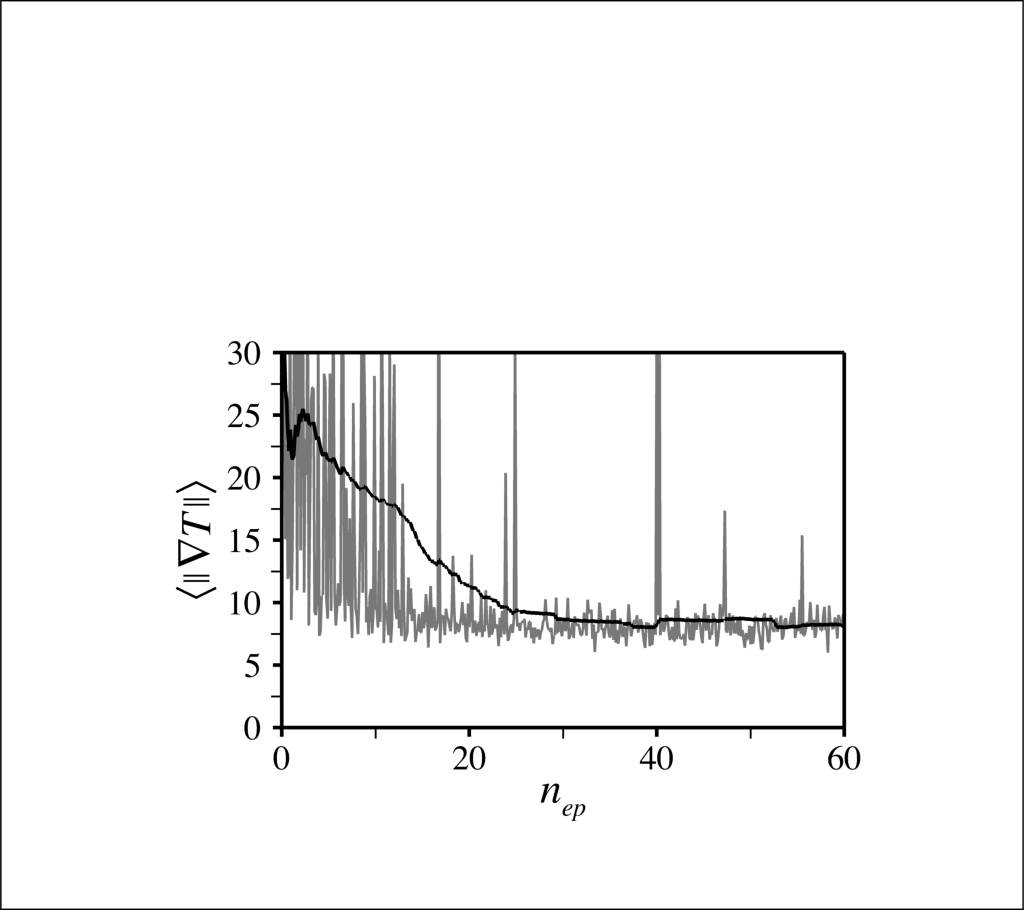}}
\put(7.65,-0.1){\includegraphics[trim=175 87.5 140 340pt,clip,height=5cm]{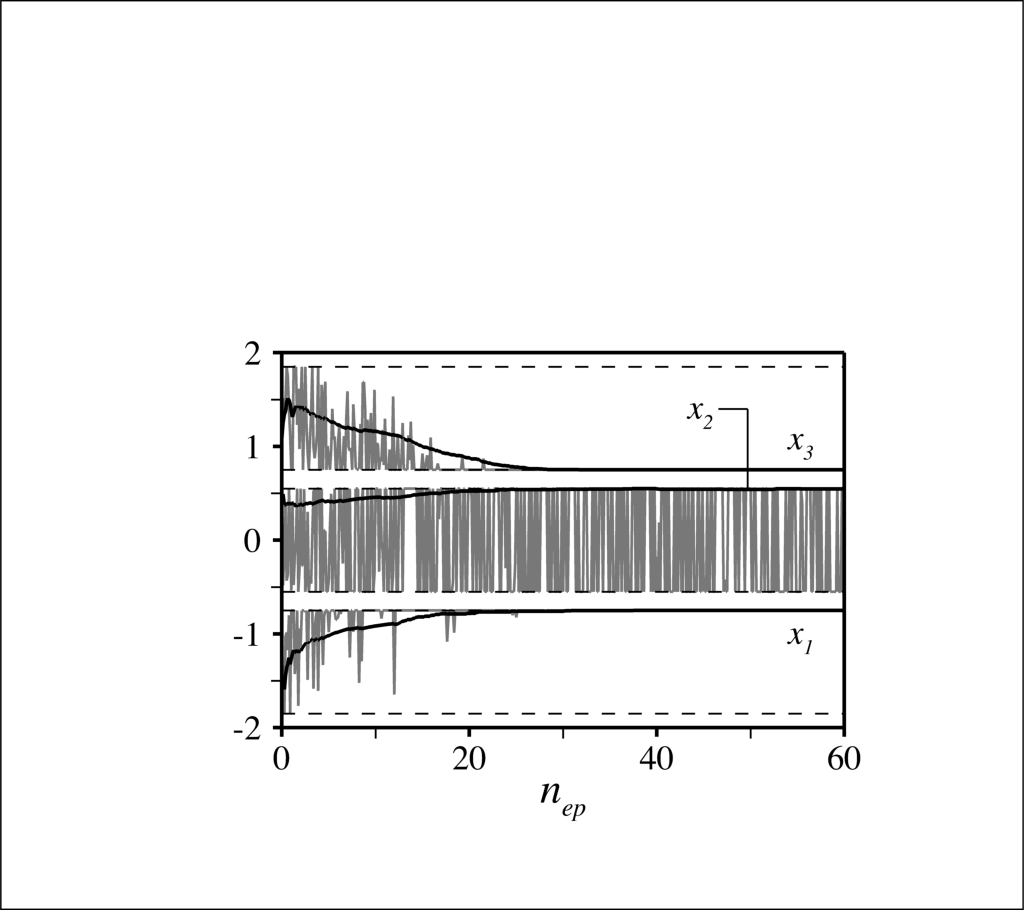}}
\put(7.85,5.2){(b)}
\put(0.55,5.2){(a)}
\end{picture}
\caption{(a) Evolution per learning episode of the instant (in grey) and moving average (in black) rewards under the fixed domain decomposition strategy $S_1$. (b) Same as (a) for the injectors center positions, with admissible values delimited by the dashed lines.}
\label{fig:forced2d_S1_reward}
\end{figure}

\begin{figure}[t!]
\setlength{\unitlength}{1cm}
\begin{picture}(20,2.9)
\put(7.2,0){\includegraphics[trim=175 110 140 575pt,clip,height=2.575cm]{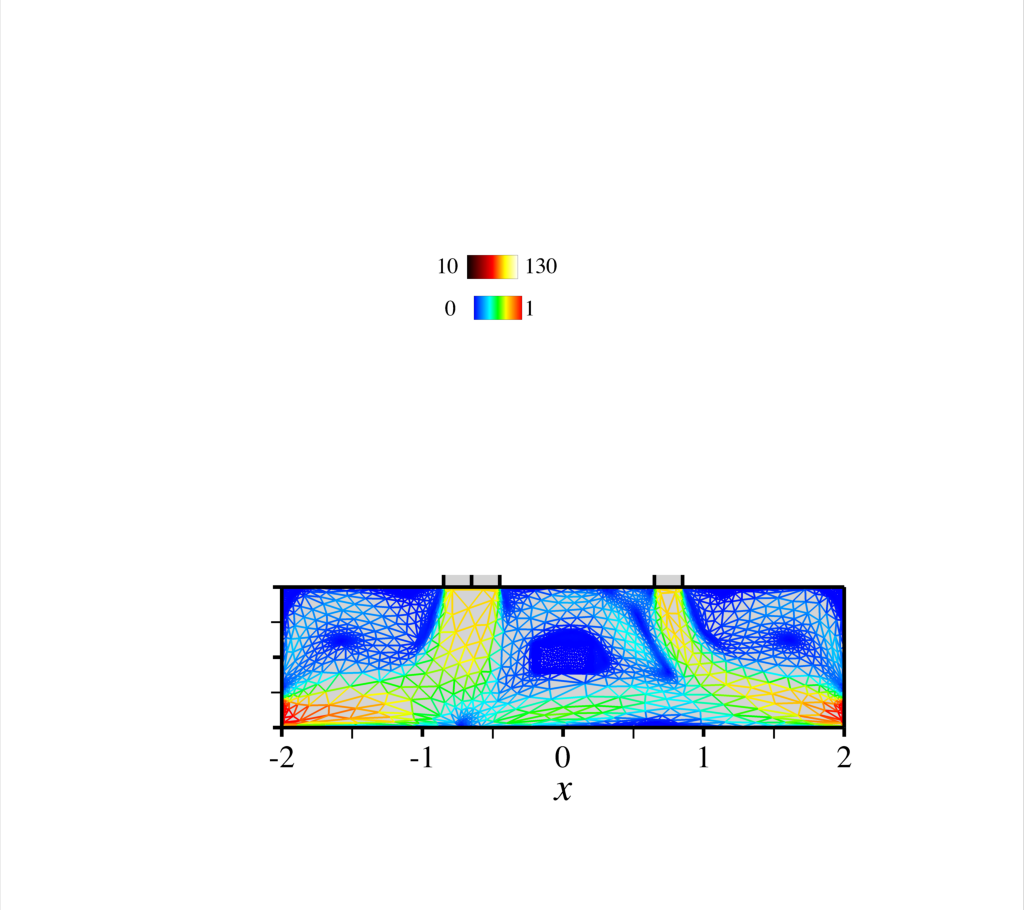}}
\put(0,0){\includegraphics[trim=175 110 140 575pt,clip,height=2.575cm]{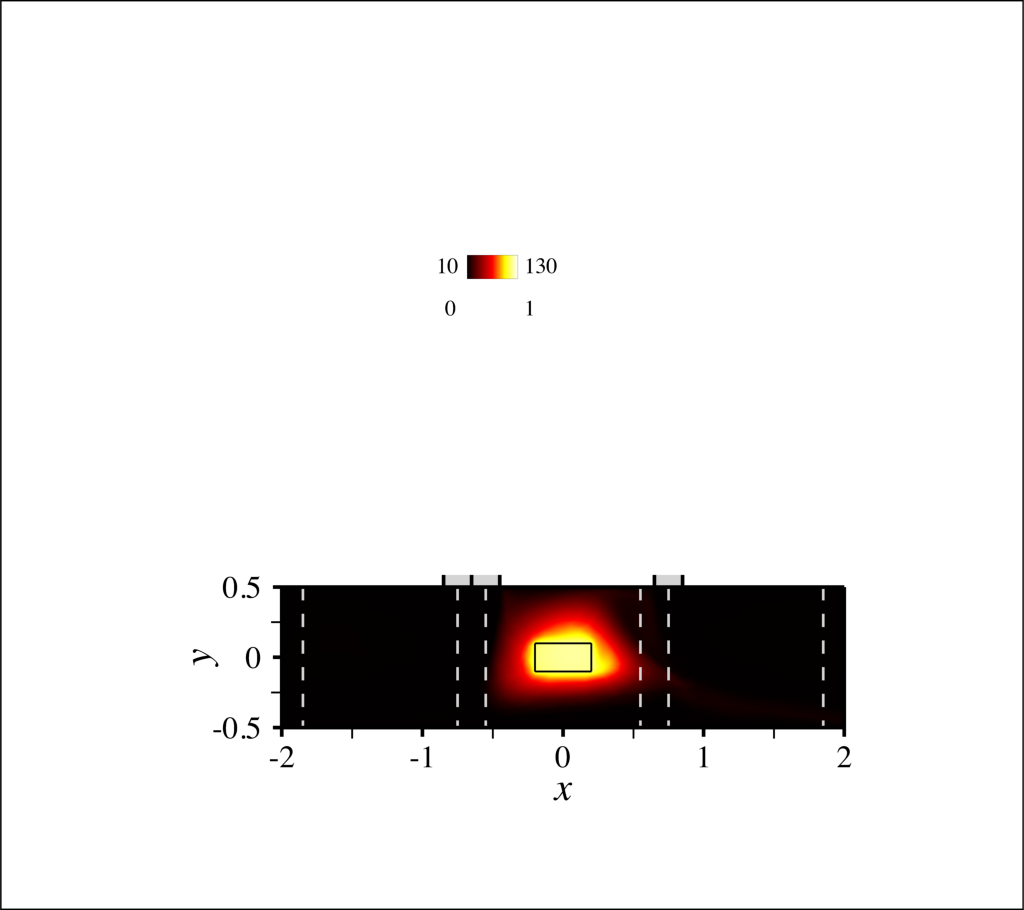}}
\put(5.6,2.7){\includegraphics[trim=420 635 440 245pt,clip,height=0.45cm]{fig11a_bar}}
\put(12.9,2.7){\includegraphics[trim=420 635 440 245pt,clip,height=0.45cm]{fig11b_bar}}
\put(7.85,2.9){(b)}
\put(0.55,2.9){(a)}
\end{picture}
\caption{Same as figure~\ref{fig:forced2d_S1_vtu} for the optimal arrangement of 3 injectors under the fixed domain decomposition strategy $S_1$.}
\label{fig:forced2d_S1_opt}
\end{figure}

One point worth mentioning is that the individual position signals are best suited to draw robust quantitative conclusion, as there is noise in the reward signal shown in figure~\ref{fig:forced2d_S1_reward}(a).
We believe the issue to be twofold: on the one hand, the reward is approximated from point-wise temperature data (similar to experimental measurements) that are more sensitive to small numerical errors (e.g., the interpolation error at the probes position) than an integral quantity. On the other hand, the mesh adaptation procedure is not a deterministic process, as 
the outcome depends on the processors and number of processors used, and any initial difference propagates over the course of the simulation because the meshes keep being adapted dynamically. In return, 
two exact same control parameters can thus yield different rewards on behalf of different interpolation errors at the probes position. This likely slows down learning and convergence, but we show in figure~\ref{fig:forced2d_S1_reward}(b) that the moving average distribution does converge to an optimal arrangement after roughly $25$ episodes. The latter consists of an injector at the right-end of the left subdomain ($x_{1}{^\star}=-0.75$) and 
two side-by-side injectors sitting astride the center and right subdomains ($x_{2}{^\star}=0.55$ and $x_{3}{^\star}=0.75$), that enclose the workpiece in a double-cell recirculation; see figure~\ref{fig:forced2d_S1_opt}. These values have been computed by averaging the instant positions of each injector over the $10$ latest episodes, with
variations $\pm0.002$ computed from the root-mean-square of the moving average over the same interval, a procedure that will be used consistently to assess convergence for all cases reported in the following. The efficiency of the control itself is estimated by computing the magnitude of tangential heat flux averaged over the same interval, found to be $\langle ||\nabla_{\|}T||\rangle{^\star}\sim8.3$.
Note, the position $x_{2}{^\star}$ is actually obtained by averaging the absolute value of the instant position $x_{2}$ (although the true, signed value is depicted in the figure), as the center injector keeps oscillating between two end positions $\pm0.55$ on behalf of reflectional symmetry with respect to the vertical centerline.

\begin{figure}[t!]
\setlength{\unitlength}{1cm}
\begin{picture}(20,5.1)
\put(7.2,2.2){\includegraphics[trim=175 110 140 575pt,clip,height=2.575cm]{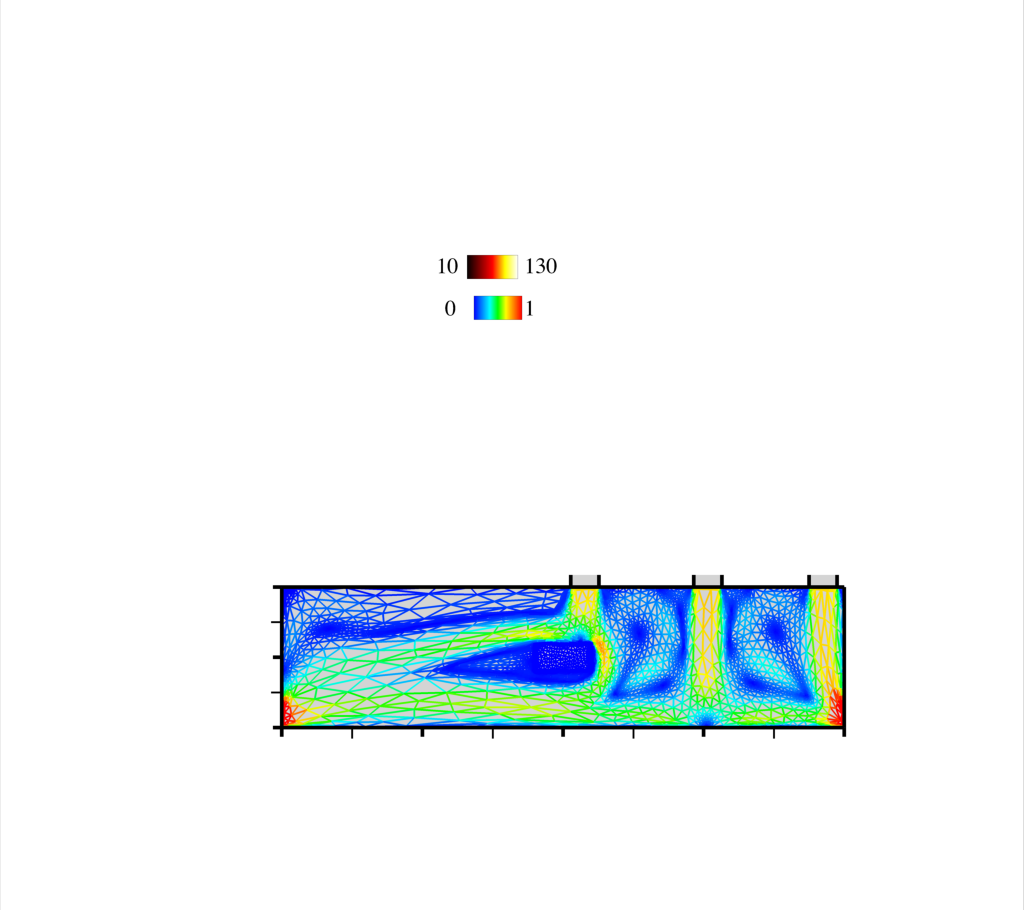}}
\put(0,2.2){\includegraphics[trim=175 110 140 575pt,clip,height=2.575cm]{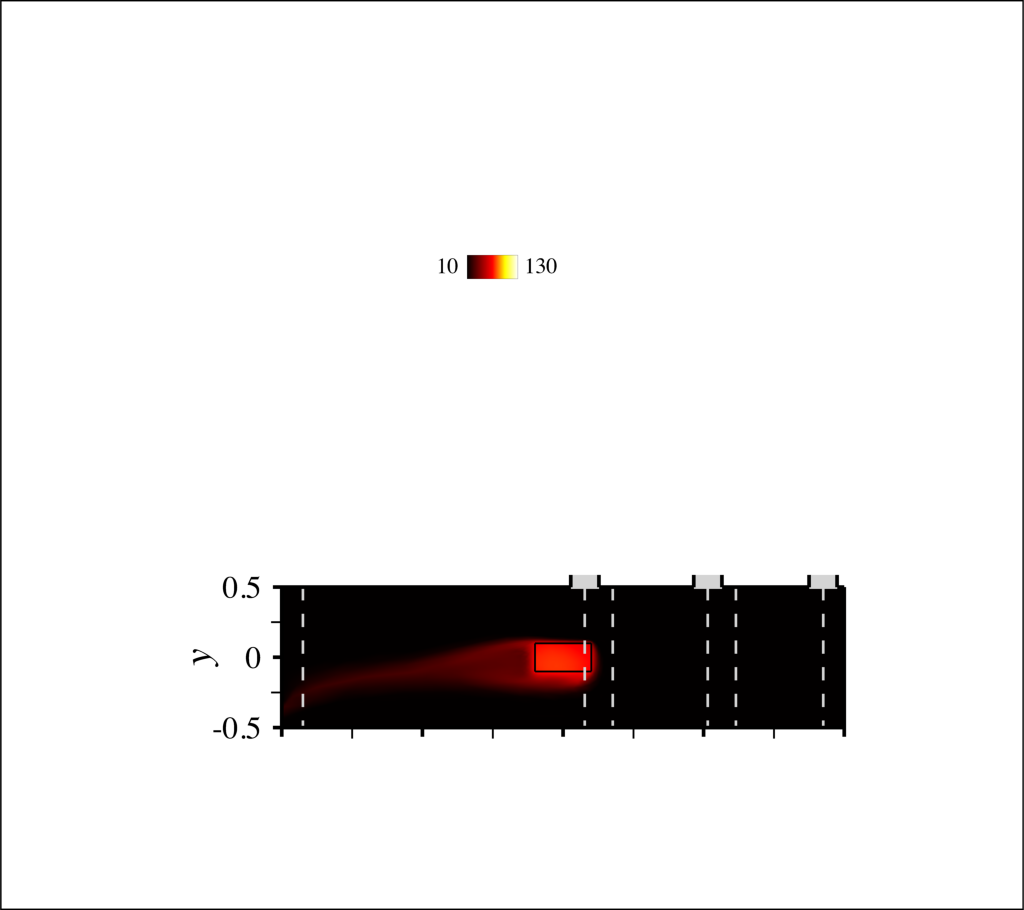}}
\put(7.2,0){\includegraphics[trim=175 110 140 575pt,clip,height=2.575cm]{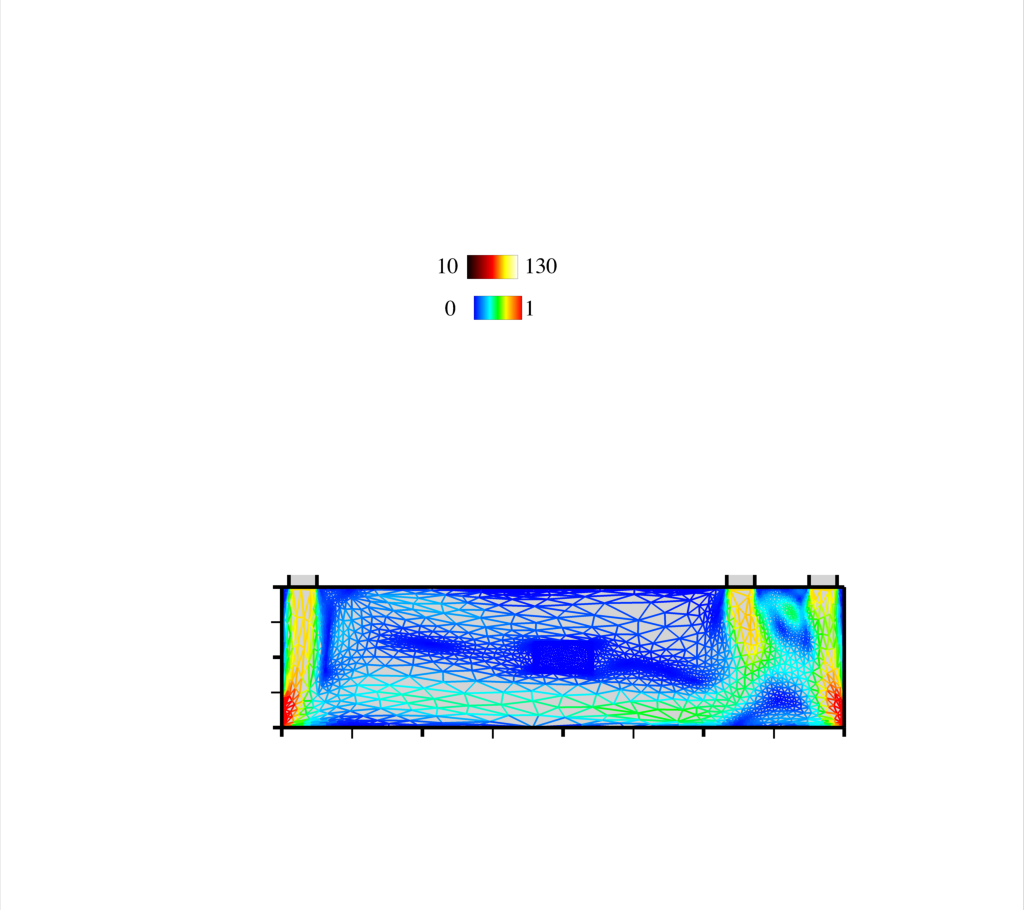}}
\put(0,0){\includegraphics[trim=175 110 140 575pt,clip,height=2.575cm]{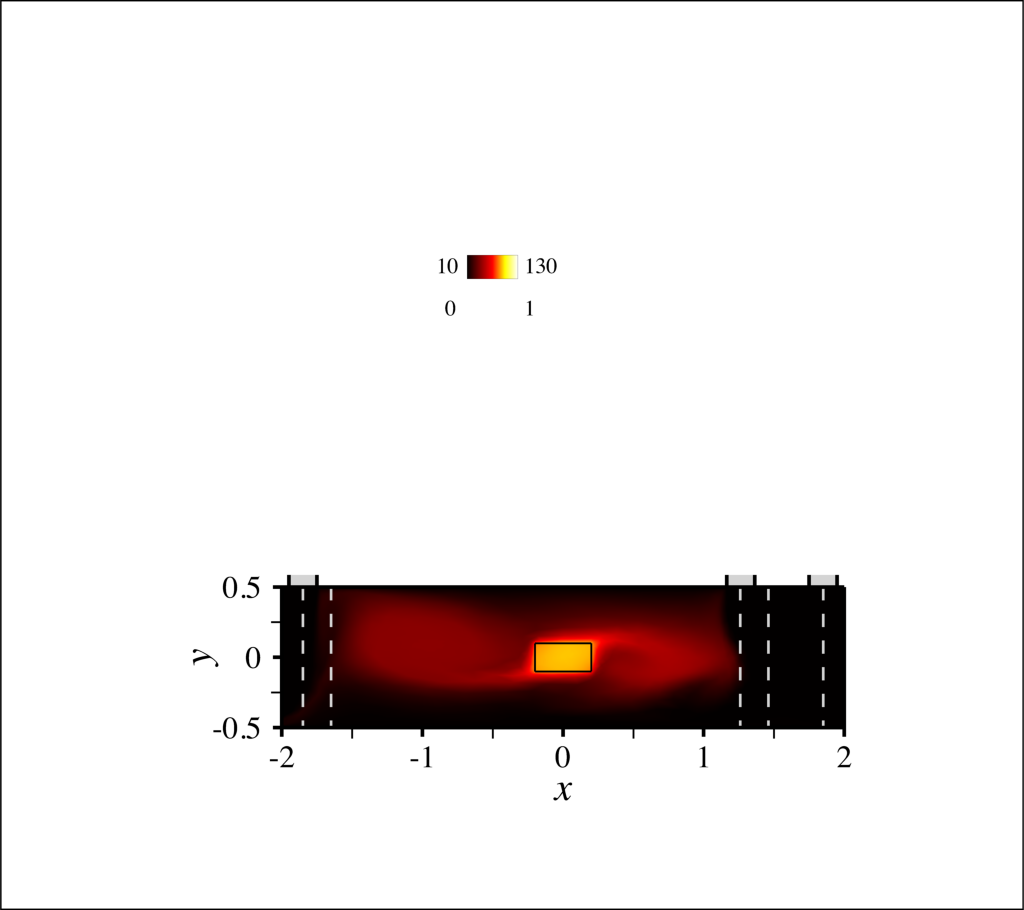}}
\put(5.6,4.9){\includegraphics[trim=420 635 440 245pt,clip,height=0.45cm]{fig11a_bar}}
\put(12.9,4.9){\includegraphics[trim=420 635 440 245pt,clip,height=0.45cm]{fig11b_bar}}
\put(7.85,5.1){(b)}
\put(0.55,5.1){(a)}
\end{picture}
\caption{(a) Steady-state temperature against arrangements of 3 injectors, with admissible values under the follow-up strategy $S_2$ delimited by the dashed lines. (b) Adapted meshes colored by the magnitude of velocity.}
\label{fig:forced2d_S2_vtu}
\end{figure}

\begin{figure}[t!]
\setlength{\unitlength}{1cm}
\begin{picture}(20,5.2)
\put(0.5,-0.1){\includegraphics[trim=235 87.5 140 340pt,clip,height=5cm]{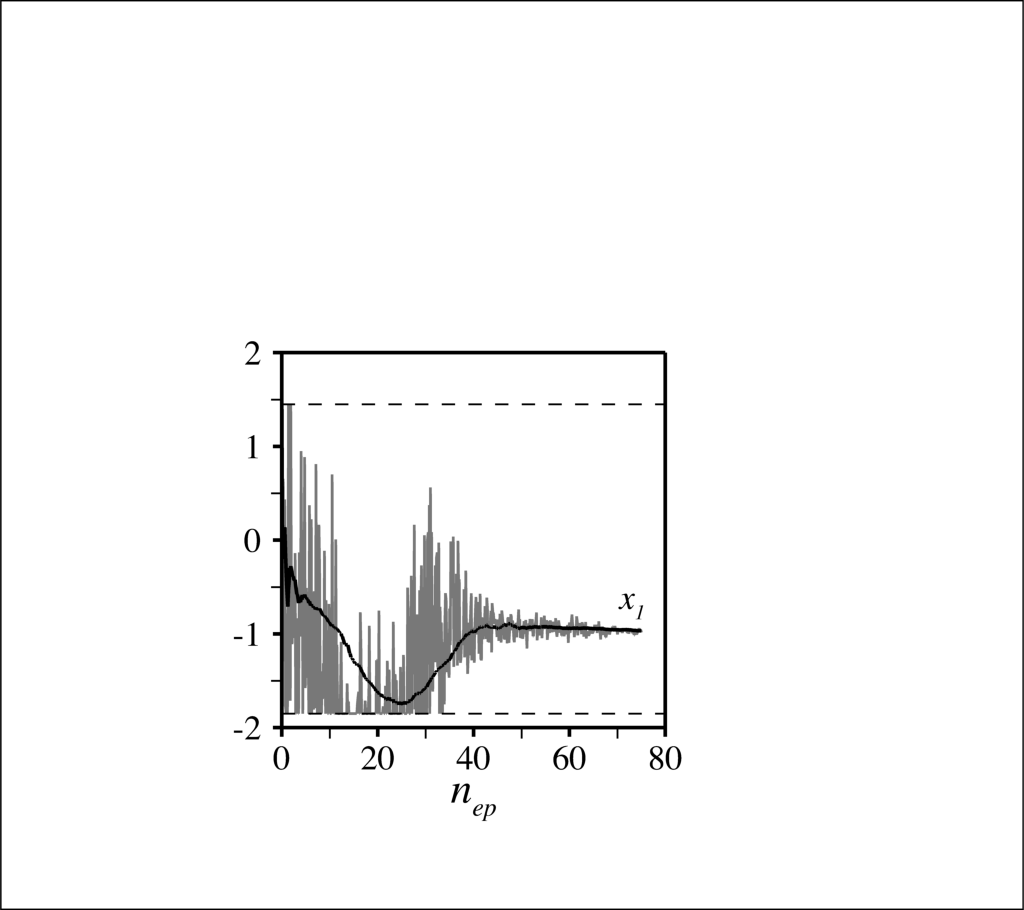}}
\put(5.3,-0.1){\includegraphics[trim=235 87.5 140 340pt,clip,height=5cm]{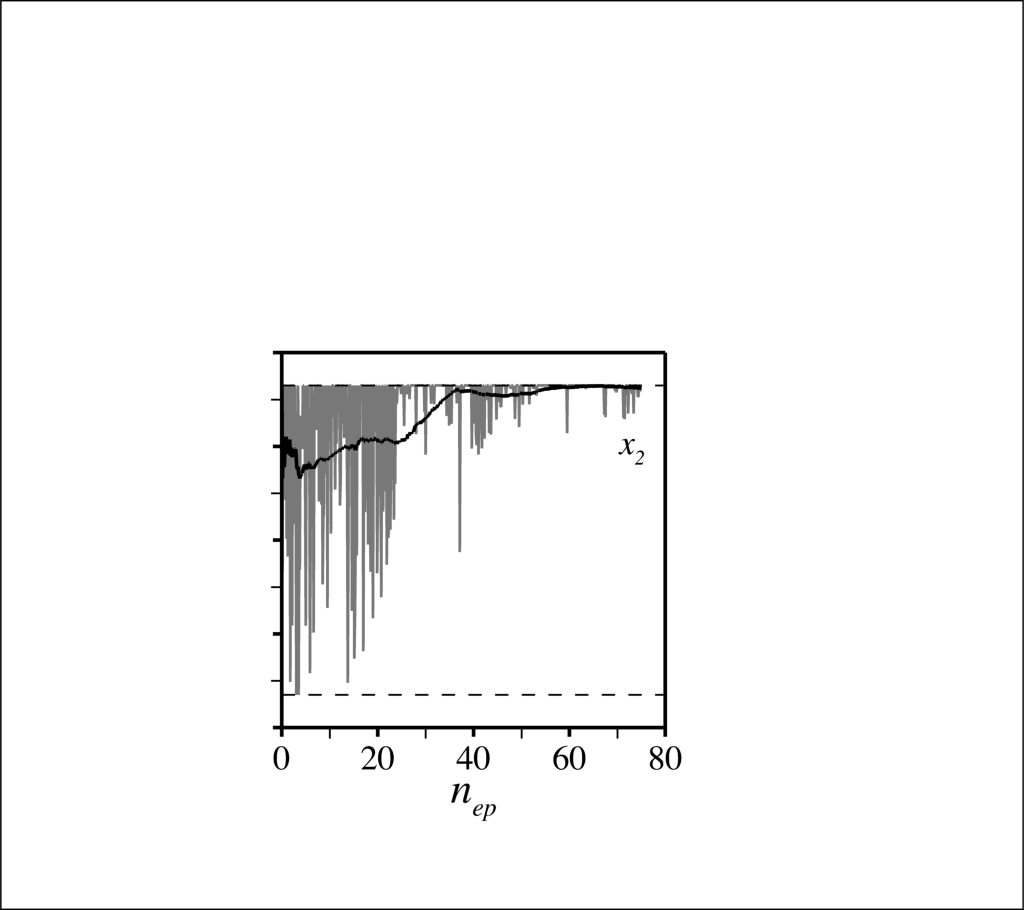}}
\put(10.1,-0.1){\includegraphics[trim=235 87.5 140 340pt,clip,height=5cm]{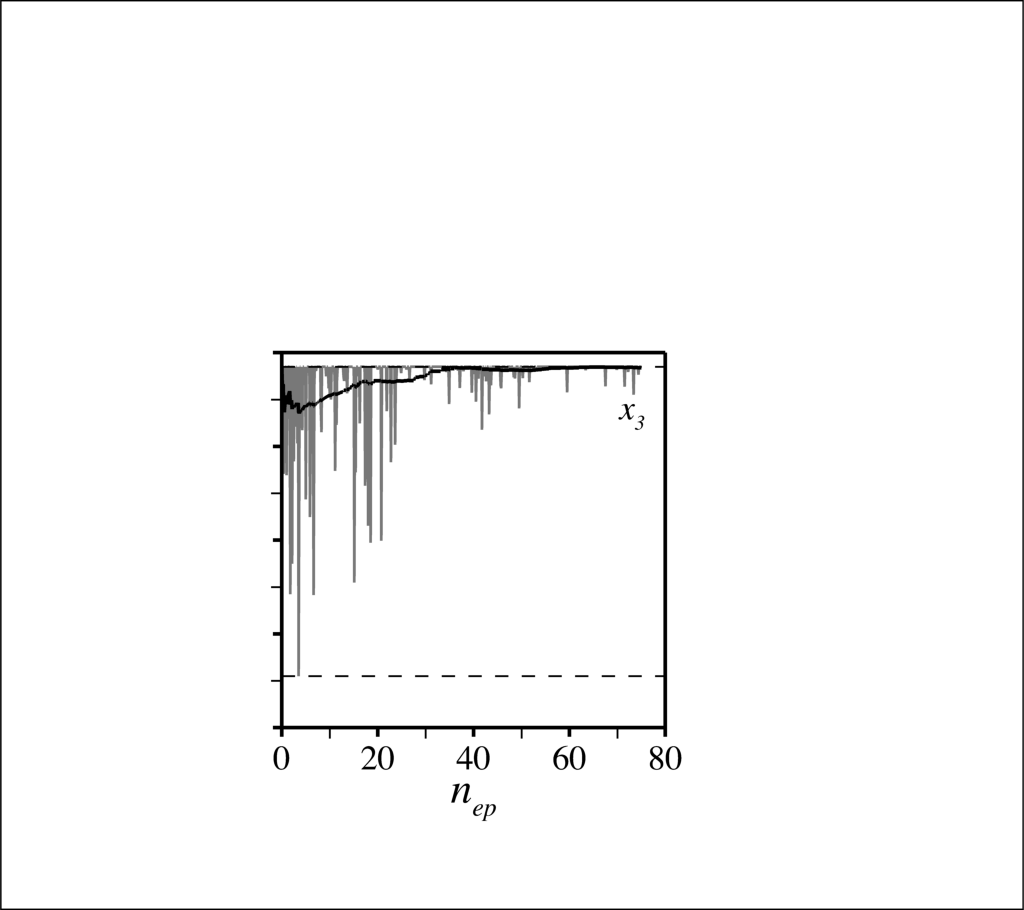}}
\put(10.15,5.2){(c)}
\put(5.3,5.2){(b)}
\put(0.55,5.2){(a)}
\end{picture}
\caption{Evolution per learning episode of the instant (in grey) and moving average (in black) injectors center positions under the follow-up strategy $S_2$, with admissible values delimited by the dashed lines.}
\label{fig:forced2d_S2_reward}
\end{figure}
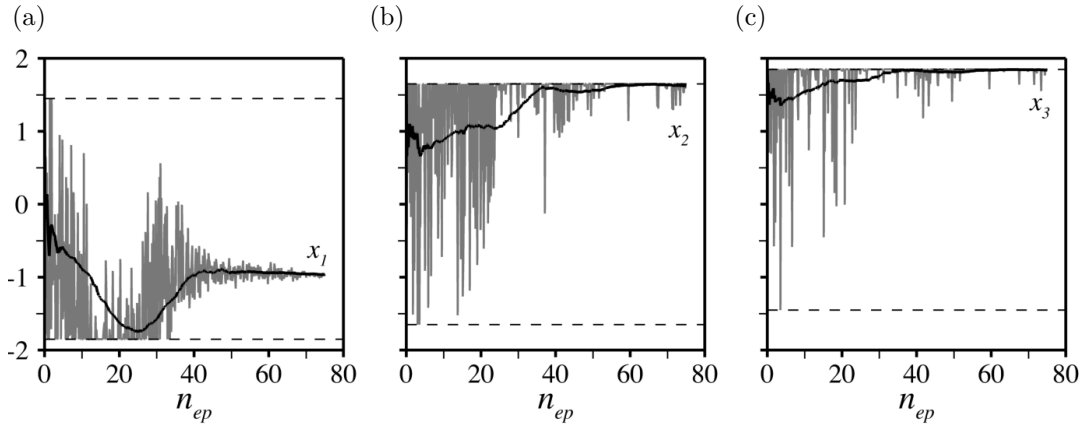

\subsubsection{Follow-up strategy}


{A less constrained problem is considered here using} the so-called follow-up strategy $S_2$, in which all injectors are distributed sequentially the ones with respect to the others. 
The corresponding edge values
\bal
x_{1}^-&=-x_m\,,& x_{1}^+&=x_m-(n_j-1)e_i\,,\\
x_{k}^-&=x_{k-1}^++e_i\,,& x_{k}^+&=x_m-(n_j-k)e_i\,,\label{eq:edgeS2}
\eal
readily express that the
$k$-th injector is forced to sit between the $k-1$-th one and the upper-right cavity edge while leaving enough space to distribute the remaining $n_j-k$ injectors, which increases the size of the control parameter space while again leaving the possibility for 
side-by-side injectors (since $x_{k}^-=x_{k-1}^++e_i$ by construction).
75 episodes have been run for this case following the exact same procedure as above, i.e., marching the zero-initial condition in time up to $t = 150$ with $\Delta t = 0.1$, hence 600 simulations, each of which is performed on 8 cores and lasts 10mn, hence 100h of total CPU cost. 

\begin{figure}[t!]
\setlength{\unitlength}{1cm}
\begin{picture}(20,2.9)
\put(7.2,0){\includegraphics[trim=175 110 140 575pt,clip,height=2.575cm]{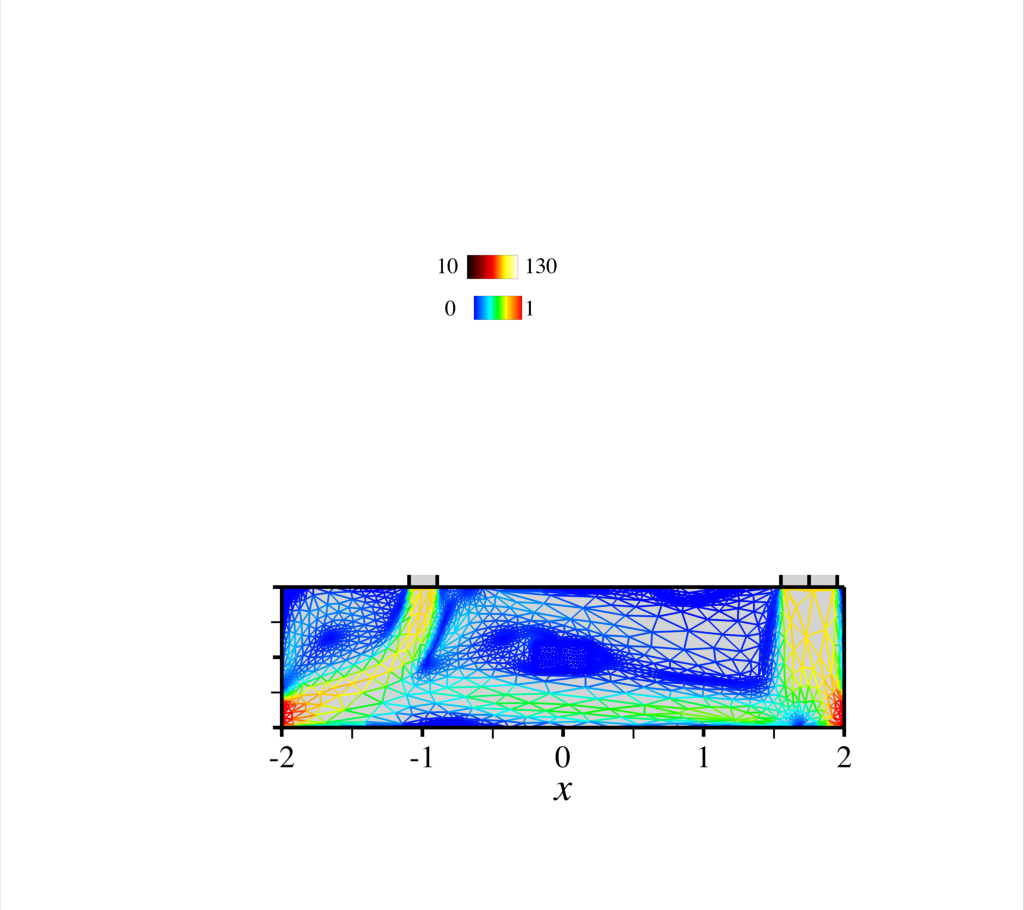}}
\put(0,0){\includegraphics[trim=175 110 140 575pt,clip,height=2.575cm]{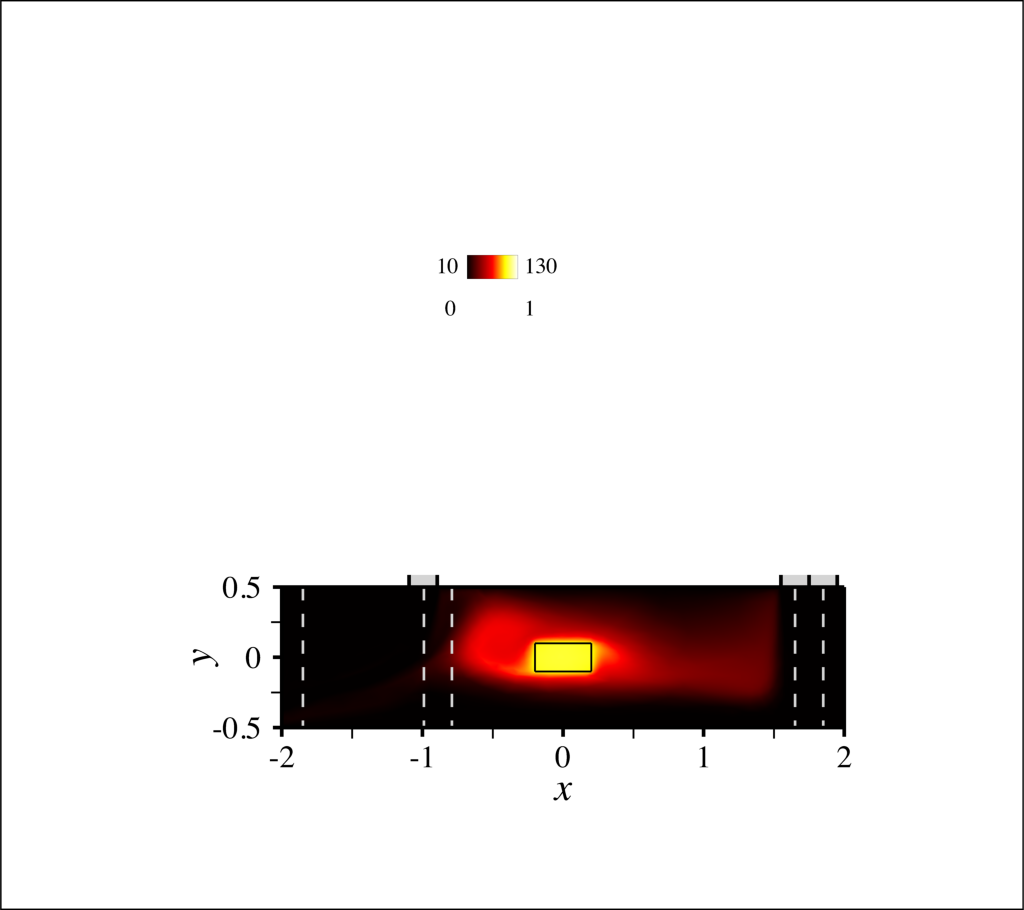}}
\put(5.6,2.7){\includegraphics[trim=420 635 440 245pt,clip,height=0.45cm]{fig11a_bar}}
\put(12.9,2.7){\includegraphics[trim=420 635 440 245pt,clip,height=0.45cm]{fig11b_bar}}
\put(7.85,2.9){(b)}
\put(0.55,2.9){(a)}
\end{picture}
\caption{Same as figure~\ref{fig:forced2d_S2_vtu} for the optimal arrangement of 3 injectors under the follow-up strategy $S_2$.}
\label{fig:forced2d_S2_opt}
\end{figure}

The computed flow patterns closely resemble those obtained under the previous fixed domain decomposition strategy, although figure~\ref{fig:forced2d_S2_vtu} exhibits increased dissymmetry when two or more injectors move simultaneously to the same side of the cavity. We show in figure~\ref{fig:forced2d_S2_reward} that the moving average distribution converges after roughly $60$ episodes, with the optimal arrangement consisting of one injector roughly midway between the left cavity sidewall and the workpiece ($x_{1}{^\star}=-0.96$), and two side-by-side injectors at the right end of the cavity ($x_{2}{^\star}=1.65$ and $x_{3}{^\star}=1.85$). The variations over the same interval are by $\pm 0.006$; see also figure~\ref{fig:forced2d_S2_opt} for the corresponding flow pattern. Convergence here is much slower than under $S_1$, as the search for an optimal is complicated by the fact that all injector positions are interdependent the ones on the others and it is up to the network to figure out exactly how. 
Another contingent matter is that the agent initially
spans a fraction of the control parameter space because the large values of $x_1$ considered limit the space available to distribute the other two injectors. This is all the more so as such configurations turn to be far from optimality, for instance the magnitude of tangential heat flux is 
 $\langle ||\nabla_{\|}T||\rangle\sim 41.3$ for $x_1=1.45$, $x_2=1.65$ and $x_3=1.85$, but $\langle ||\nabla_{\|}T||\rangle{^\star}\sim6.3$ at optimality. The latter value is smaller than the optimal achieved under $S_1$, consistently with the fact that all positions spanned under $S_1$ are admissible under $S_2$, hence the $S_1$ optimal is expected to be a $S_2$ sub-optimal. 

\subsubsection{Free strategy}

We examine now a third strategy $S_3$ referred to as the free strategy, in which all injectors are independent and free to move along the top cavity wall {(a mildly constrained optimization problem, relevant for cases where the design is flexible and the
practitioner has great freedom to act).} 
The edge values for the position $x_k$ of the $k$-th injector read
\bal
x_{k}^-=-x_m\,,\qquad \qquad x_{k}^+=x_m\,,\label{eq:edgeS3}
\eal
so two injectors can end up side-by side and even overlapping one another if $|x_l-x_m|<e_i$. If so, we 
implement a single injector of width $e_i+|x_l-x_m|$ and maintain the blowing velocity (not the flow rate) for the purpose of automating the set-up design process, meaning that having $n_j$ injectors, two of which overlap exactly (i.e., $|x_l-x_m|=0$) is rigorously equivalent to having $n_j-1$ injectors.
60 episodes have been run for this case following the exact same procedure as above.

\begin{figure}[t!]
\setlength{\unitlength}{1cm}
\begin{picture}(20,5.1)
\put(7.2,2.2){\includegraphics[trim=175 110 140 575pt,clip,height=2.575cm]{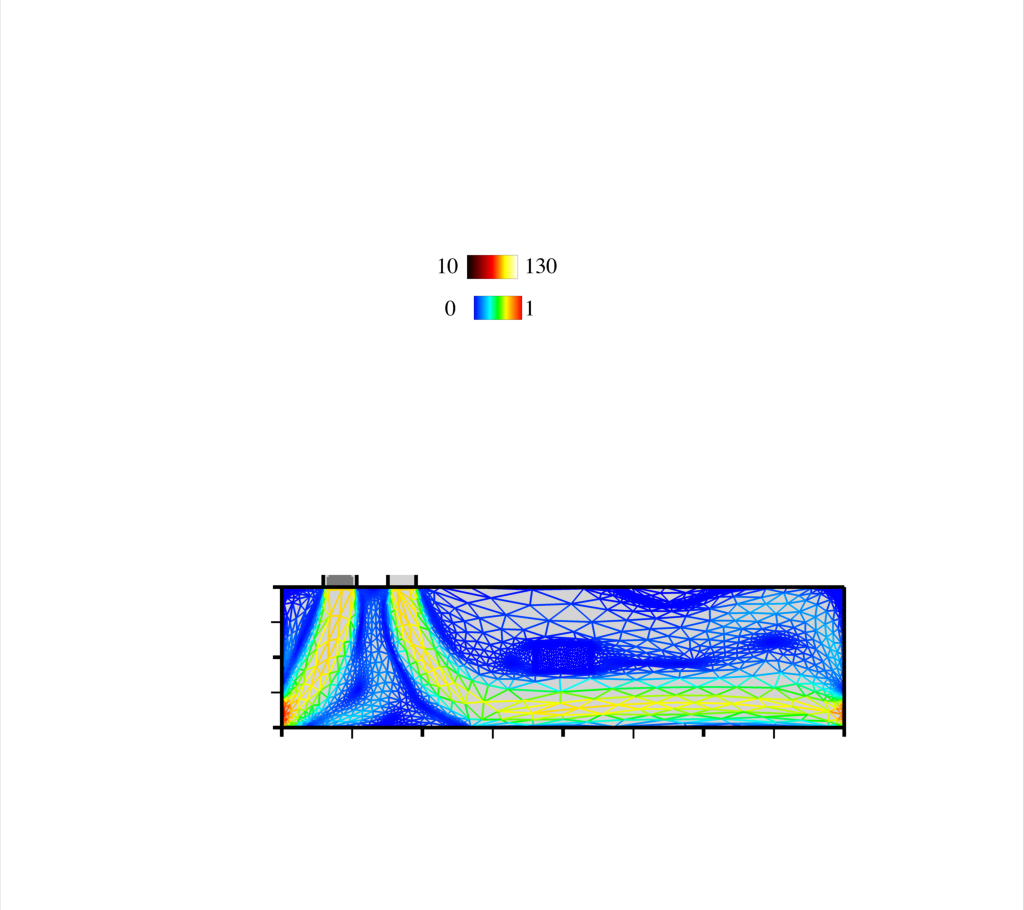}}
\put(0,2.2){\includegraphics[trim=175 110 140 575pt,clip,height=2.575cm]{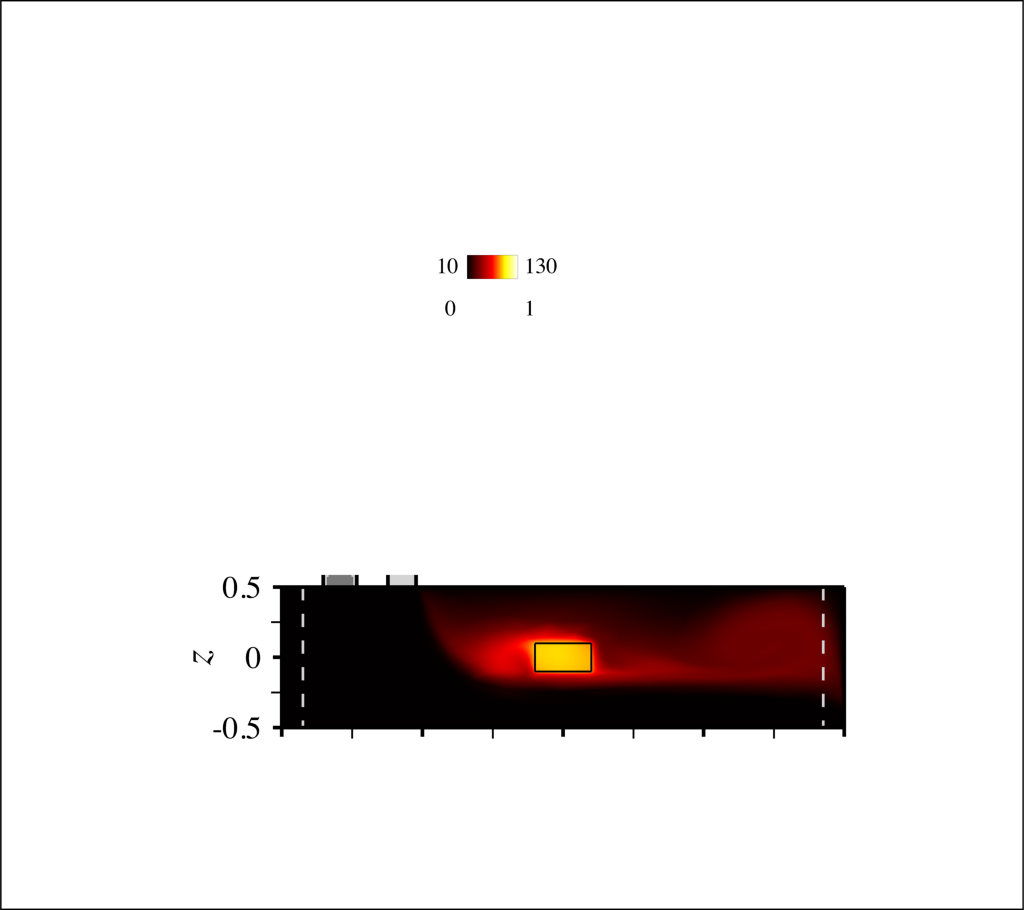}}
\put(7.2,0){\includegraphics[trim=175 110 140 575pt,clip,height=2.575cm]{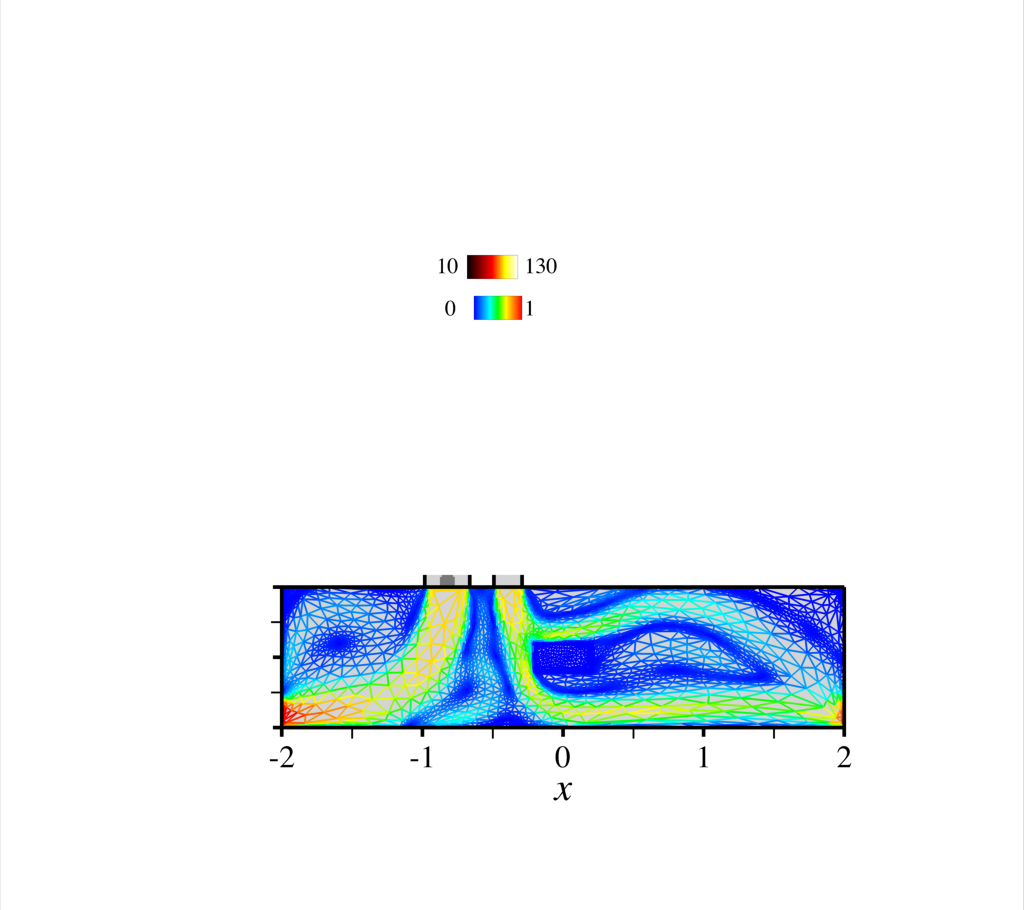}}
\put(0,0){\includegraphics[trim=175 110 140 575pt,clip,height=2.575cm]{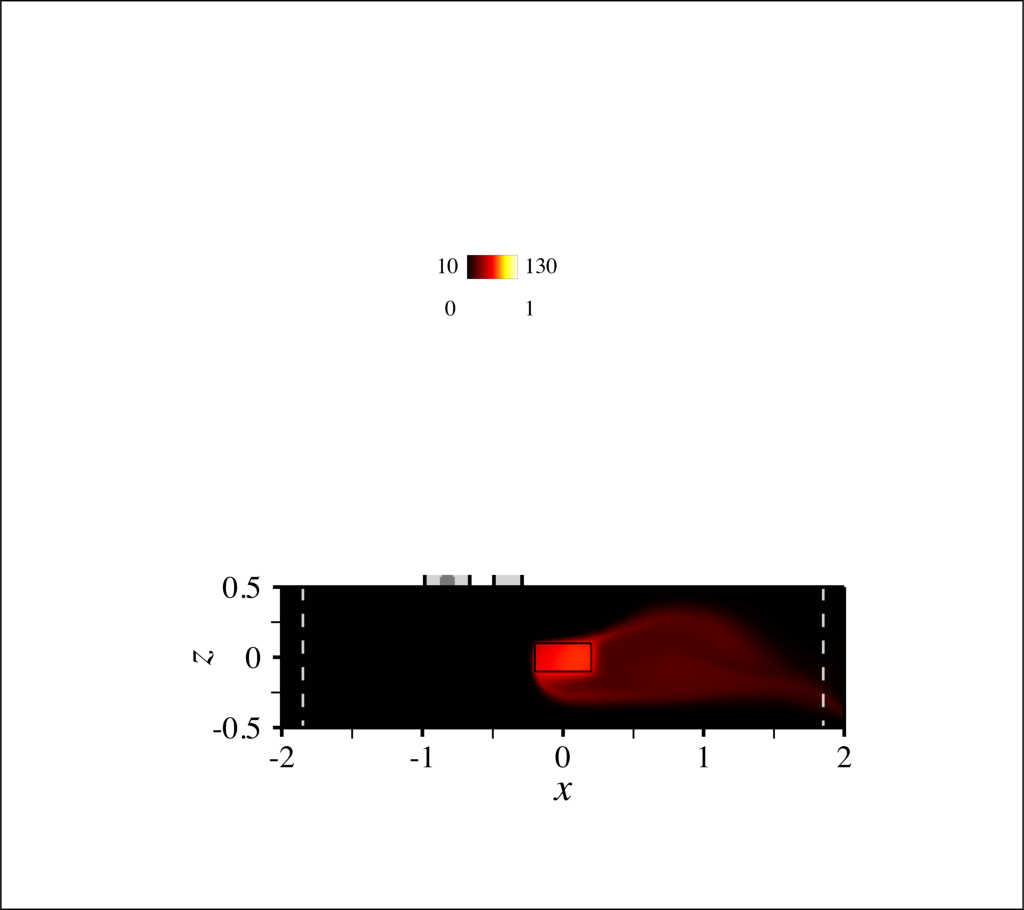}}
\put(5.6,4.9){\includegraphics[trim=420 635 440 245pt,clip,height=0.45cm]{fig11a_bar}}
\put(12.9,4.9){\includegraphics[trim=420 635 440 245pt,clip,height=0.45cm]{fig11b_bar}}
\put(7.85,5.1){(b)}
\put(0.55,5.1){(a)}
\end{picture}
\caption{(a) Steady-state temperature against arrangements of 3 injectors, with admissible values under the free strategy $S_3$ delimited by the dashed lines and overlaps marked by the dark grey shade. (b) Adapted meshes colored by the magnitude of velocity.}
\label{fig:forced2d_S3_vtu}
\end{figure}

\begin{figure}[t!]
\setlength{\unitlength}{1cm}
\begin{picture}(20,5.2)
\put(0.5,-0.1){\includegraphics[trim=235 87.5 140 340pt,clip,height=5cm]{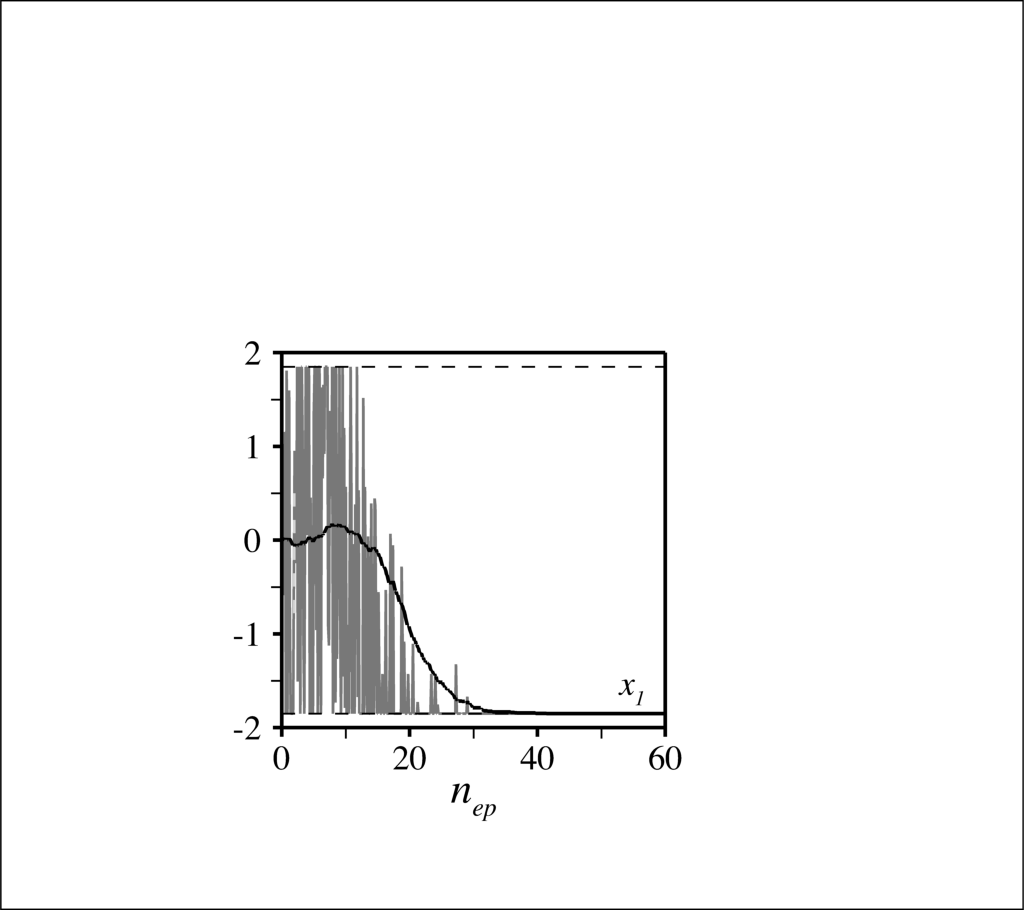}}
\put(5.3,-0.1){\includegraphics[trim=235 87.5 140 340pt,clip,height=5cm]{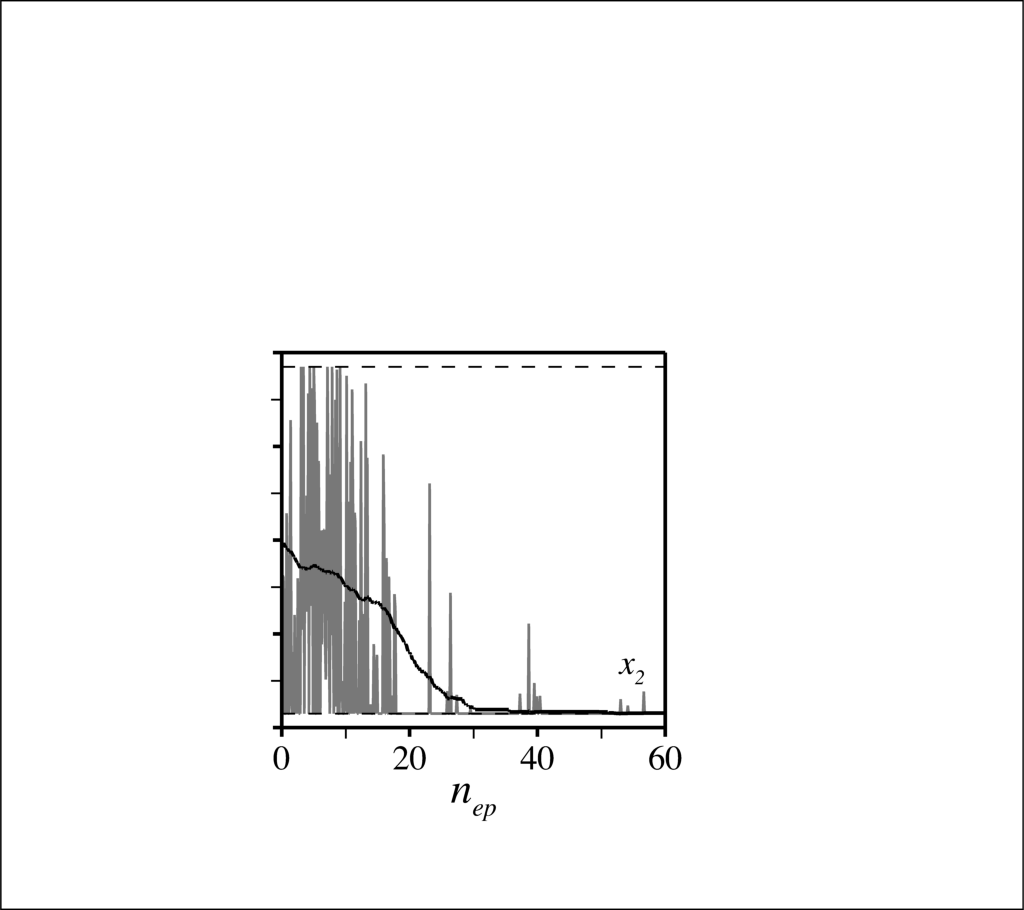}}
\put(10.1,-0.1){\includegraphics[trim=235 87.5 140 340pt,clip,height=5cm]{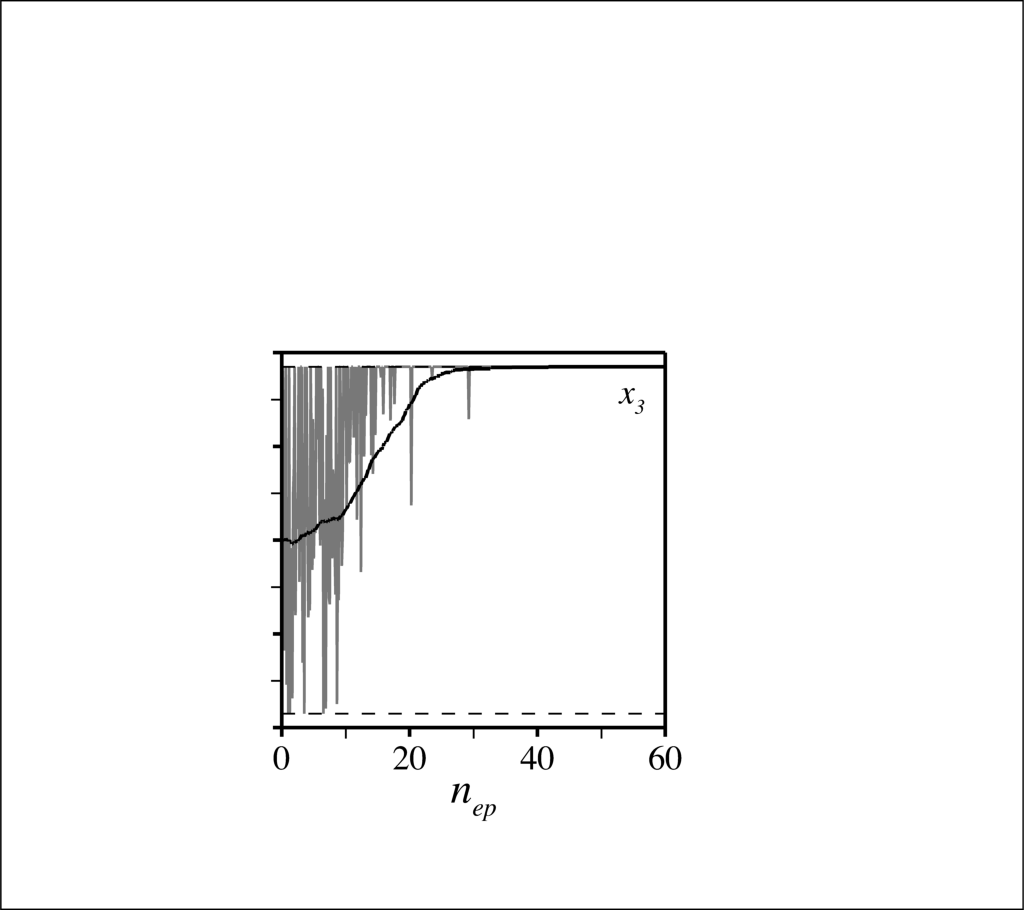}}
\put(10.15,5.2){(c)}
\put(5.3,5.2){(b)}
\put(0.55,5.2){(a)}
\end{picture}
\caption{Evolution per learning episode of the instant (in grey) and moving average (in black) injectors center positions under the free strategy $S_3$, with admissible values delimited by the dashed lines.}
\label{fig:forced2d_S3_reward}
\end{figure}

\begin{figure}[t!]
\setlength{\unitlength}{1cm}
\begin{picture}(20,2.9)
\put(7.2,0){\includegraphics[trim=175 110 140 575pt,clip,height=2.575cm]{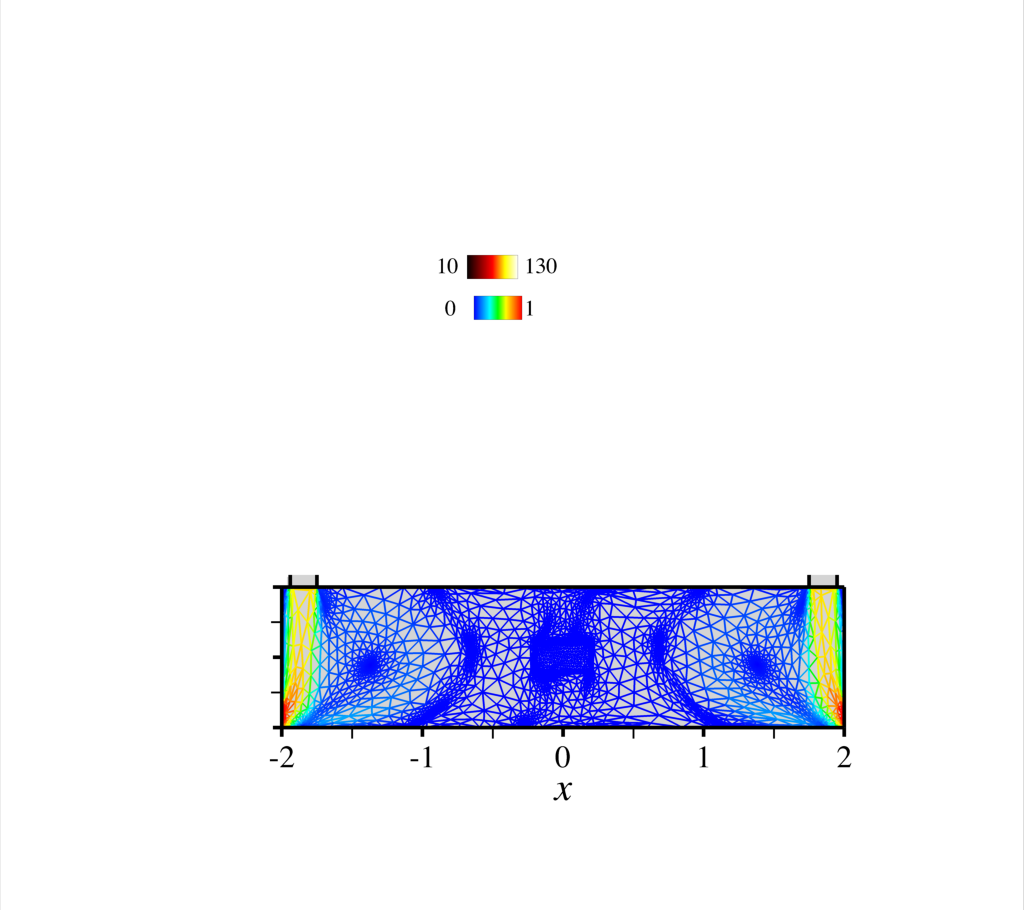}}
\put(0,0){\includegraphics[trim=175 110 140 575pt,clip,height=2.575cm]{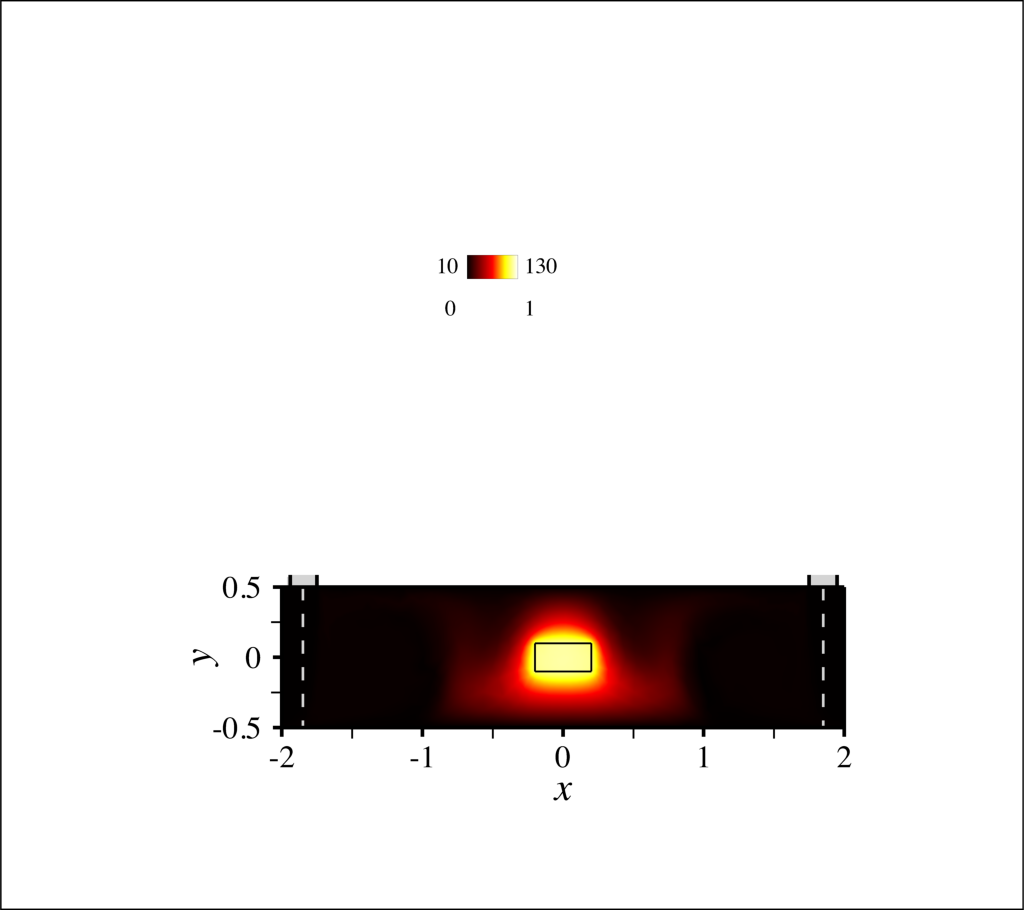}}
\put(5.6,2.7){\includegraphics[trim=420 635 440 245pt,clip,height=0.45cm]{fig11a_bar}}
\put(12.9,2.7){\includegraphics[trim=420 635 440 245pt,clip,height=0.45cm]{fig11b_bar}}
\put(7.85,2.9){(b)}
\put(0.55,2.9){(a)}
\end{picture}
\caption{Same as figure~\ref{fig:forced2d_S3_vtu} for the optimal arrangement of 3 injectors under the free strategy $S_3$.}
\label{fig:forced2d_S3_opt}
\end{figure}

All flow patterns are reminiscent of those obtained under the previous fix decomposition $S_1$ and follow-up $S_2$ strategies, even when two injectors overlap; see figure~\ref{fig:forced2d_S3_vtu}. Other than that, we show in figures~\ref{fig:forced2d_S3_reward} that the moving average distribution converges to an optimal consisting of two injectors almost perfectly overlapping one another at the left end of the cavity ($x_{1}{^\star}=-1.85$ and $x_{2}{^\star}=-1.82$), and a third injector at the right end of the cavity ($x_{3}{^\star}=1.85$). The variations over the same interval are by $\pm 0.007$, and the associated flow pattern shown in figure~\ref{fig:forced2d_S3_opt} is symmetrical and features two large recirculation regions on either side of the workpiece. Convergence occurs after roughly $40$ episodes, i.e., faster than under $S_2$ (consistently with the fact that there is no need to learn anymore about how the network outputs depend the ones on the others) but slower than under $S_1$ (consistently with the fact that the size of the control parameter space has increased substantially). It is worth noticing that the system is invariant by permutations of the network outputs, meaning that there exist $2^{n_j}-2$ distributions (hence 6 for $n_j=3$) associated with the same reward. Nonetheless, a single optimal is selected, which is essentially fortuitous since the agent does not learn about symmetries under the optimization process (otherwise $S_1$ would have similarly selected a single optimal).
The magnitude of tangential heat flux is $\langle ||\nabla_{\|}T||\rangle{^\star}\sim11.2$ at optimality, i.e., larger than that achieved under $S_2$. This can seem surprising at first, because all positions spanned under $S_2$ are admissible under $S_3$, and the $S_2$ optimal is thus expected to be a $S_3$ sub-optimal. However, the argument does not hold here because the overlap in the $S_3$ optimal reduces the flow rate to that of a two-injectors set-up, so the comparison should be with the $S_2$ optimal with $n_j=2$.

\subsubsection{Inverse strategy}\label{section:forced2d:inverse}

Finally, we propose here to make the most of the numerical framework flexibility to solve a different optimization problem consisting in selecting first an injector distribution, then in finding the position $x_0$ of the solid center of mass {minimizing} the magnitude of tangential heat flux {(which is relevant for cases where the practitioner simply cannot act on the design)}. The so-called inverse strategy $S_4$ considered herein features two injectors at each end of the cavity ($x_1=-1.85$ and $x_2=1.85$), identical to the  optimal arrangement of 3 injectors under the free strategy $S_3$. 
The center of mass can take any value in $[-x_{0m};x_{0m}]$ where we set $x_{0m}=2(H-h)$ to avoid numerical issues at the sidewalls. The same coordinate system as above is used, but with reference frame attached to the cavity, not the moving solid (hence all results obtained under the previous strategies pertain to $x_0=0$ in the new system). 

\begin{figure}[t!]
\setlength{\unitlength}{1cm}
\begin{picture}(20,9.6)
\put(7.2,6.6){\includegraphics[trim=175 110 140 575pt,clip,height=2.575cm]{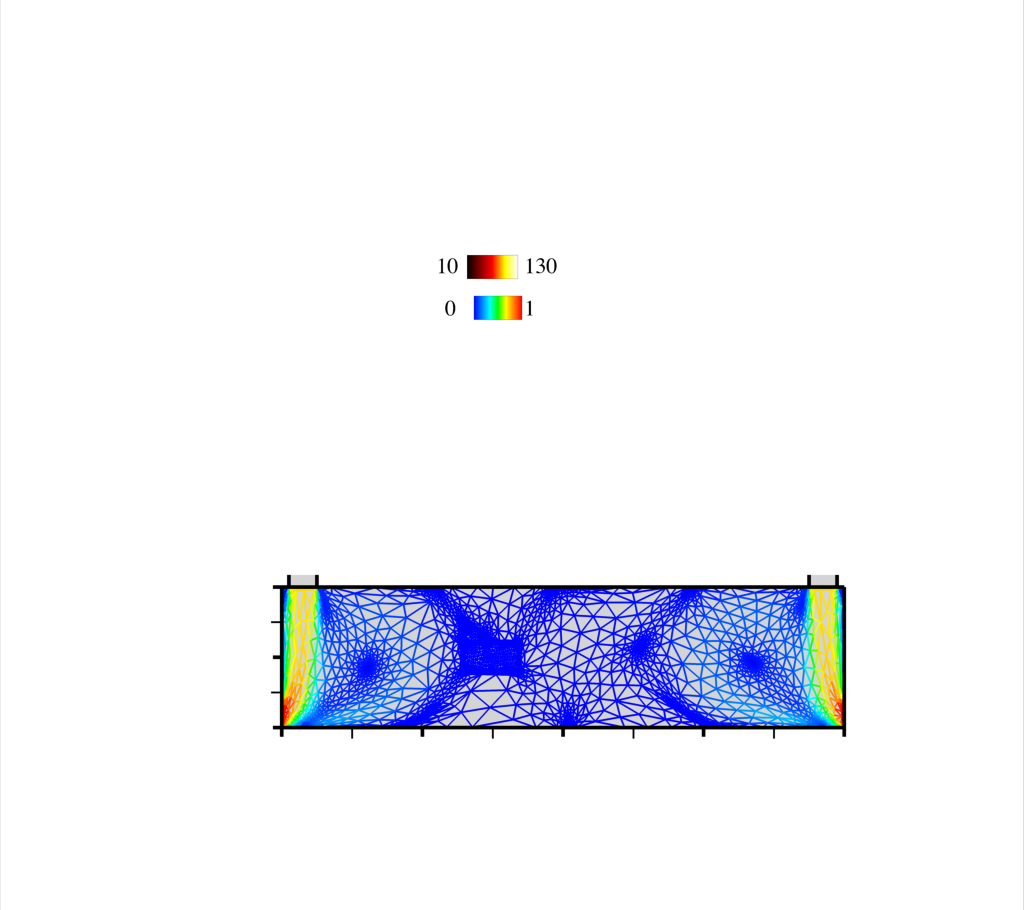}}
\put(0,6.6){\includegraphics[trim=175 110 140 575pt,clip,height=2.575cm]{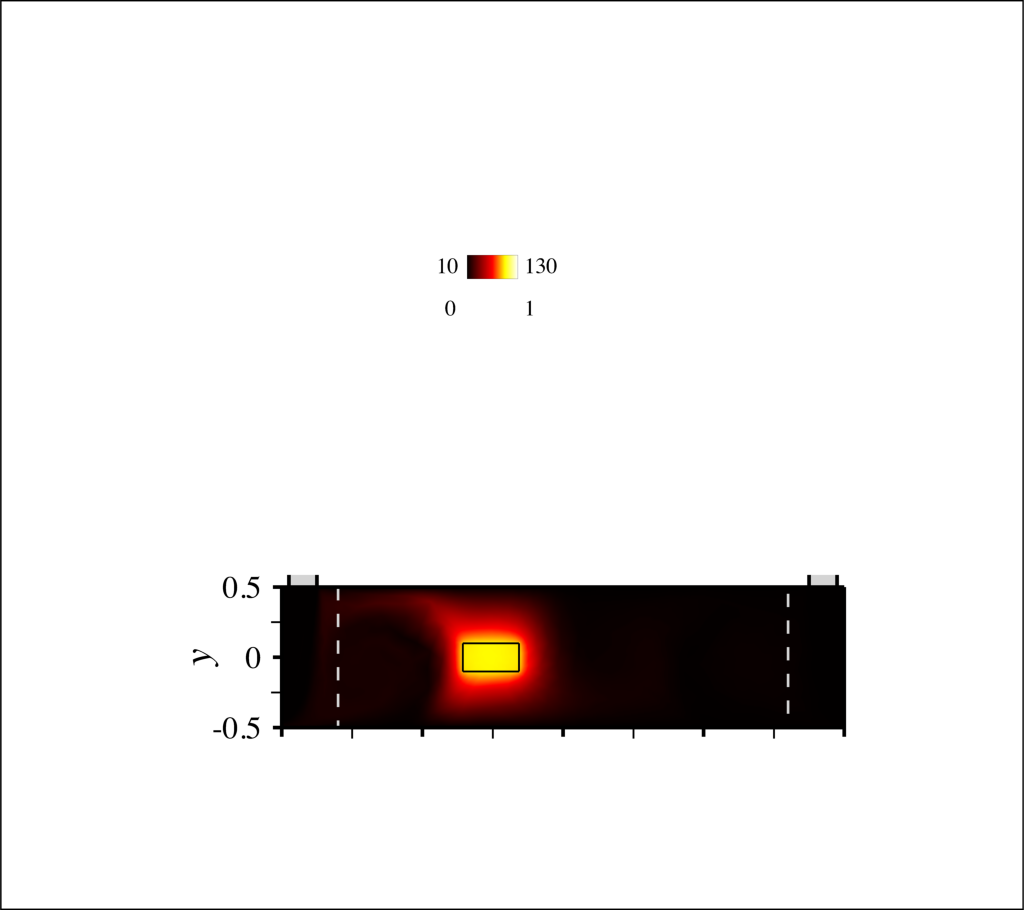}}
\put(7.2,4.4){\includegraphics[trim=175 110 140 575pt,clip,height=2.575cm]{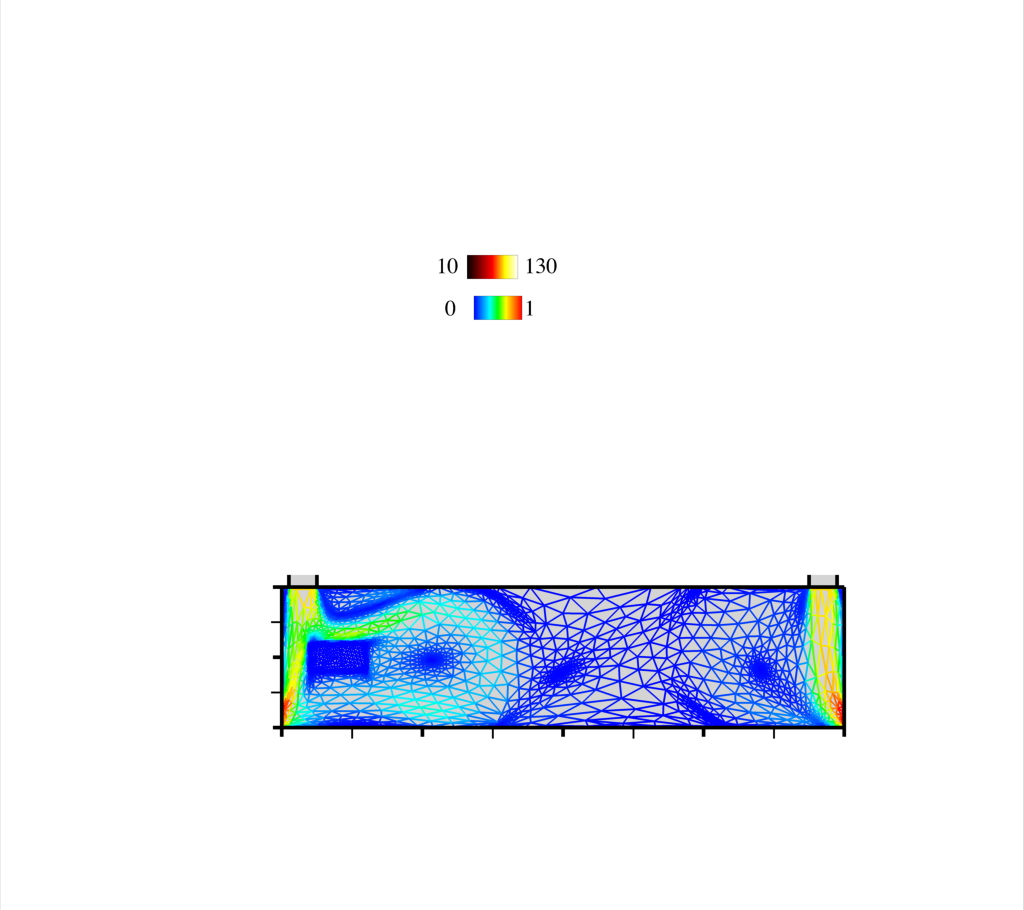}}
\put(0,4.4){\includegraphics[trim=175 110 140 575pt,clip,height=2.575cm]{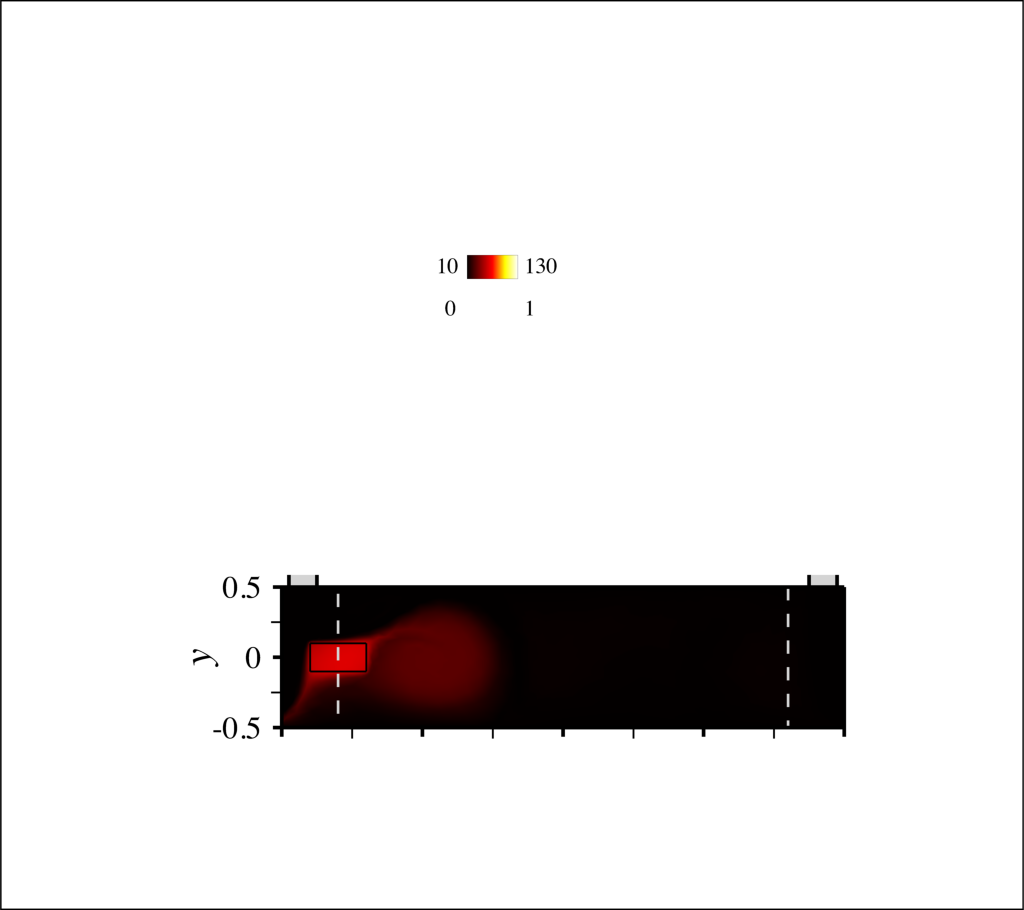}}
\put(7.2,2.2){\includegraphics[trim=175 110 140 575pt,clip,height=2.575cm]{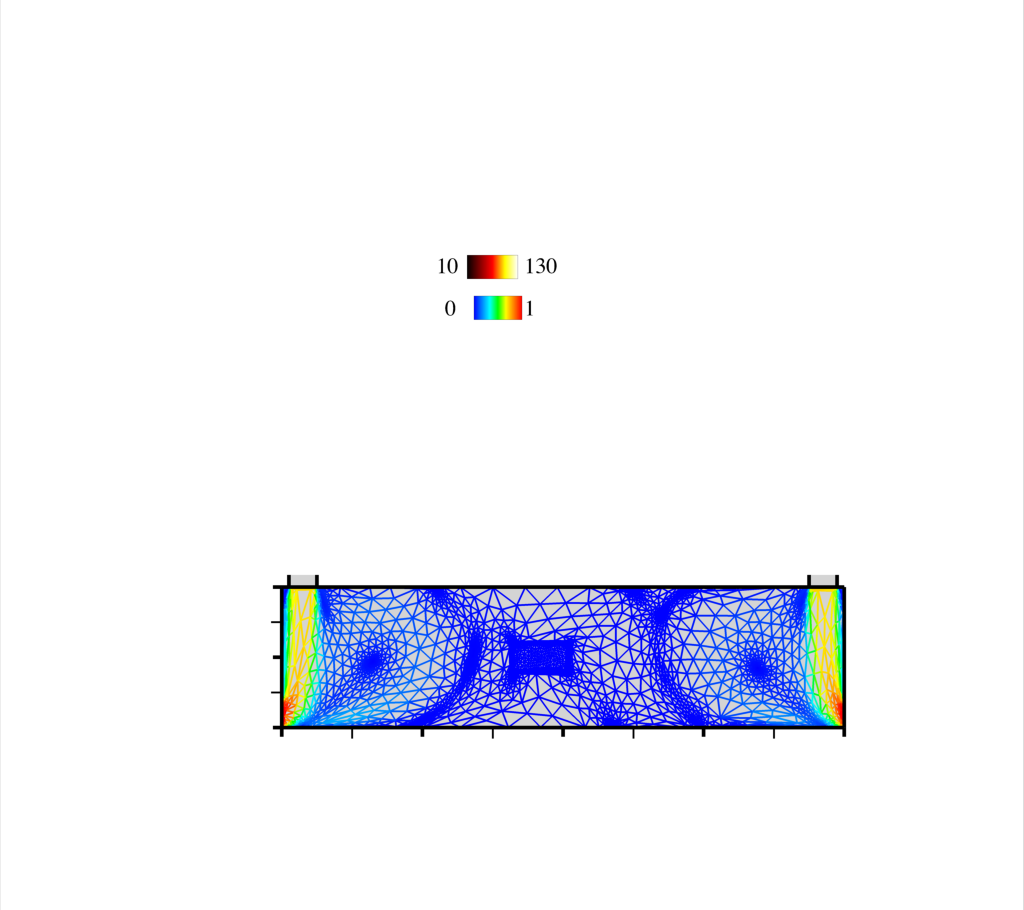}}
\put(0,2.2){\includegraphics[trim=175 110 140 575pt,clip,height=2.575cm]{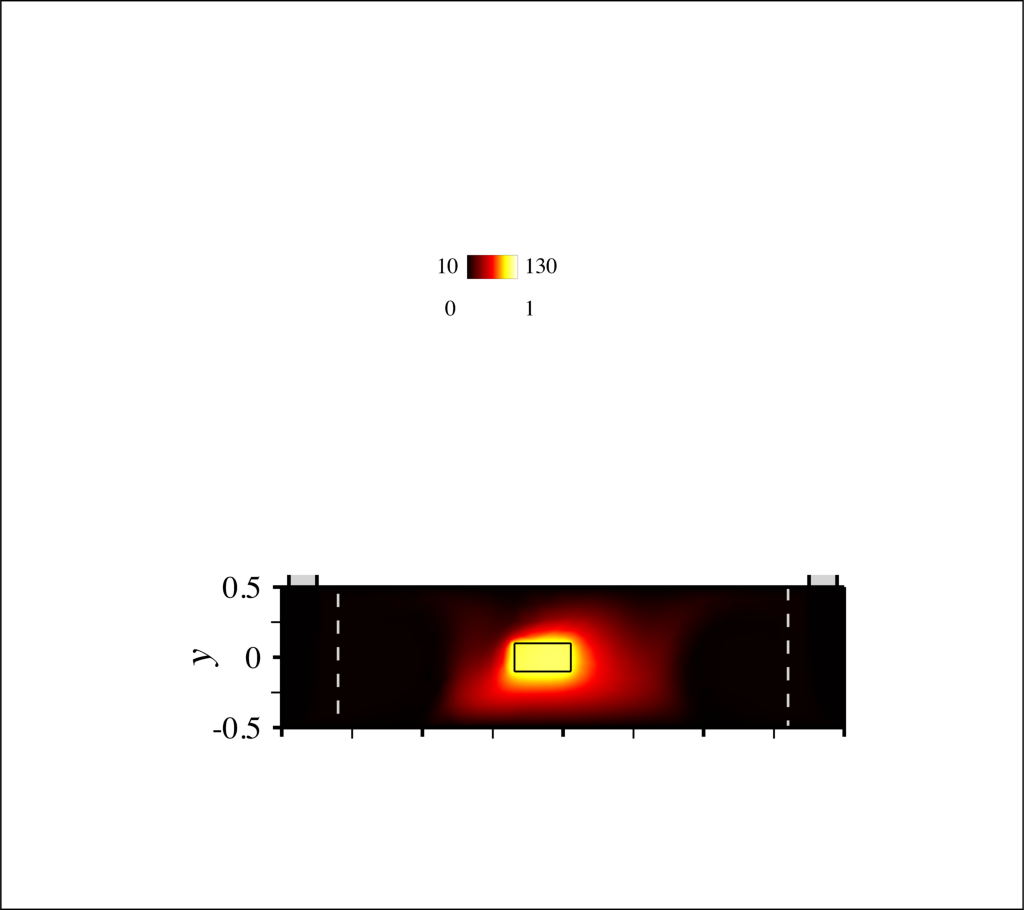}}
\put(7.2,0){\includegraphics[trim=175 110 140 575pt,clip,height=2.575cm]{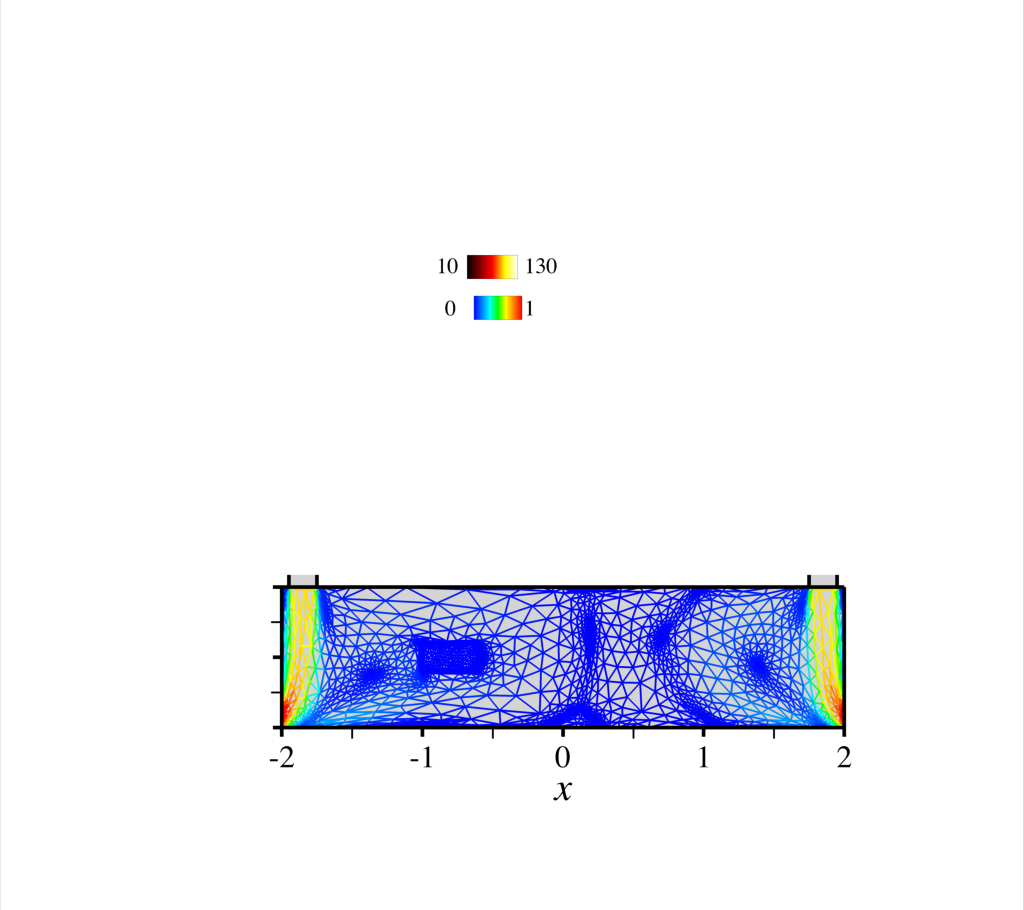}}
\put(0,0){\includegraphics[trim=175 110 140 575pt,clip,height=2.575cm]{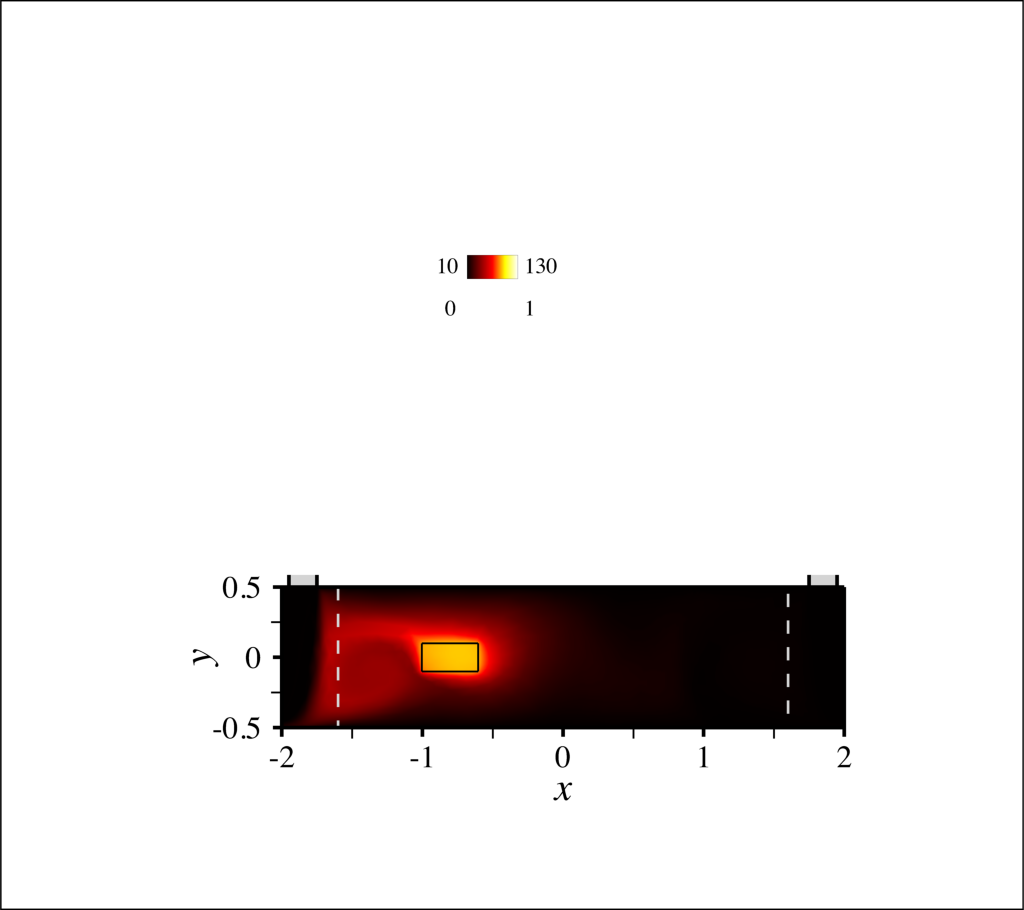}}
\put(5.6,9.3){\includegraphics[trim=420 635 440 245pt,clip,height=0.45cm]{fig11a_bar}}
\put(12.9,9.3){\includegraphics[trim=420 635 440 245pt,clip,height=0.45cm]{fig11b_bar}}
\put(7.85,9.6){(b)}
\put(0.55,9.6){(a)}
\end{picture}
\caption{(a) Steady-state temperature against solid center of mass position, with admissible domains under the inverse strategy $S_4$ marked by the dashed lines. (b) Adapted meshes colored by the norm of velocity.}
\label{fig:forced2d_Sinv_vtu}
\end{figure}

\begin{figure}[t!]
\setlength{\unitlength}{1cm}
\begin{picture}(20,5)
\put(4,-0.1){\includegraphics[trim=175 87.5 140 340pt,clip,height=5cm]{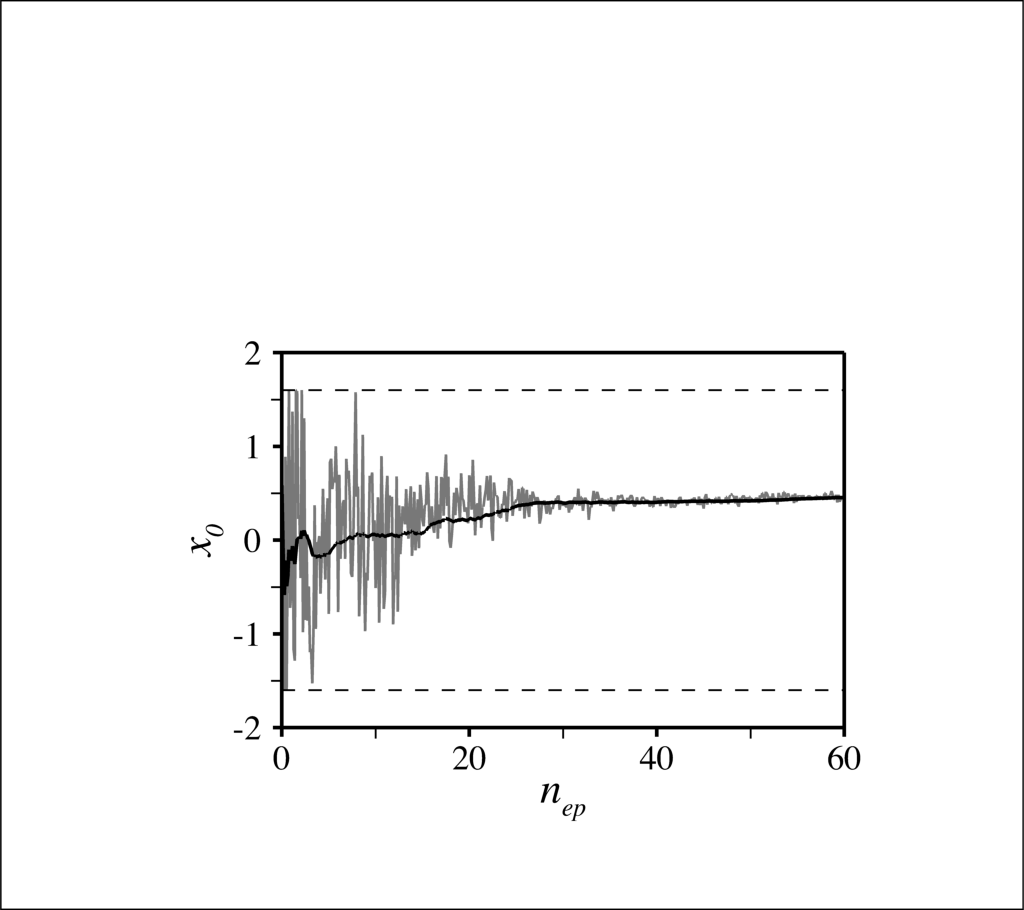}}
\end{picture}
\caption{Evolution per learning episode of the instant (in grey) and moving average (in black) center of mass positions under the inverse strategy $S_4$, with admissible values delimited by the dashed lines.}
\label{fig:forced2d_Sinv_reward}
\end{figure}

\begin{figure}[t!]
\setlength{\unitlength}{1cm}
\begin{picture}(20,3)
\put(7.2,0){\includegraphics[trim=175 110 140 575pt,clip,height=2.575cm]{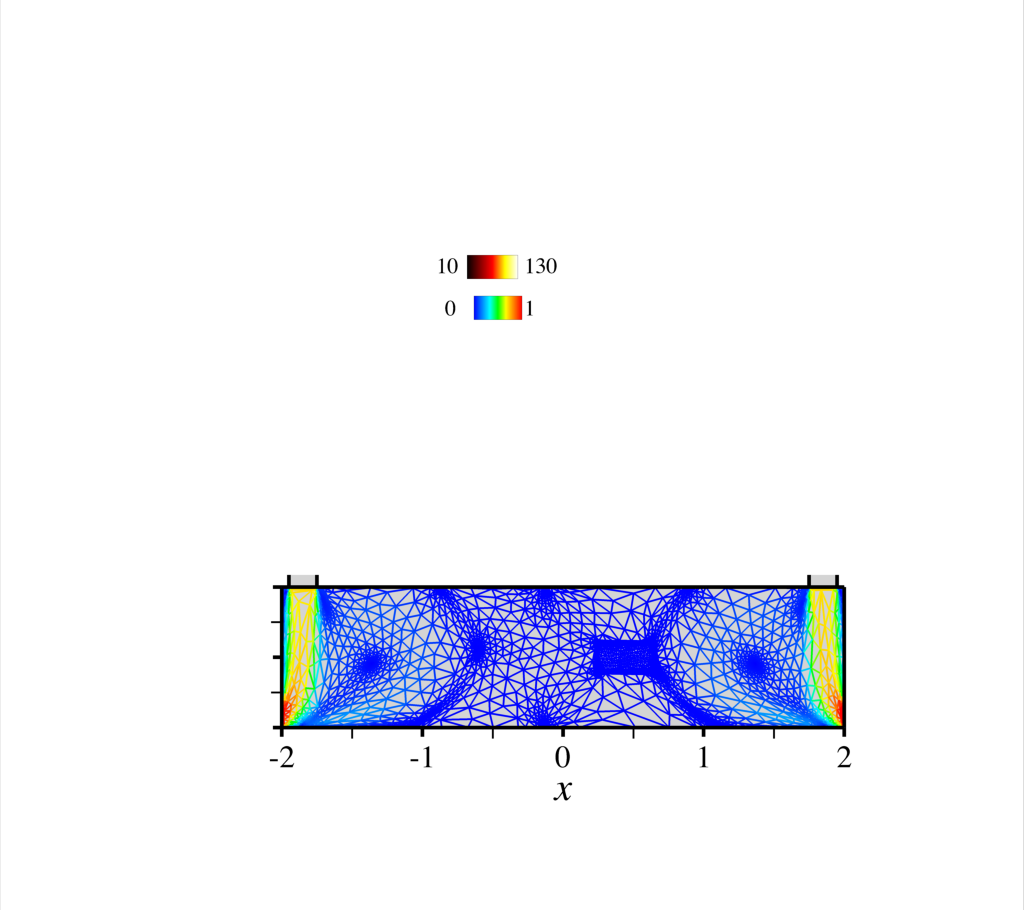}}
\put(0,0){\includegraphics[trim=175 110 140 575pt,clip,height=2.575cm]{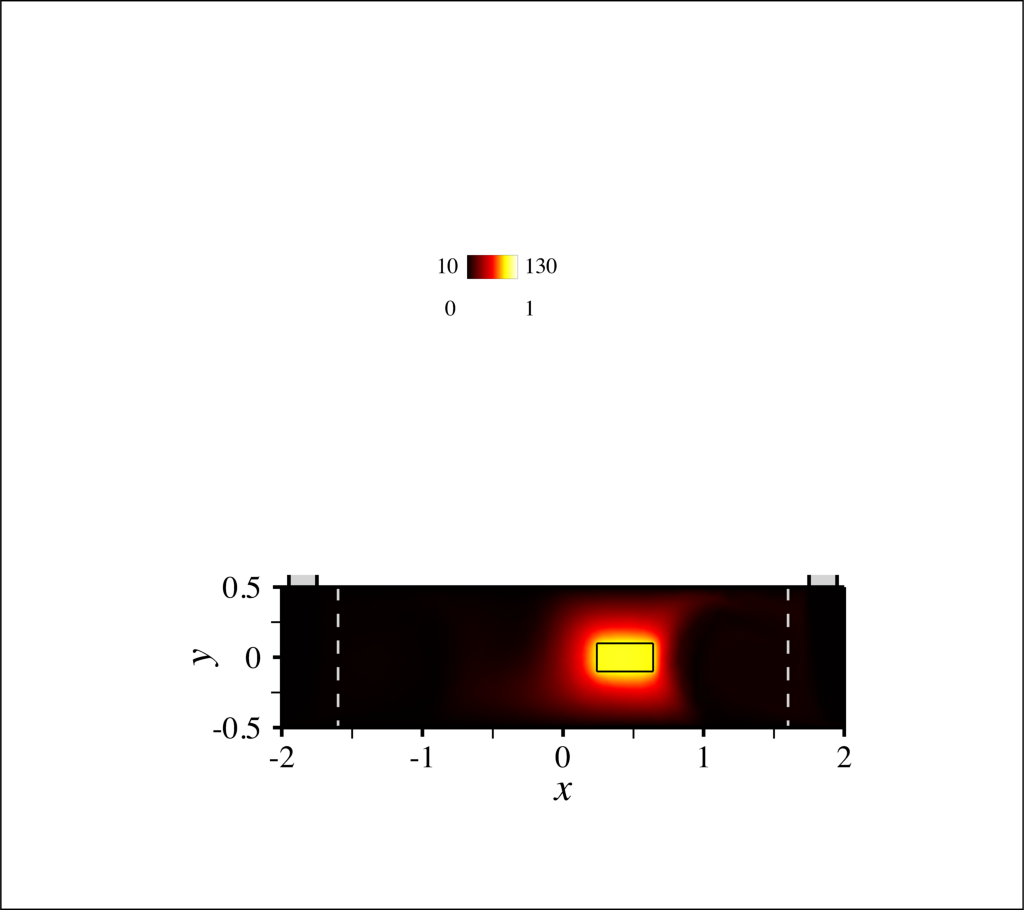}}
\put(5.6,2.7){\includegraphics[trim=420 635 440 245pt,clip,height=0.45cm]{fig11a_bar}}
\put(12.9,2.7){\includegraphics[trim=420 635 440 245pt,clip,height=0.45cm]{fig11b_bar}}
\put(7.85,3){(b)}
\put(0.55,3){(a)}
\end{picture}
\caption{Same as figure~\ref{fig:forced2d_Sinv_vtu} for the optimal center of mass position under the inverse strategy $S_4$.}
\label{fig:forced2d_Sinv_opt}
\end{figure}

\begin{figure}[t!]
\setlength{\unitlength}{1cm}
\begin{picture}(20,10.5)
\put(0.4,0){\includegraphics[trim=175 92.5 140 335pt,clip,height=5cm]{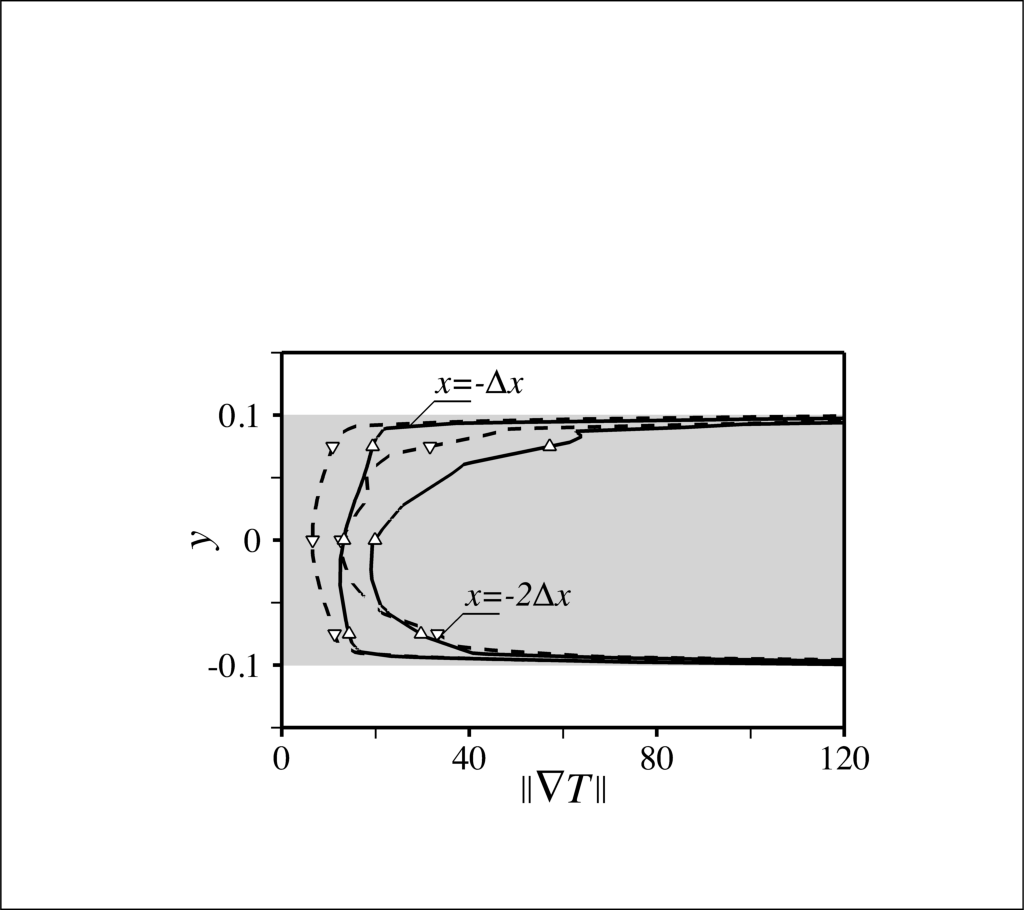}}
\put(7.65,0){\includegraphics[trim=175 92.5 140 335pt,clip,height=5cm]{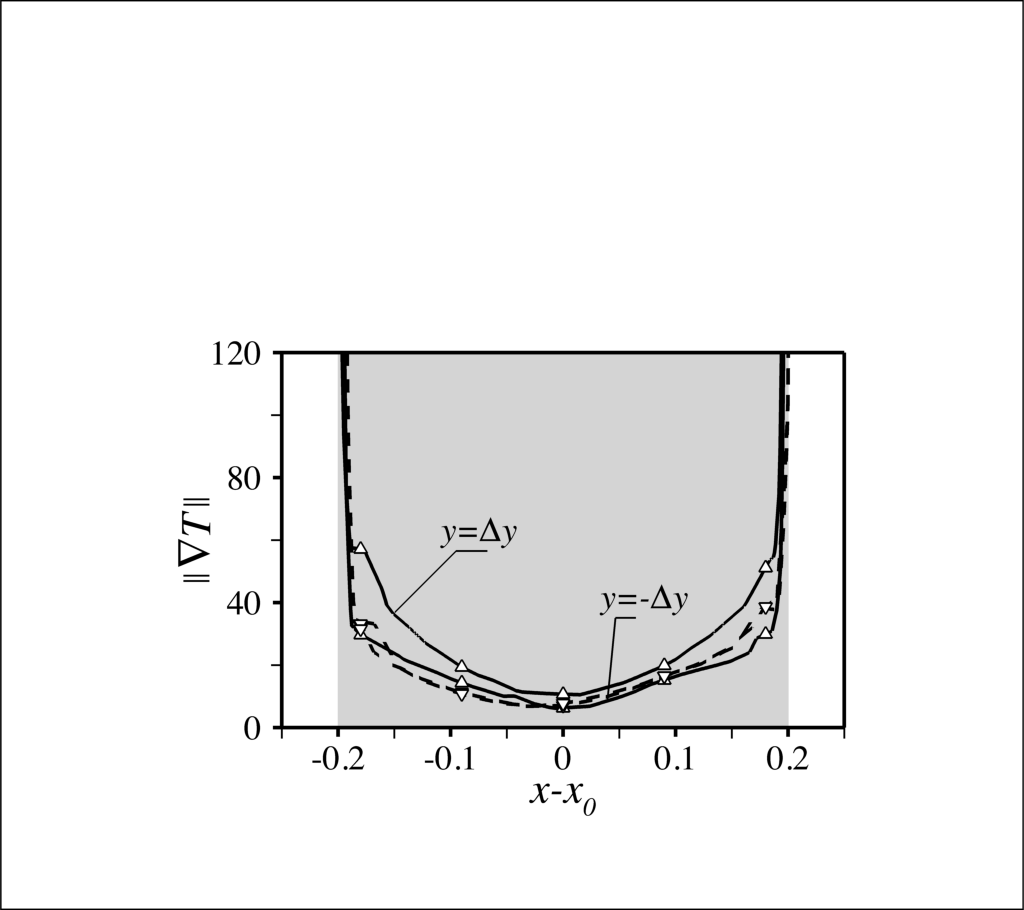}}
\put(0.4,5.2){\includegraphics[trim=175 92.5 140 335pt,clip,height=5cm]{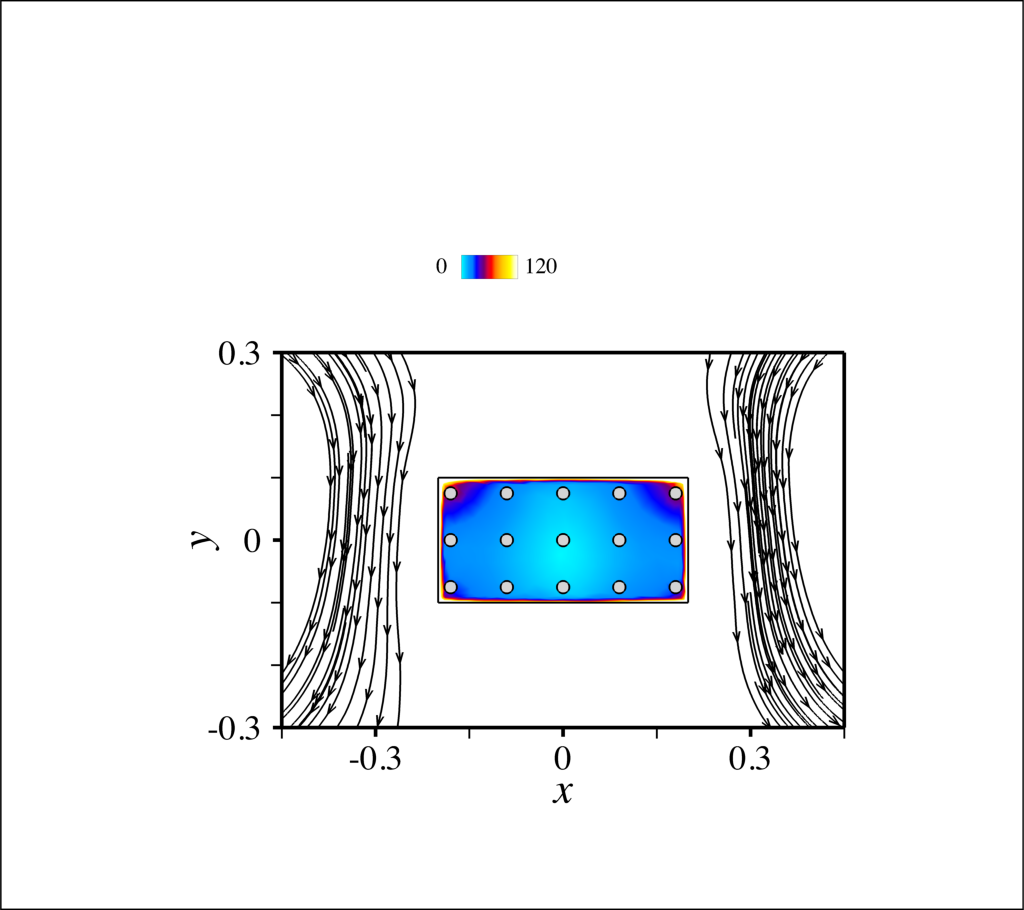}}
\put(7.65,5.2){\includegraphics[trim=175 92.5 140 335pt,clip,height=5cm]{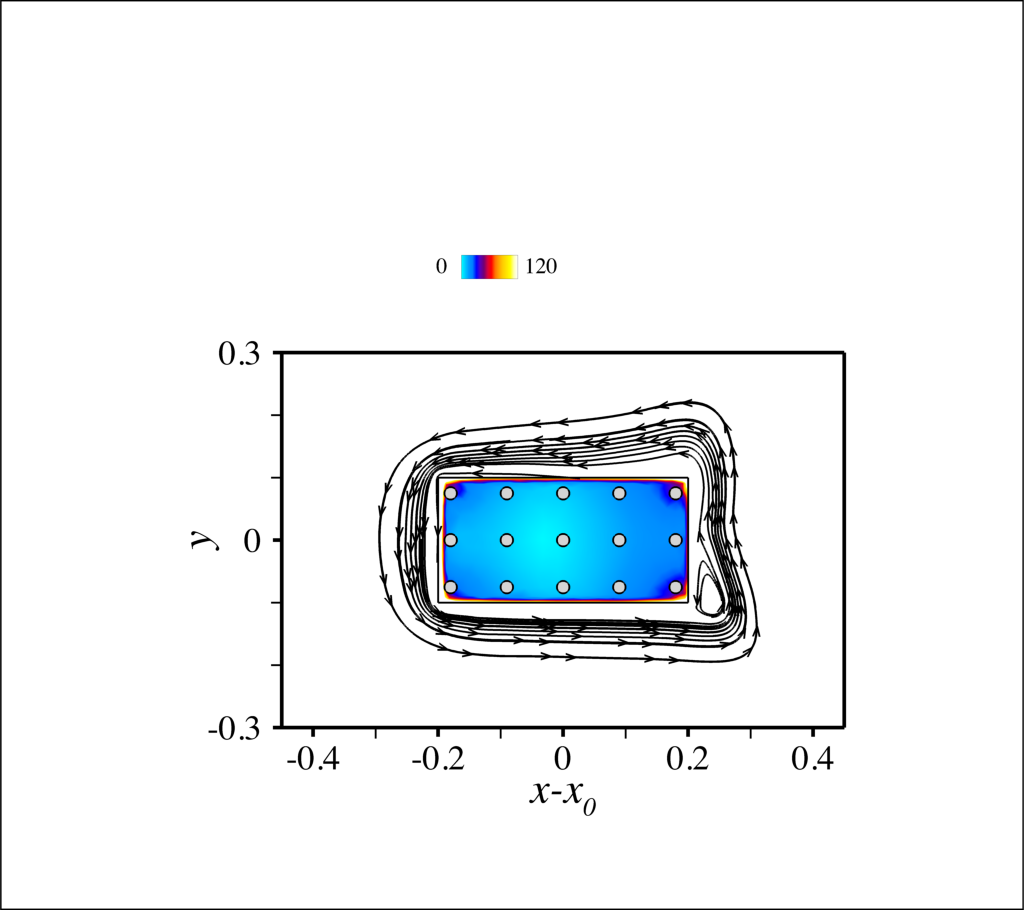}}
\put(5.2,10.15){\includegraphics[trim=420 635 440 245pt,clip,height=0.45cm]{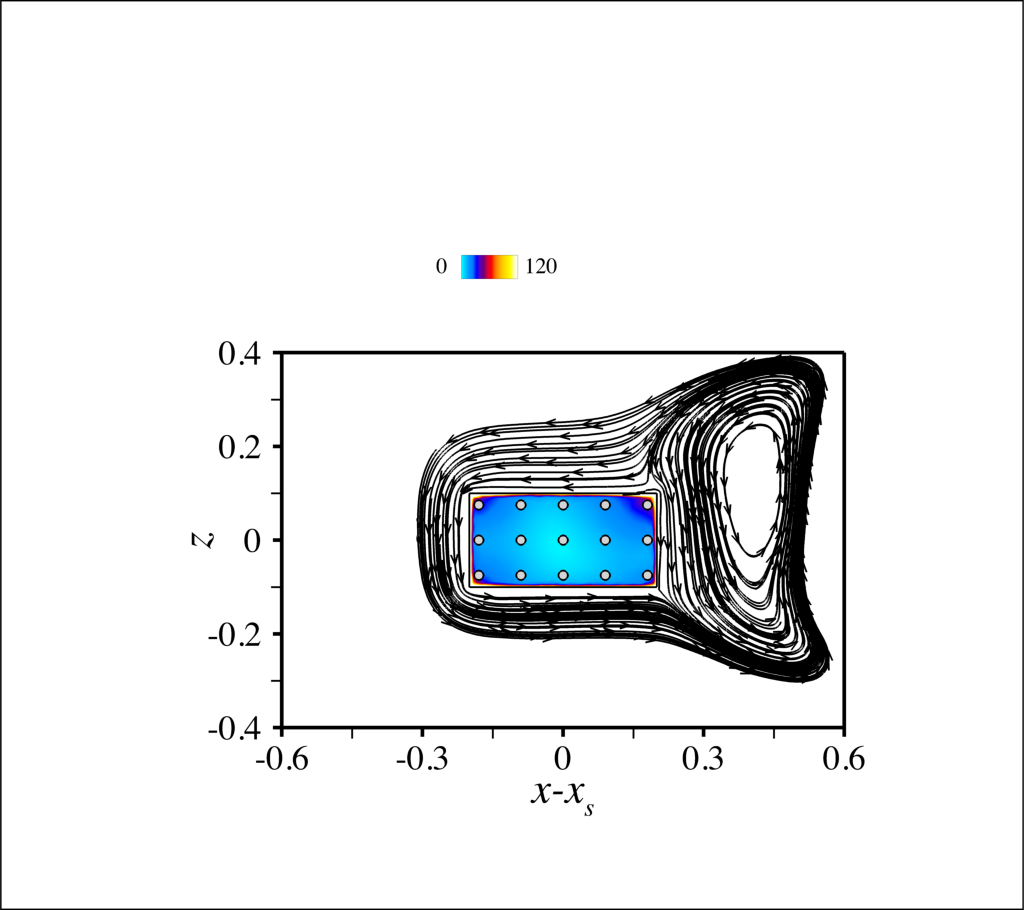}}
\put(12.45,10.15){\includegraphics[trim=420 635 440 245pt,clip,height=0.45cm]{fig23_bar}}
\put(0.55,5.3){(c)}
\put(7.85,5.3){(d)}
\put(0.55,10.5){(a)}
\put(7.85,10.5){(b)}
\end{picture}
\caption{(a,b) Norm of the temperature gradient in the solid domain with superimposed streamlines of the underlying velocity field, as computed for (a) $x_0=0$, and (b) $x_{0}{^\star}=0.45$, i.e., the optimal position selected under the inverse strategy $S_4$. (c) Cuts along the two leftmost columns of probes. The solid and dashed lines refer to $x_0=0$ and $x_{0}{^\star}=0.42$, respectively, and the symbols mark the probe values. (d) Same as (c) for cuts along the lower and upper rows of columns.} 
\label{fig:gradT}
\end{figure}

A total of $60$ episodes have been run for this case using the exact same DRL agent, the only difference being in the network action output, now made up of a single value $\hat{x}_{0}\in[-1;1]$, mapped into the actual position using
\bal
x_0=x_{0m}\hat{x}_0\,.
\eal
A large variety of flow patterns is obtained by doing so, that closely resemble those computed under the previous strategies, only the outcome is now also altered by the width of the gap between the cavity sidewalls and the workpiece, as illustrated in figure~\ref{fig:forced2d_Sinv_vtu}.
We show in figure~\ref{fig:forced2d_Sinv_reward} that the position of the solid center of mass converges to an optimal $x_{0}{^\star}= 0.42$ (the variations over the same interval being by $\pm 0.005$),  the associated magnitude of tangential heat flux $\langle ||\nabla_{\|}T||\rangle{^\star}\sim4.1$, being smaller than that achieved under $S_3$ using a centered workpiece. The fact that the optimal position is offset from the {vertical} centerline is a little surprising at first, because intuition suggests that the simplest way to achieve homogeneous heat transfer is by having symmetrically distributed injectors.
Nonetheless, examining carefully the norm of the temperature gradient in the solid domain shows that $x_0=0$ achieves close to perfect horizontal symmetry but vertical asymmetry, owing to the formation 
of two large-scale, small velocity end vortices entraining heat laterally downwards; see figure~\ref{fig:gradT}(a).
Conversely, for $x\sim x_{0}{^\star}$, the workpiece it almost at the core of the closest recirculation region, hence the surrounding fluid particles have small velocities and wrap almost perfectly around its surface, as illustrated in figure~\ref{fig:gradT}(b). This restores excellent vertical symmetry, as evidenced by relevant cuts along the two leftmost columns of probes in figure~\ref{fig:gradT}(c), and along the lower and upper rows in figure~\ref{fig:gradT}(d), which explains the improved the reward.


\subsection{Discussion}

\begin{figure}[t!]
\setlength{\unitlength}{1cm}
\begin{picture}(20,5.9)
\put(7.2,3){\includegraphics[trim=175 110 140 575pt,clip,height=2.575cm]{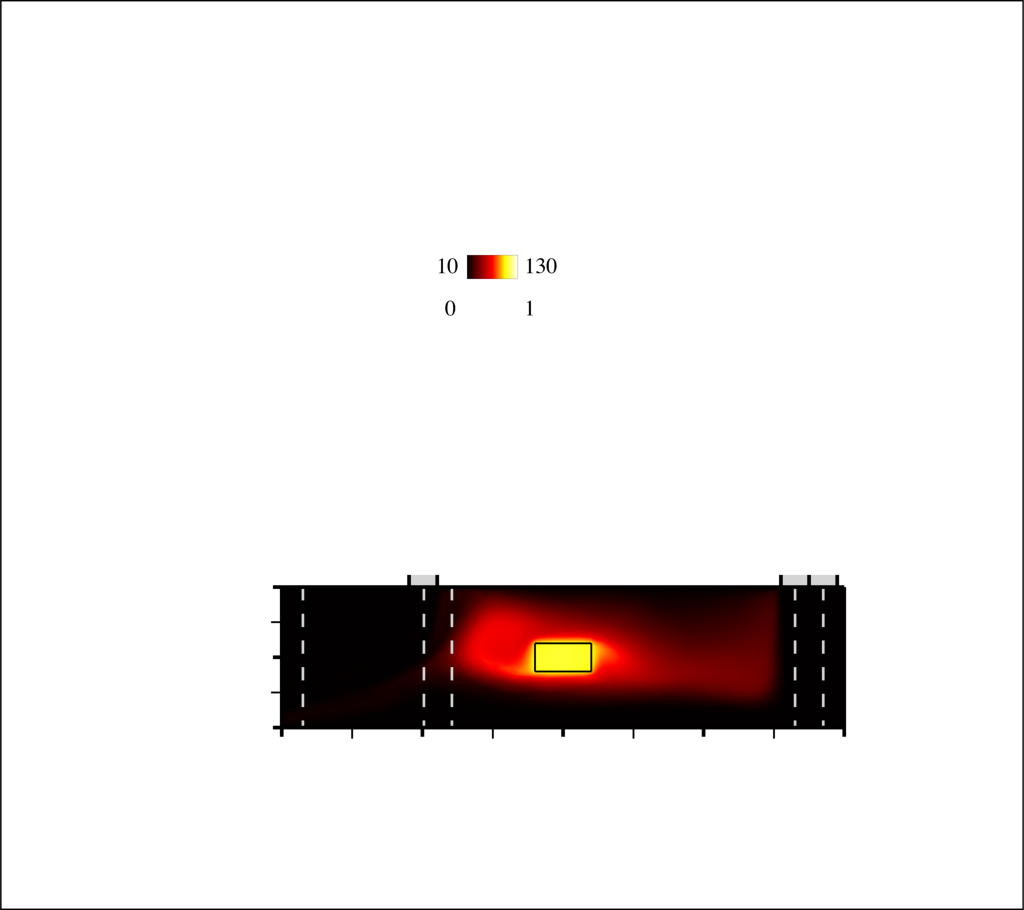}}
\put(0,3){\includegraphics[trim=175 110 140 575pt,clip,height=2.575cm]{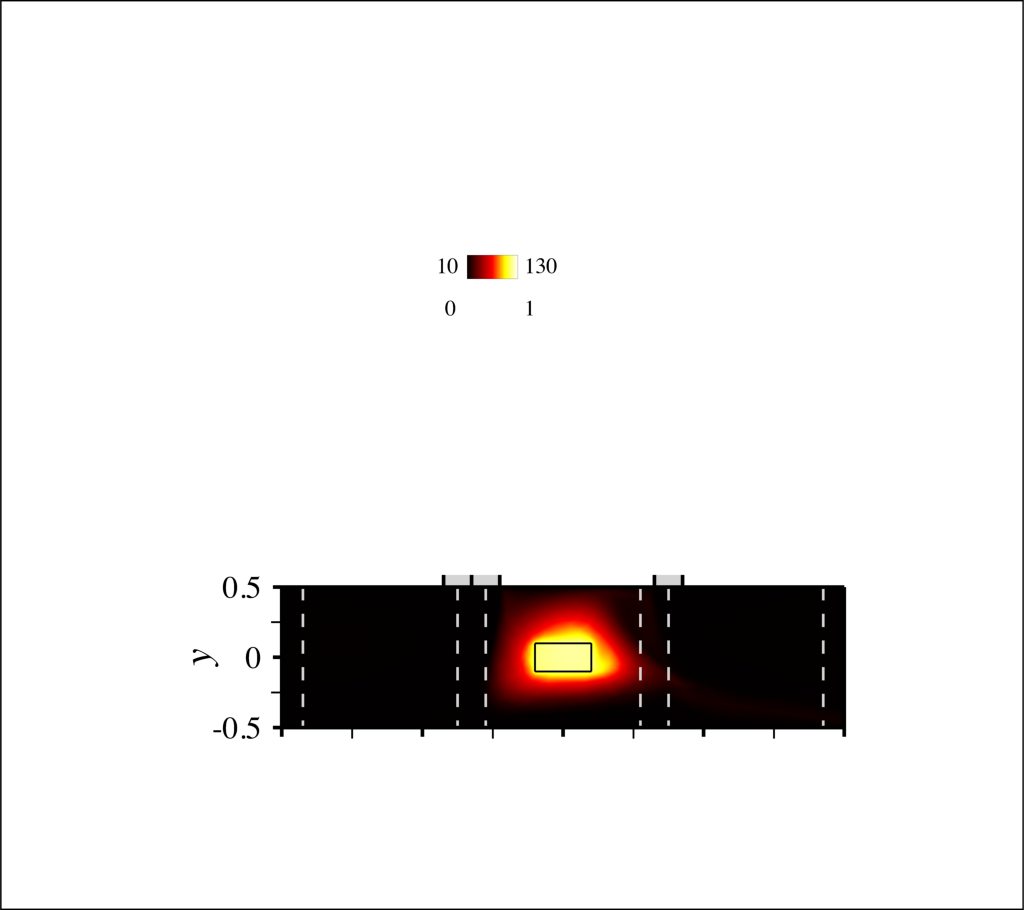}}
\put(7.2,0){\includegraphics[trim=175 110 140 575pt,clip,height=2.575cm]{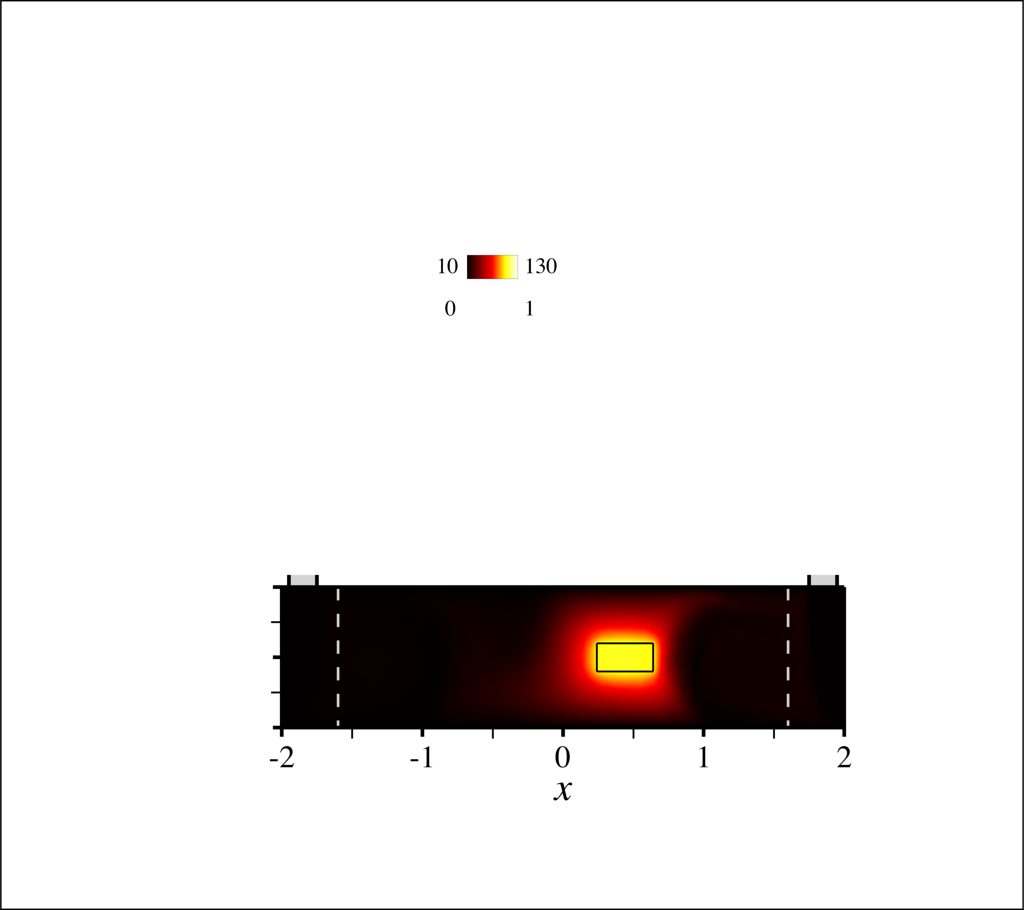}}
\put(0,0){\includegraphics[trim=175 110 140 575pt,clip,height=2.575cm]{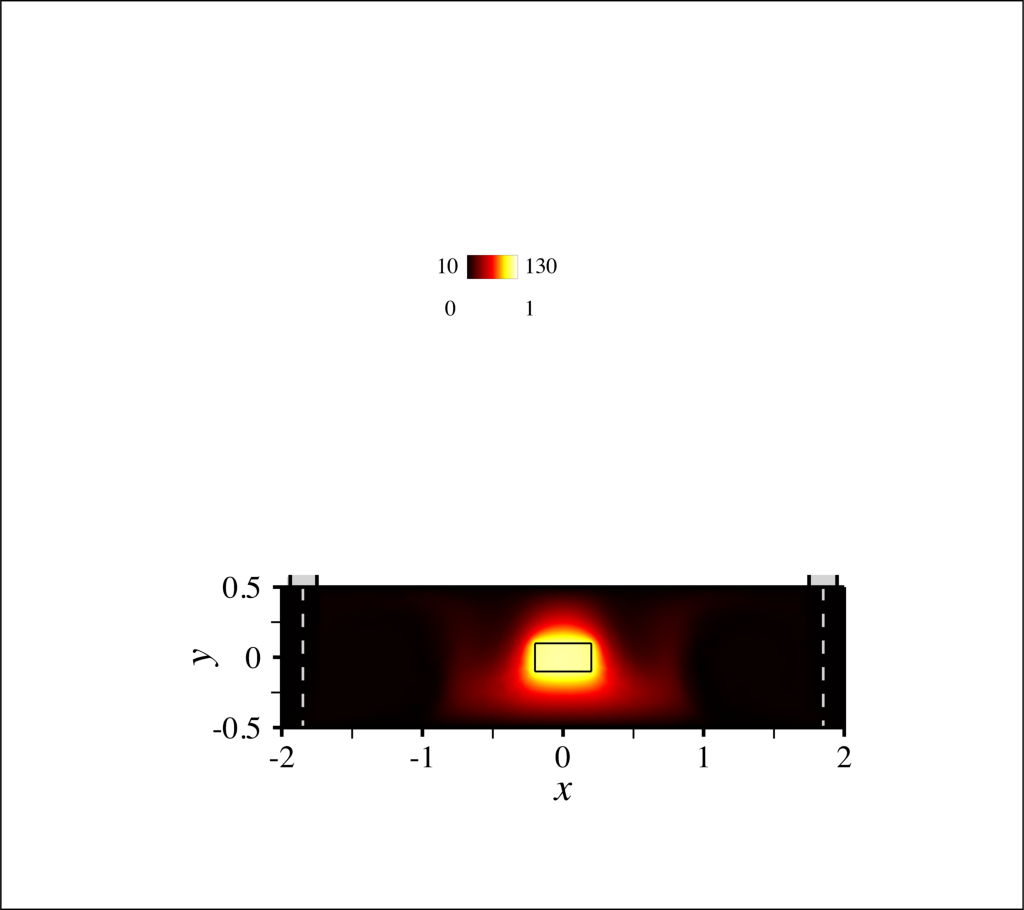}}
\put(8.15,5.9){(b)}
\put(1,5.9){(a)}
\put(8.15,2.9){(d)}
\put(1,2.9){(c)}
\put(5.6,5.7){\includegraphics[trim=420 635 440 245pt,clip,height=0.45cm]{fig11a_bar}}
\end{picture}
\caption{(a-c) Optimal arrangements of 3 injectors under the (a) fixed decomposition domain strategy $S_1$, (b)
follow-up strategy $S_2$ and (c) free strategy $S_3$.
(b) Optimal position of the workpiece under the inverse strategy $S_4$.}
\label{fig:forced2d_opt_vtu}
\end{figure}

Figure~\ref{fig:forced2d_opt_vtu} reproduces the optimal temperature distributions computed under the various strategies considered above. For benchmarking purposes, we also provide in table~\ref{table:forced2drecap} relevant convergence data computed over the $10$ latest episodes. To recap, the most homogeneous cooling is achieved under the follow-up strategy $S_2$, but the DRL agent seems more easily trained under the fixed decomposition domain strategy $S_1$ and the free strategy $S_3$. 
Another interesting point is the extent to which the workpiece is actually cooled, for which $S_2$ seems more relevant, on behalf of the dissymmetry in the left and right flow rates that creates order one velocities at the bottom of the cavity. This stresses $S_2$ as a possible compromise to achieve efficient \textit{and} homogeneous cooling, although a true optimal with this regard can be computed rigorously by applying the same approach to compound functionals weighing, e.g., the magnitude of the tangential heat flux and the solid center temperature (which we defer to future work).

\begin{table}[t!]
\begin{center}
\begin{tabular}{cccccccc}
\toprule
& \multicolumn{1}{p{1cm}}{\makecell[r]{$n_j$}} & \multicolumn{1}{p{1cm}}{\makecell[r]{$n_{ep}$}} & \multicolumn{1}{p{1.1cm}}{\makecell[r]{$x_0$}} & \multicolumn{1}{p{1.1cm}}{\makecell[r]{$x_1$}} & \multicolumn{1}{p{1.1cm}}{\makecell[r]{$x_2$}} & \multicolumn{1}{p{1.1cm}}{\makecell[r]{$x_3$}} & \multicolumn{1}{p{2cm}}{\makecell[r]{$\langle ||\nabla_{\|}T||\rangle$}}\\
\cmidrule(lr){1-8}
\multicolumn{1}{r}{$S_1$} & \multicolumn{1}{r}{$3$} & \multicolumn{1}{r}{$60$} & \multicolumn{1}{r}{$0$} & \multicolumn{1}{r}{$-0.75$} &\multicolumn{1}{r}{$\pm0.55$} & \multicolumn{1}{r}{$0.75$} & \multicolumn{1}{r}{$8.3$}\\
\multicolumn{1}{r}{$S_2$} & \multicolumn{1}{r}{$3$} & \multicolumn{1}{r}{$75$} & \multicolumn{1}{r}{$0$} & \multicolumn{1}{r}{$-0.96$} &\multicolumn{1}{r}{$1.65$} & \multicolumn{1}{r}{$1.85$} & \multicolumn{1}{r}{$6.3$}\\
\multicolumn{1}{r}{$S_3$} & \multicolumn{1}{r}{$3$} & \multicolumn{1}{r}{$60$} & \multicolumn{1}{r}{$0$} & \multicolumn{1}{r}{$-1.85$} &\multicolumn{1}{r}{$-1.82$} & \multicolumn{1}{r}{$1.85$} & \multicolumn{1}{r}{$11.2$}\\
\multicolumn{1}{r}{$S_4$} & \multicolumn{1}{r}{$2$} & \multicolumn{1}{r}{$60$} & \multicolumn{1}{r}{$0.42$} & \multicolumn{1}{r}{$-1.85$} &\multicolumn{1}{r}{$1.85$} & \multicolumn{1}{r}{--} & \multicolumn{1}{r}{$4.1$}\\
\bottomrule
\end{tabular}
\caption{Numerical data for the optimal arrangements computed under strategies $S_{1-4}$. All values computed by averaging the instant signal over the 10 latest learning episodes.}
\label{table:forced2drecap}
\end{center}	
\end{table}

These results provide a basis for future self-assessment of the method and identifies potential for improvement regarding the convergence efficiency. The approach can certainly benefit from a fine tuning of the reward computation, as having sufficient spatial resolution on the relevant state of the system is an obvious requirement to allow a successful control. Adjusting the trade-off between exploration and exploitation is also worth consideration to better handle the existence of multiple global optima (whether they stem from symmetries of from the topology of the reward itself) which could be done using non-normal probability density functions.

\section{Extension to 3-D forced convection}\label{section:forced3d} 

\subsection{Case description}

\begin{figure}[t!]
\setlength{\unitlength}{1cm}
\begin{picture}(20,6.5)
\put(1,0){\includegraphics[trim=10 200 10 200pt,clip,height=6.5cm]{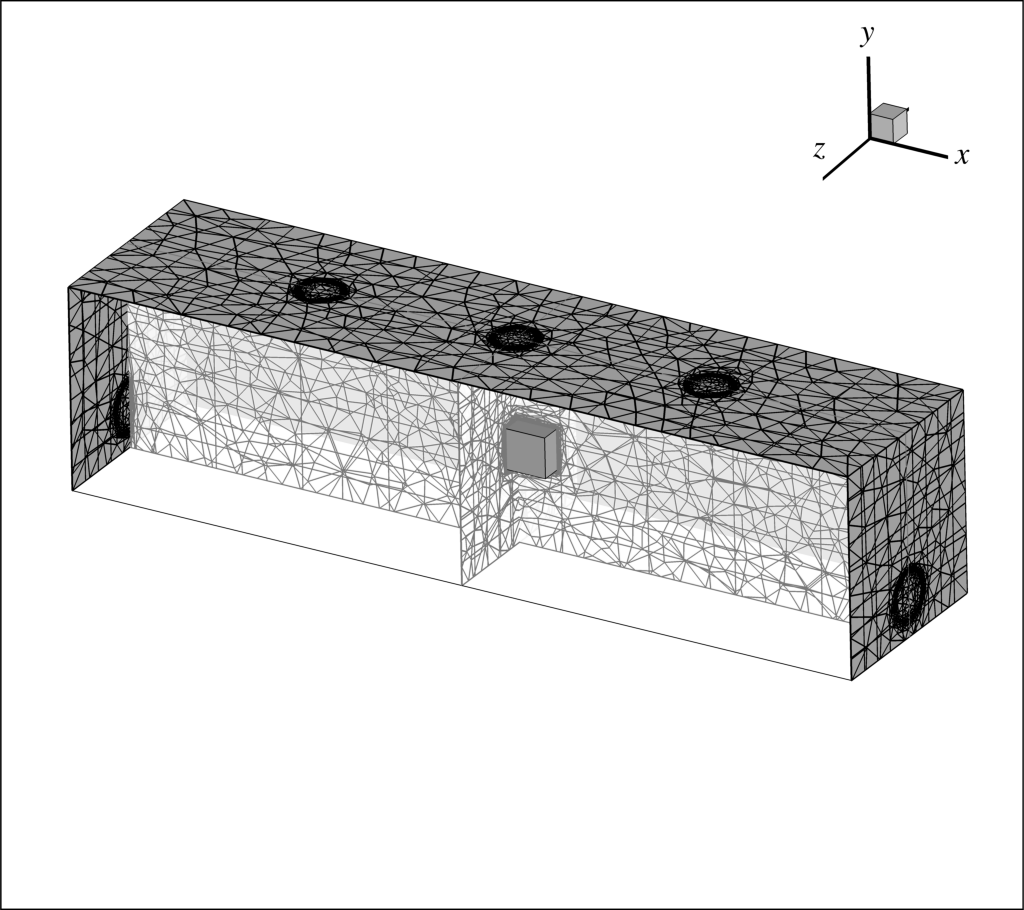}}
\put(9.5,5){\includegraphics[trim=650 715 10 30pt,clip,height=1.75cm]{fig25}}
\end{picture}
\caption{Schematic of the 3-D forced convection set-up.}
\label{fig:forced3d_conf}
\end{figure}

The model cooling set up considered in section~\ref{section:forced2d}  is extended here to 3-D to assess the extent to which the approach carries over to three-dimensional conjugate heat transfer. The main differences between 2-D and 3-D are as follows: a Cartesian coordinate system is used with origin at the center of mass of the solid, horizontal $x$-axis, vertical $y$-axis, and the $z$-axis completes the direct triad; see figure~\ref{fig:forced3d_conf}. The solid is a rectangular prism with aspect ratio 2:1:1, and is fixed at the center of a rectangular cavity with height $H$ and aspect ratio 4:1:1. We consider $n_j$ circular-shaped injectors with diameter $d_i$, whose exit planes are forced to be symmetrical with respect to $z=0$, hence each injector is identified by the horizontal position of its center $x_{k\in\{1\dots n_j\}}$. We also use circular-shaped exhaust areas with diameter $d_o$, offset by a distance $\delta_o$ from the bottom of the cavity, and
whose exit planes are also symmetrical with respect to $z=0$, hence each exhaust area is identified by the vertical position of its center $(d_0+\delta_o-H)/2$. The governing equations are solved with the exact same boundary conditions as in section~\ref{section:forced2d}. 
All parameters are provided in Table~\ref{table:forced3d}, including the material properties used to
model the composite fluid, that yield fluid values of the Reynolds and Prandtl numbers
\bal
\rey=\frac{\rho V_i d_i}{\mu}=20
\,,\qquad\qquad \pr=20\,.
\eal

\begin{table}[t!]
\begin{center}
\begin{tabular}{cccccccccccccc}
\toprule
\multicolumn{1}{p{0.6cm}}{\makecell[r]{$H$}} & \multicolumn{1}{p{0.6cm}}{\makecell[r]{$h$}} & \multicolumn{1}{p{0.6cm}}{\makecell[r]{$d_i$}} & \multicolumn{1}{p{0.6cm}}{\makecell[r]{$d_0$}}
& \multicolumn{1}{p{0.6cm}}{\makecell[r]{$\delta_0$}}
 & \multicolumn{1}{p{0.6cm}}{\makecell[r]{$V_i$}} & \multicolumn{1}{p{0.6cm}}{\makecell[r]{$T_w$}} & \multicolumn{1}{p{0.6cm}}{\makecell[r]{$T_c$}} & \multicolumn{1}{p{0.6cm}}{\makecell[r]{$T_h$}} & \multicolumn{1}{p{0.85cm}}{\makecell[r]{$\mu$}} & \multicolumn{1}{p{0.6cm}}{\makecell[r]{$\rho$}} & \multicolumn{1}{p{0.6cm}}{\makecell[r]{$\lambda$}} & \multicolumn{1}{p{0.85cm}}{\makecell[r]{$c_p$}}\\
\cmidrule(lr){1-14}
\multicolumn{1}{r}{\multirow{2}{*}{$1$}} & \multicolumn{1}{r}{\multirow{2}{*}{$0.2$}} & \multicolumn{1}{r}{\multirow{2}{*}{$0.2$}} & \multicolumn{1}{r}{\multirow{2}{*}{$0.24$}} & \multicolumn{1}{r}{\multirow{2}{*}{$0.16$}} & \multicolumn{1}{r}{\multirow{2}{*}{$1$}} & \multicolumn{1}{r}{\multirow{2}{*}{$10$}} & \multicolumn{1}{r}{\multirow{2}{*}{$10$}} & \multicolumn{1}{r}{\multirow{2}{*}{$150$}} & \multicolumn{1}{r}{\makecell[r]{$0.01$}} & \multicolumn{1}{r}{\makecell[r]{$1$}} & \multicolumn{1}{r}{\makecell[r]{$0.5$}} & \multicolumn{1}{r}{\makecell[r]{$1000$}} & \multicolumn{1}{r}{\makecell[r]{Fluid}}\\
\cmidrule(lr){10-14}
\multicolumn{9}{r}{\makecell[r]{}} & \multicolumn{1}{r}{\makecell[r]{$1000$}} & \multicolumn{1}{r}{\makecell[r]{$100$}} & \multicolumn{1}{r}{\makecell[r]{$15$}} & \multicolumn{1}{r}{\makecell[r]{$300$}} & \multicolumn{1}{r}{\makecell[r]{Solid}}\\
\bottomrule
\end{tabular}
\caption{Numerical parameters used in the 3-D forced convection problem. All values in SI units, with the exception of temperatures given in Celsius.}
\label{table:forced3d}
\end{center}	
\end{table}

\subsection{Control strategy}

We keep here the same control objective and compute the reward fed to the DRL from $45$ probes arranged symmetrically into $n_z=3$ transverse layers with resolution $\Delta z=0.075$, each of which distributes uniformly $15$ probes into $n_x=5$ columns and $n_y=3$ rows with resolutions $\Delta x=0.09$ and $\Delta y=0.075$.
In practice, the 3-D reward is simply the average over $z$ of the 2-D reward defined  in section~\ref{section:forced2d}, hence
$r_t=-\langle ||\nabla_{\|}T||\rangle$ with
\bal
\langle ||\nabla_{\|}T||\rangle=\frac{1}{(n_x+n_y)n_z}\sum_{i,j,k}\langle ||\nabla_{\|}T||\rangle_{ik}+\langle ||\nabla_{\|}T||\rangle_{jk}\,,\label{eq:forced3dreward2}
\eal
with
\bal
\langle ||\nabla_{\|}T||\rangle_{ik}=&\frac{2}{n_y-1}|\sum_{j\neq 0}\text{sgn}(j)||\nabla T||_{ijk}|\,,\quad\quad\langle ||\nabla_{\|}T||\rangle_{jk}=&\frac{2}{n_x-1}|\sum_{i\neq 0}\text{sgn}(i)||\nabla T||_{ijk}|\,,\label{eq:forced3dreward}
\eal
and the subscripts $ik$, $jk$ and $ijk$ denote quantities evaluated at $(x,z)=(i\Delta x,k\Delta z)$, $(y,z)=(j\Delta y,k\Delta z)$ and $(x,y,z)=(i\Delta x,j\Delta y,k\Delta z)$, respectively.

\begin{figure}[t!]
\setlength{\unitlength}{1cm}
\begin{picture}(20,19)
\put(1.75,13.5){\includegraphics[trim=10 275 10 250pt,clip,height=4.5cm]{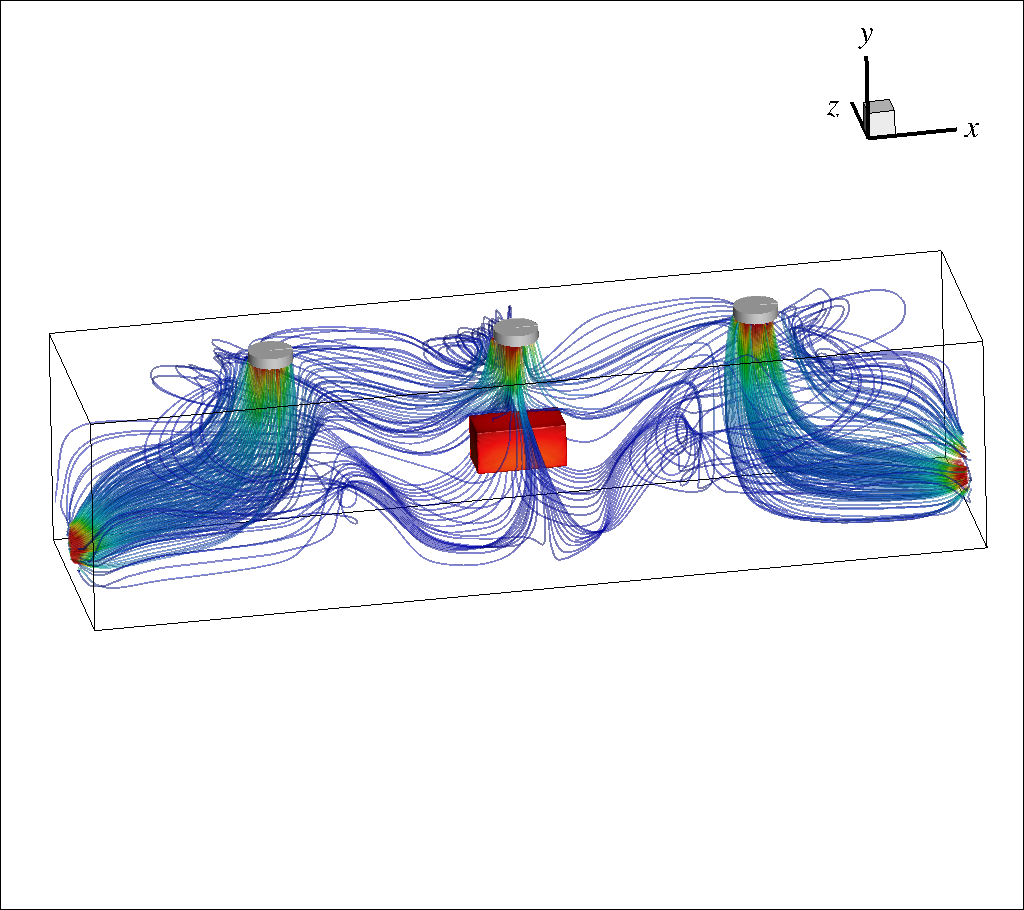}}
\put(1.75,9){\includegraphics[trim=10 275 10 250pt,clip,height=4.5cm]{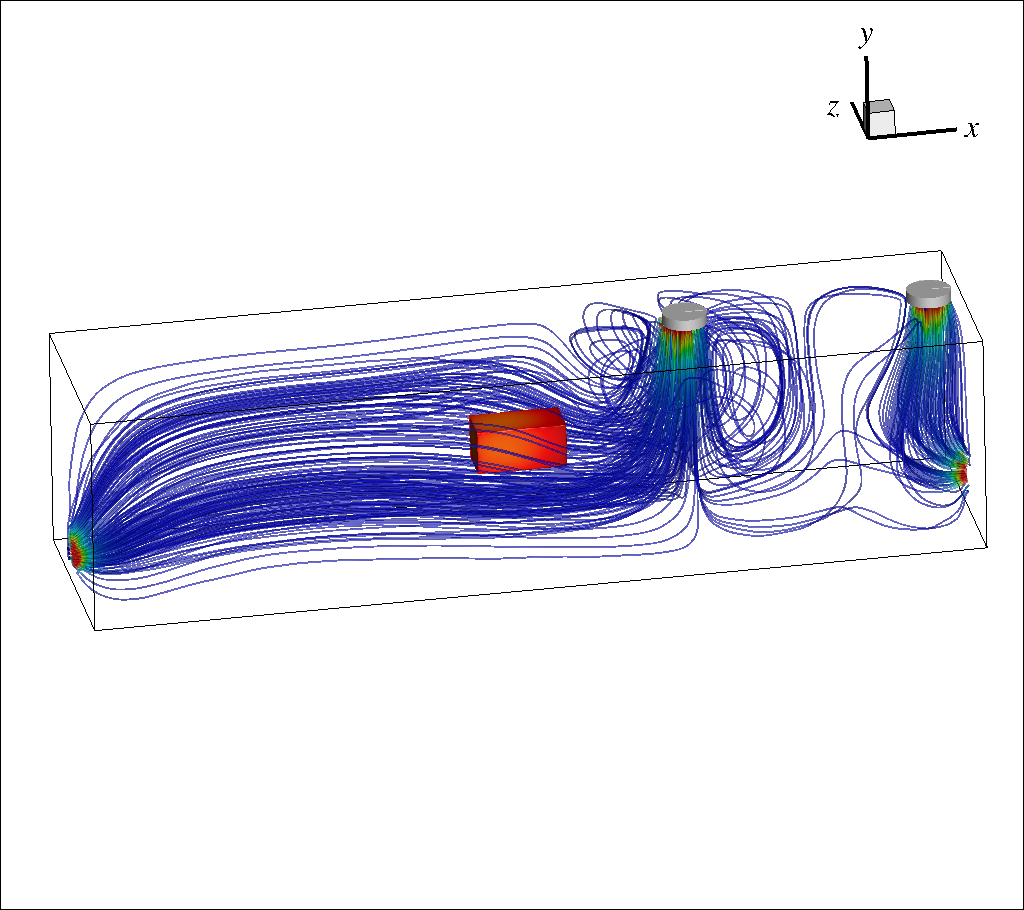}}
\put(1.75,4.5){\includegraphics[trim=10 275 10 250pt,clip,height=4.5cm]{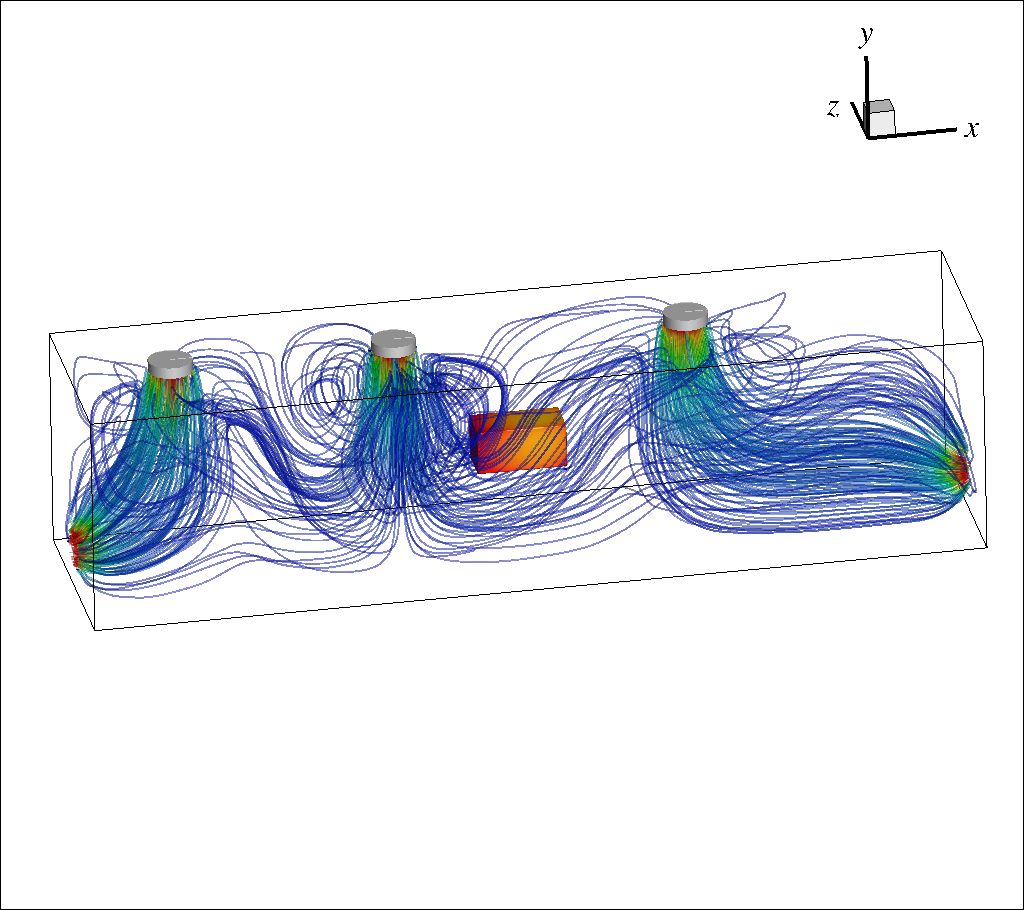}}
\put(1.75,0){\includegraphics[trim=10 275 10 250pt,clip,height=4.5cm]{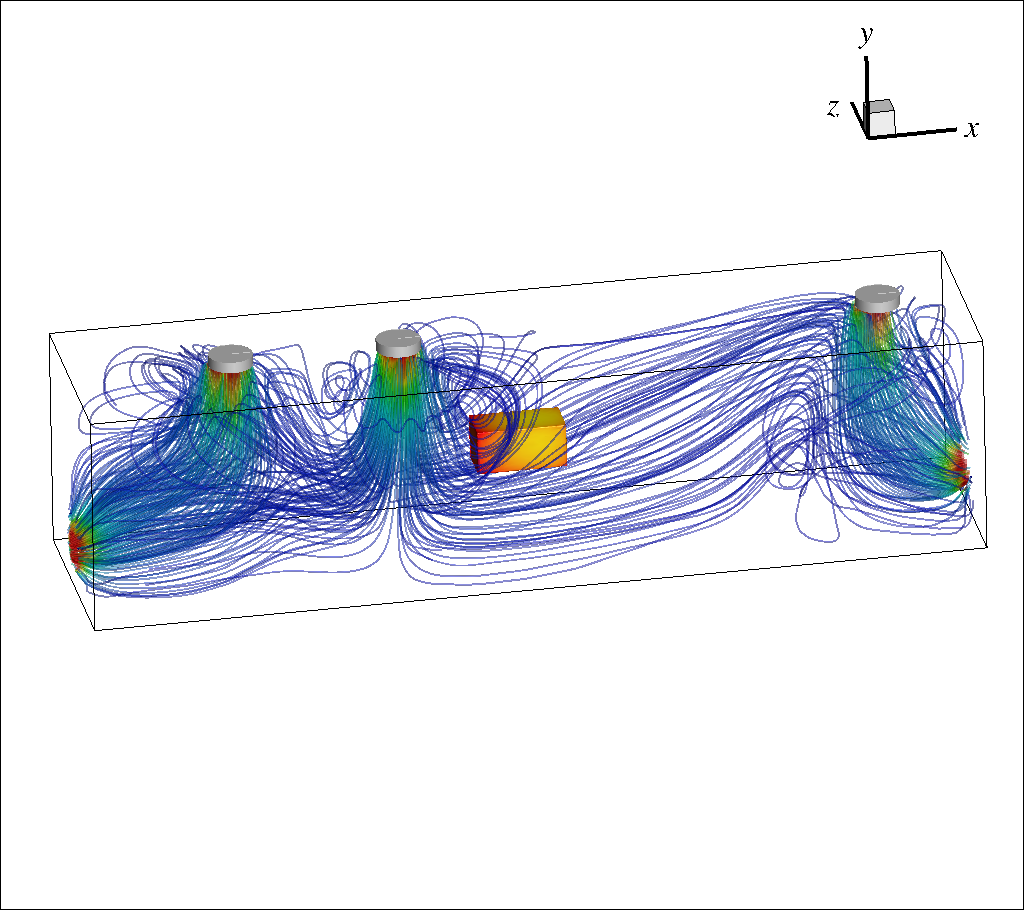}}
\put(1.5,17.5){\includegraphics[trim=650 715 10 30pt,clip,height=1.75cm]{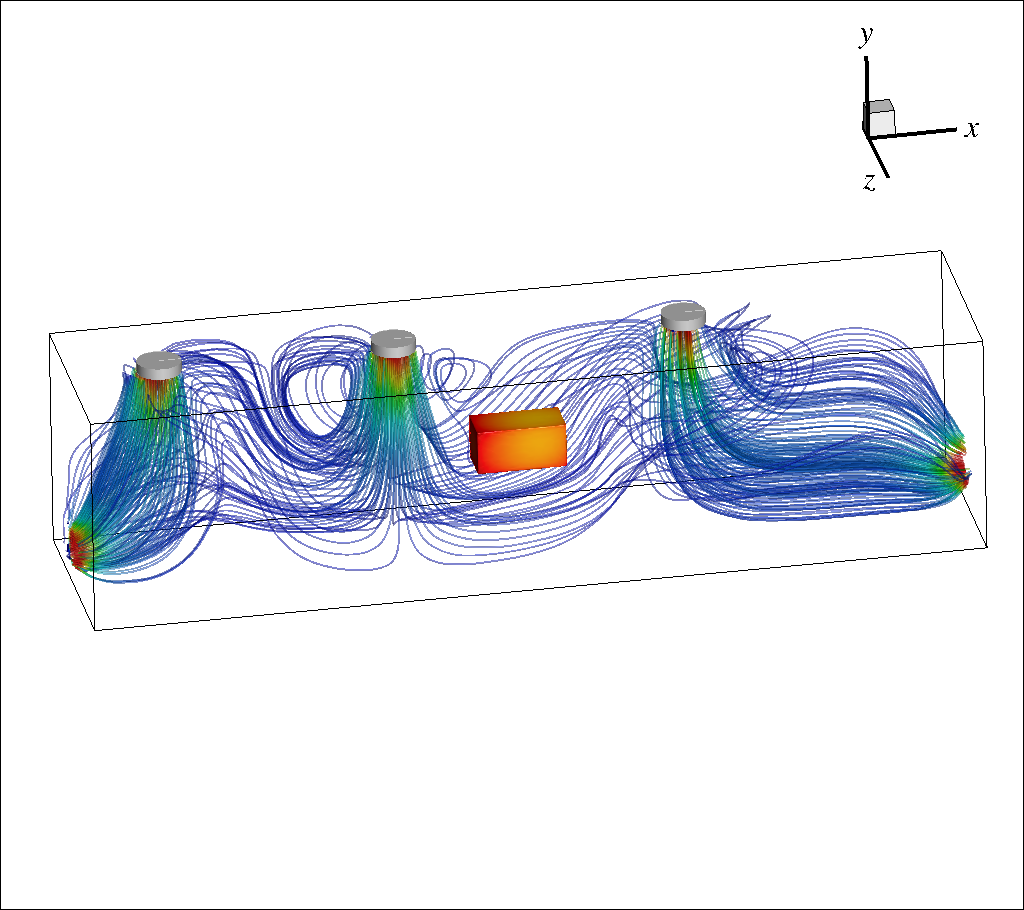}}
\put(8,18.5){\includegraphics[trim=420 635 440 245pt,clip,height=0.45cm]{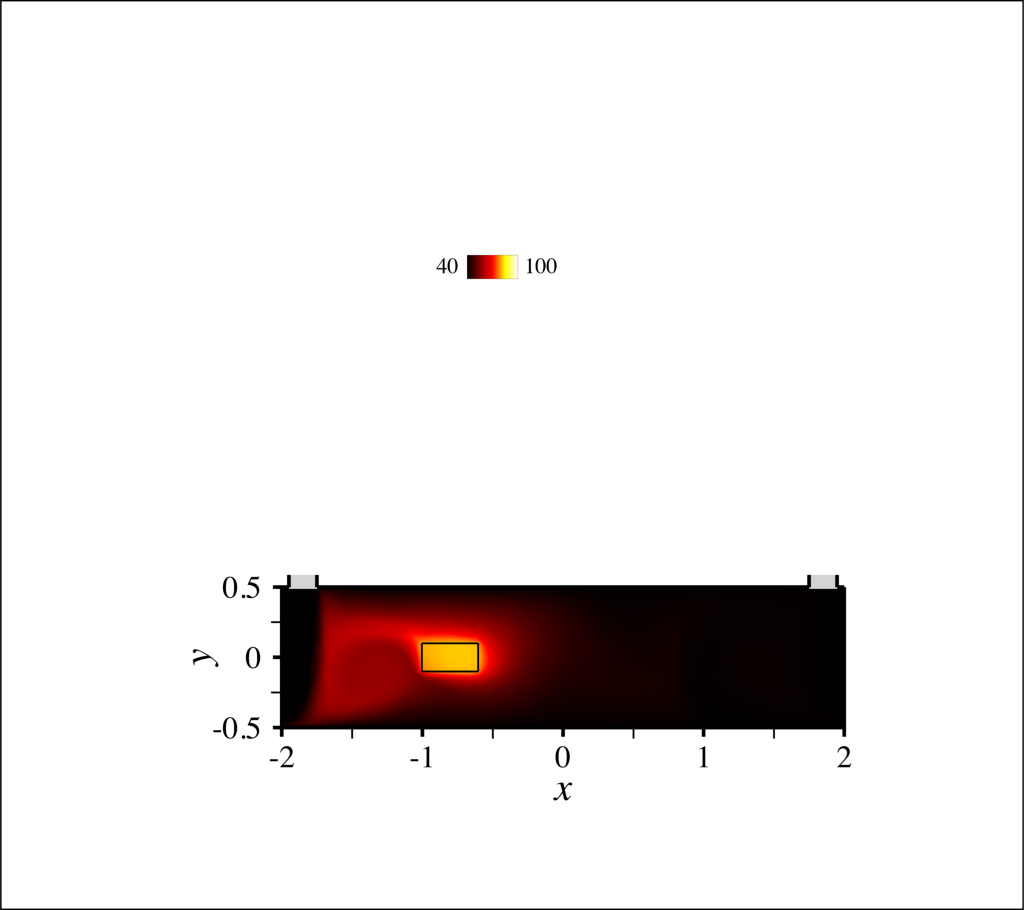}}
\put(10.5,18.5){\includegraphics[trim=420 635 440 245pt,clip,height=0.45cm]{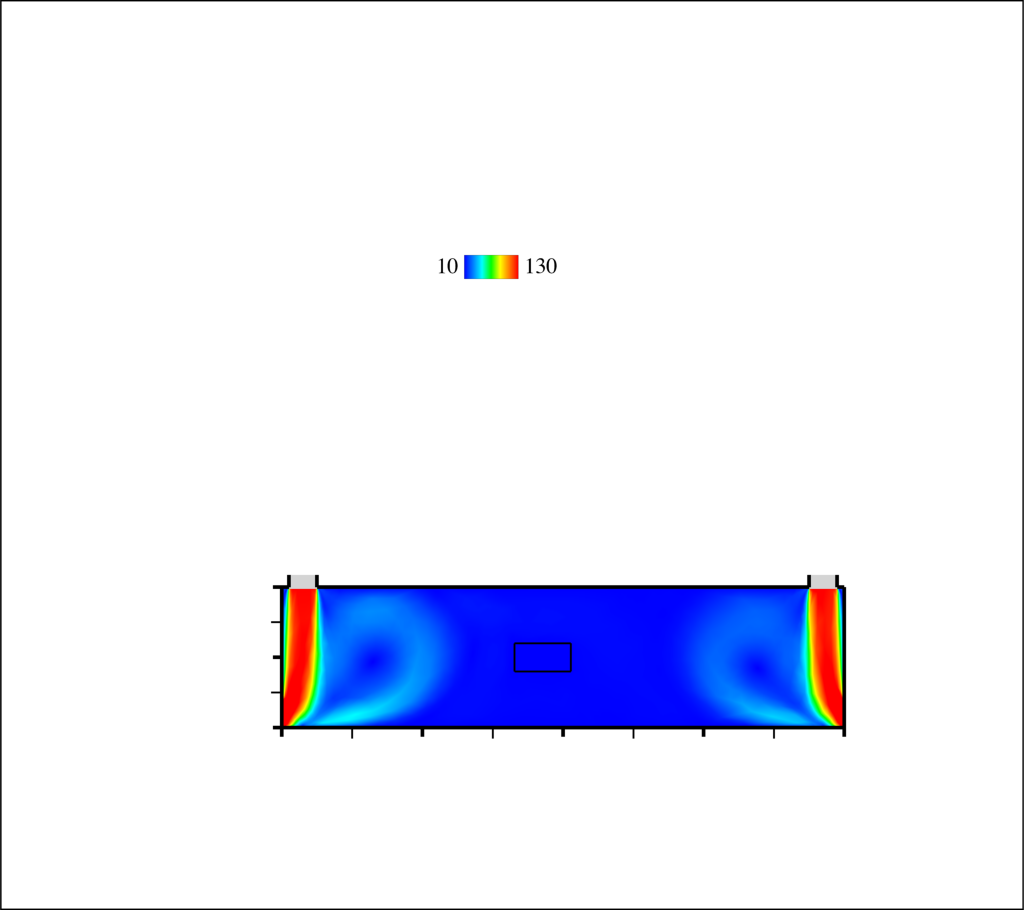}}
\end{picture}
\caption{Representative steady-state temperature distributions at the solid/fluid interface together with 3-D streamlines colored by the magnitude of velocity.}
\label{fig:forced3d_control_vtu}
\end{figure}

All results reported in the following are for $n_j=3$ injectors. 
The edge values needed to map the network action output into the actual injectors positions deduce straightforwardly from~\myrefeq{eq:edgeS1}-\myrefeq{eq:edgeS3} substituting the diameter $d_i$ of the 3-D injectors for the length $e_i$ of the 2-D injectors. 
The same DRL agent is used, that consists of two hidden layers, each holding 2 neurons, and 
the resolution process uses 8 environments and 2 steps mini-batches to update the network for 32 epochs. 
Each environment performs $1250$ iterations with time step $\Delta t = 0.1$ to march the same initial condition (consisting of zero velocity and uniform temperature, except in the solid domain) to steady state, using the level set, velocity and temperature as multiple-component criterion to adapt the mesh (initially pre-adapted using the sole level set) every 10 time steps under the constraint of a fixed number of elements $n_{el} = 120000$. This is likely insufficient to claim true numerical accuracy, but given the numerical cost
($320$ 3-D simulations per strategy, each of which is performed on 8 cores and lasts 2h30, hence 800h of total CPU cost), we believe this is a reasonable compromise to assess feasibility while producing qualitative results to build on.

\subsection{Results}

Only the fixed domain decomposition $S_1$ strategy (in which the top cavity wall is split into $n_j$ equal subdomains and each injector is forced to sit in a different subdomain) and the free $S_3$ strategy (in which the injectors are entirely independent and free to move along the top cavity wall) are considered here to save computational resources, as learning has been seen to be slower in 2-D under the follow-up $S_2$ strategy. 

\begin{figure}[t!]
\setlength{\unitlength}{1cm}
\begin{picture}(20,5)
\put(4,-0.1){\includegraphics[trim=175 87.5 140 340pt,clip,height=5cm]{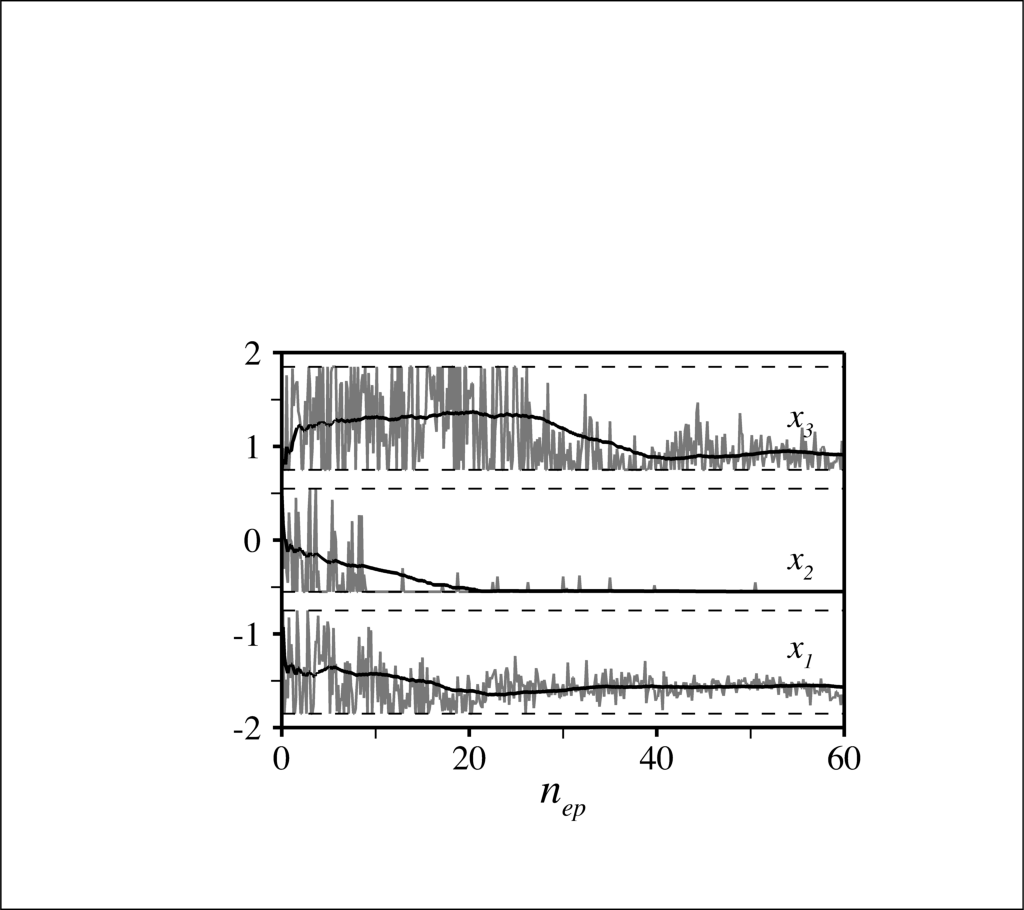}}
\end{picture}
\caption{Evolution per learning episode of the instant (in grey) and moving average (in black) injectors center positions under the three-dimensional fixed domain decomposition strategy $S_1$, with admissible values delimited by the dashed lines.}
\label{fig:forced3d_S1_reward}
\end{figure}

\begin{figure}[t!]
\setlength{\unitlength}{1cm}
\begin{picture}(20,5.5)
\put(1.75,0){\includegraphics[trim=10 275 10 250pt,clip,height=4.5cm]{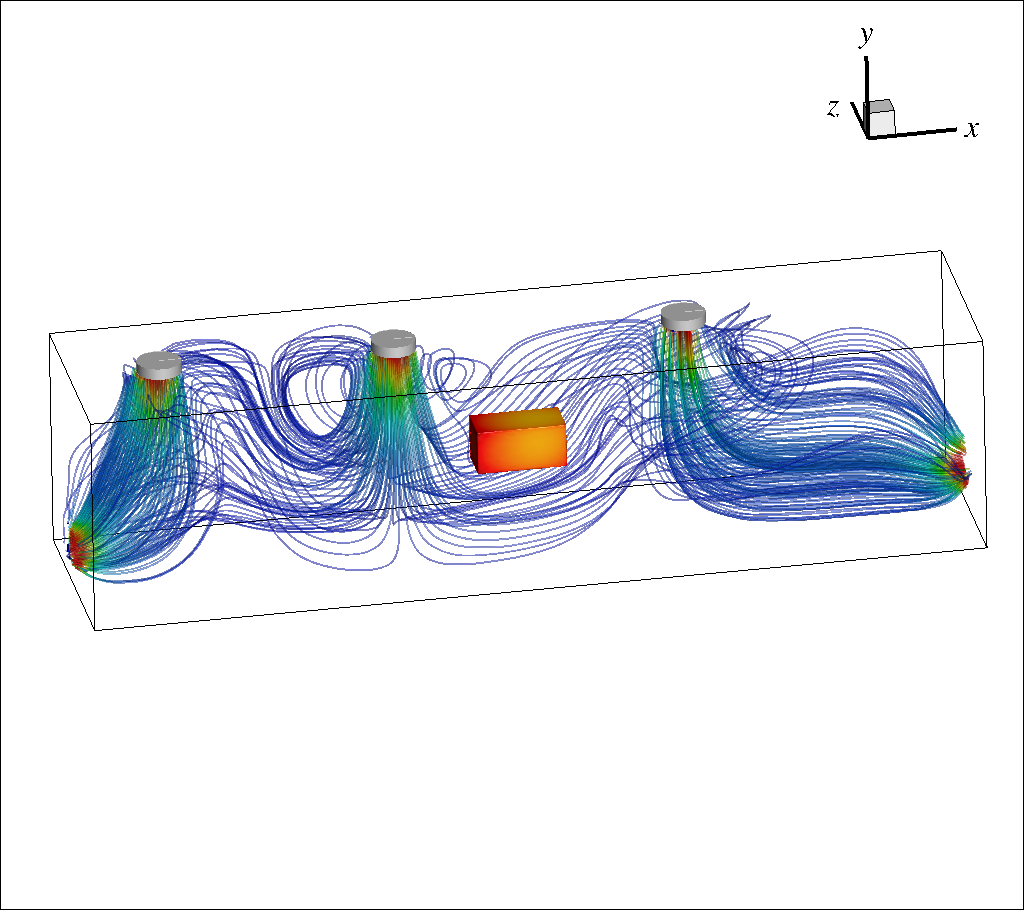}}
\put(1.5,4){\includegraphics[trim=650 715 10 30pt,clip,height=1.75cm]{fig26_axes}}
\put(8,5){\includegraphics[trim=420 635 440 245pt,clip,height=0.45cm]{fig26_bar_a}}
\put(10.5,5){\includegraphics[trim=420 635 440 245pt,clip,height=0.45cm]{fig26_bar_b}}
\end{picture}
\caption{Optimal 3 injector arrangement under the three-dimensional fixed decomposition domain strategy $S_1$.}
\label{fig:forced3d_S1_opt}
\end{figure}

A total of $60$ episodes have been run under the fixed domain decomposition strategy $S_1$.
Several representative flow patterns computed over the course of optimization are shown in figure~\ref{fig:forced3d_control_vtu} via iso-contours of the steady-state temperature at the fluid-solid interface and 3-D streamlines colored by the magnitude of velocity, to put special emphasis on transverse inhomogeneities and display the increased degree of complexity due to the formation of large-scale horseshoe vortices wrapped around the nozzle jets. We show in figure~\ref{fig:forced3d_S1_reward} that 
the distribution slowly converges to an optimal arrangement consisting of one injector 
at the left end of the left subdomain ($x_{1}{^\star}=-1.63$), another one 
at the left end of the center subdomain ($x_{2}{^\star}=-0.55$), and a third one at the left end of the right subdomain ($x_{3}{^\star}=0.87)$, as has been 
determined by averaging the instant positions of
each injector over the latest 10 learning episodes, with variations by roughly $\pm 0.04$ computed from the root-mean-square of the moving average over the same interval. This is larger by one order of magnitude than the variations reported in 2-D,  as the agent keeps exploring slightly sub-optimal positions of the lateral injectors, which likely simply reflects the challenging nature of performing three-dimensional optimal control. The 3-D $S_1$ optimal somehow resemble its 2-D counterpart, namely the center injector is at the exact same position, while the lateral injectors (especially the leftmost one) have been pushed towards the cavity sidewalls. The associated flow pattern is reported in figure~\ref{fig:forced3d_S1_opt}. The associated optimal reward computed over the same interval is $\langle ||\nabla_{\|}T||\rangle{^\star}\sim19.5$, i.e. twice as large than in 2-D, although it is difficult to compare further because of the difference in the Reynolds and Prandtl number. 

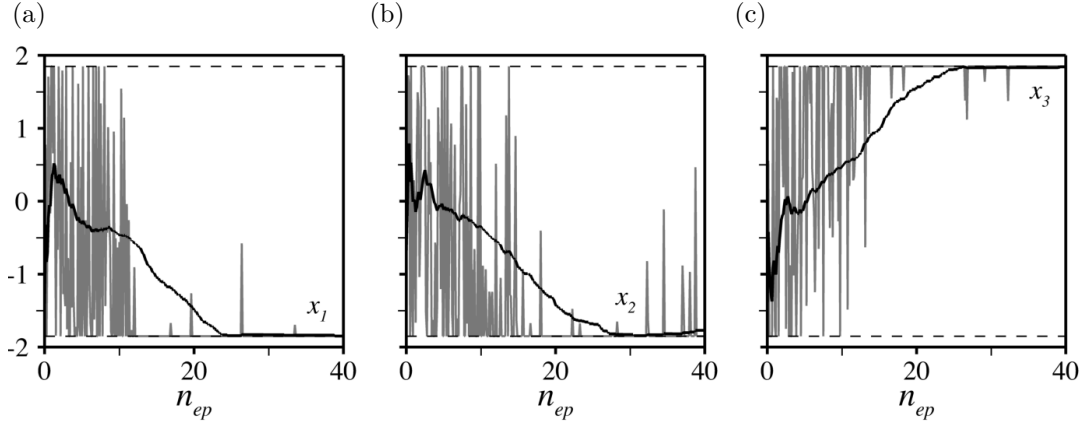
\begin{figure}[t!]
\setlength{\unitlength}{1cm}
\begin{picture}(20,5.2)
\put(0.5,-0.1){\includegraphics[trim=235 87.5 140 340pt,clip,height=5cm]{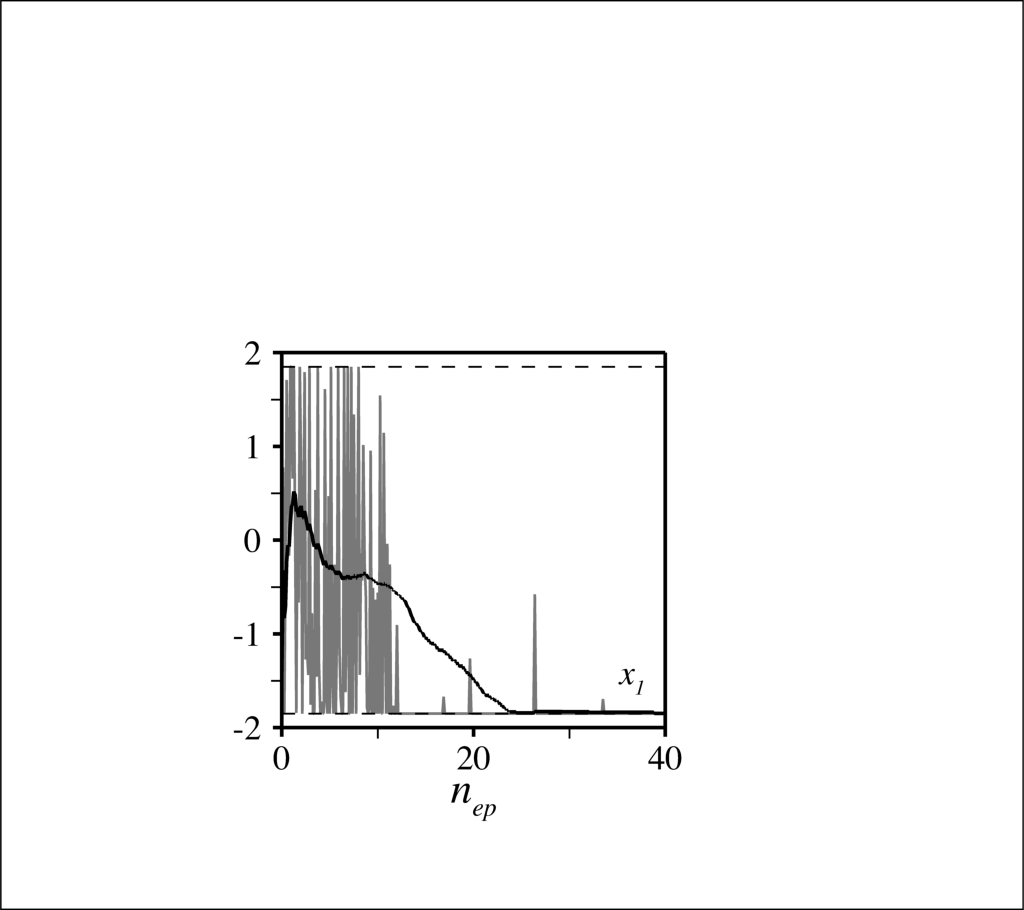}}
\put(5.3,-0.1){\includegraphics[trim=235 87.5 140 340pt,clip,height=5cm]{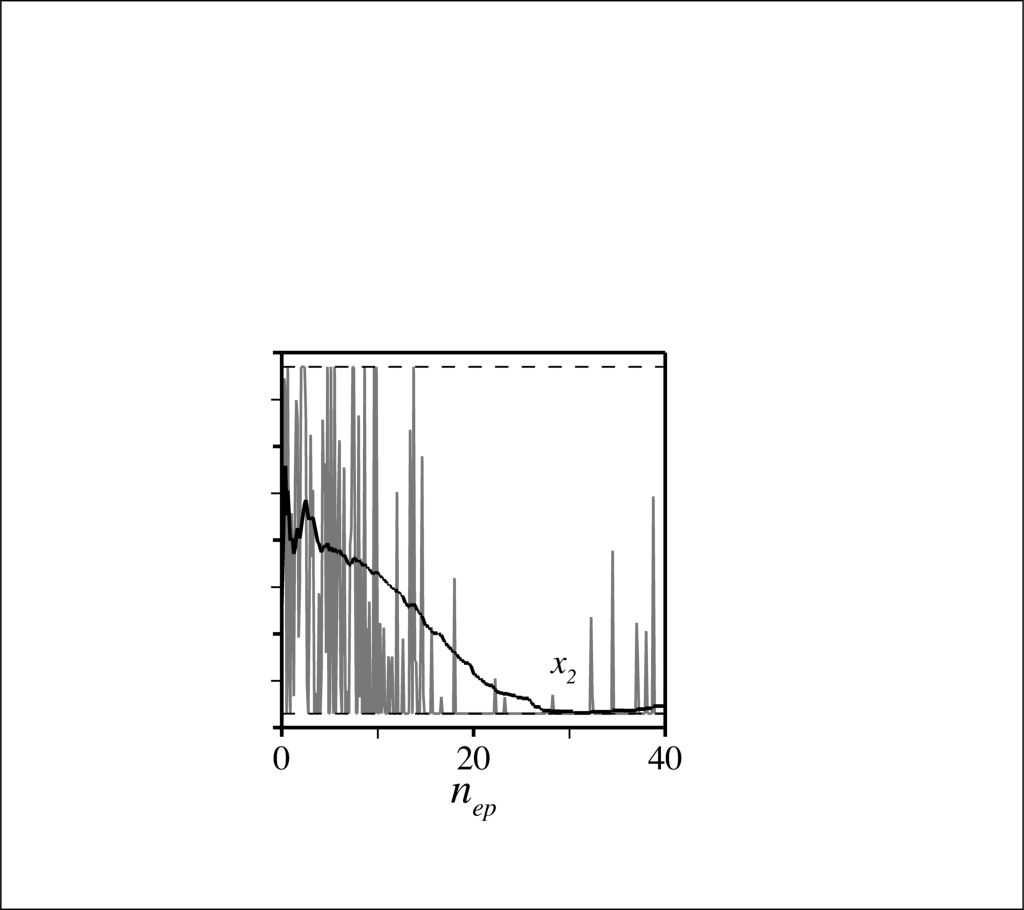}}
\put(10.1,-0.1){\includegraphics[trim=235 87.5 140 340pt,clip,height=5cm]{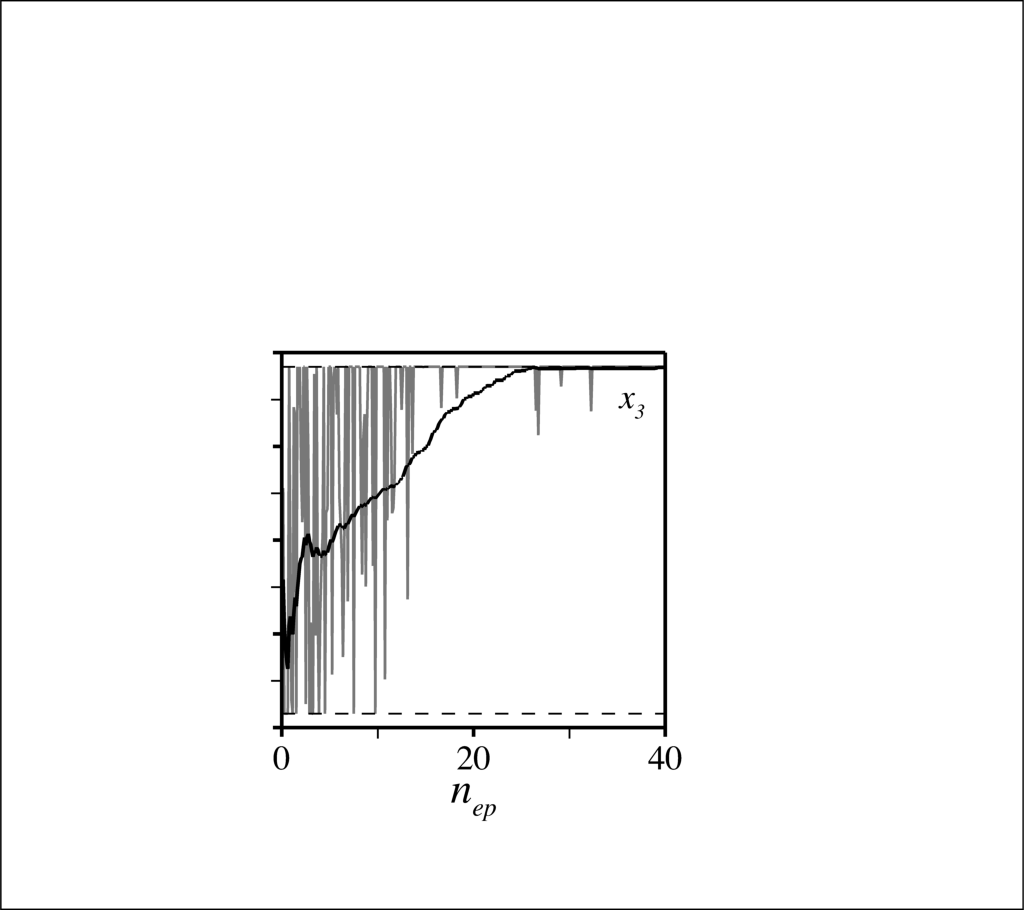}}
\put(10.15,5.2){(c)}
\put(5.3,5.2){(b)}
\put(0.55,5.2){(a)}
\end{picture}
\caption{Evolution per learning episode of the instant (in grey) and moving average (in black) injectors center positions under the three-dimensional free strategy $S_3$, with admissible values delimited by the dashed lines.}
\label{fig:forced3d_S3_reward}
\end{figure}

\begin{figure}[t!]
\setlength{\unitlength}{1cm}
\begin{picture}(20,5.5)
\put(1.75,0){\includegraphics[trim=10 275 10 250pt,clip,height=4.5cm]{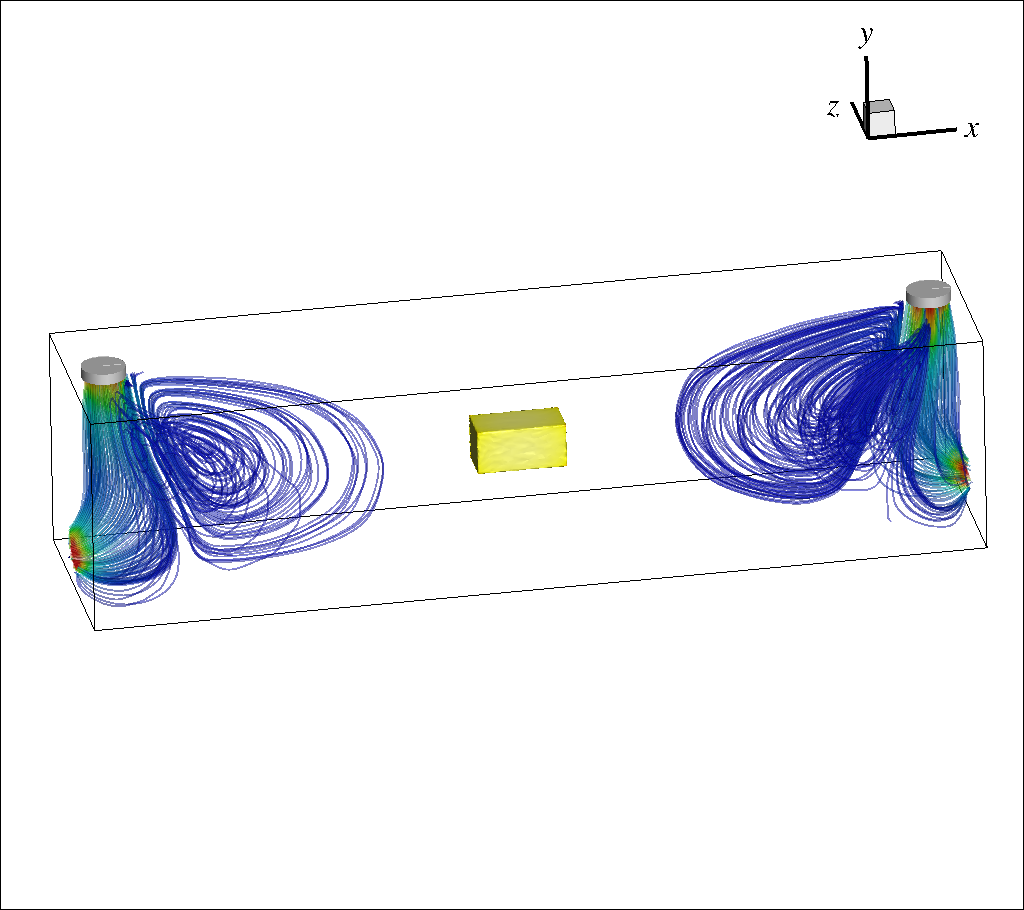}}
\put(1.5,4){\includegraphics[trim=650 715 10 30pt,clip,height=1.75cm]{fig26_axes}}
\put(8,5){\includegraphics[trim=420 635 440 245pt,clip,height=0.45cm]{fig26_bar_a}}
\put(10.5,5){\includegraphics[trim=420 635 440 245pt,clip,height=0.45cm]{fig26_bar_b}}
\end{picture}
\caption{Optimal 3 injector arrangement under the three-dimensional free strategy $S_3$.}
\label{fig:forced3d_S3_opt}
\end{figure}

Another $40$ episodes have been run under the free strategy $S_3$, for which the results are almost identical to their 2-D counterparts, as the distribution converges in figure~\ref{fig:forced3d_S3_reward} to an optimal arrangement consisting of two overlapping injectors at the left end of the cavity ($x_{1}{^\star}=-1.83$ and $x_{2}{^\star}=-1.82$), and a third injector at the right end ($x_{3}{^\star}=1.83$),
with variations by with $\pm 0.01$ for the lateral injectors, but $\pm0.03$ for the center injector, for which the agent keeps occasionally exploring sub-optimal positions. The corresponding flow pattern shown in figure~\ref{fig:forced3d_S3_opt} is thus again symmetrical with two large, 3-D recirculation regions on either side of the workpiece. The associated optimal reward computed over the same interval is $\langle ||\nabla_{\|}T||\rangle{^\star}\sim 4.7$ substantially smaller than that achieved under the 3-D $S_1$ strategy, which again demonstrates the feasibility to improve performances by allowing overlaps. All relevant numerical data are reported in table~\ref{table:forced3drecap} for the sake of completeness.

\begin{table}[t!]
\begin{center}
\begin{tabular}{cccccccc}
\toprule
& \multicolumn{1}{p{1cm}}{\makecell[r]{$n_j$}} & \multicolumn{1}{p{1cm}}{\makecell[r]{$n_{ep}$}} & \multicolumn{1}{p{1.1cm}}{\makecell[r]{$x_0$}} & \multicolumn{1}{p{1.1cm}}{\makecell[r]{$x_1$}} & \multicolumn{1}{p{1.1cm}}{\makecell[r]{$x_2$}} & \multicolumn{1}{p{1.1cm}}{\makecell[r]{$x_3$}} & \multicolumn{1}{p{2cm}}{\makecell[r]{$\langle ||\nabla_{\|}T||\rangle$}}\\
\cmidrule(lr){1-8}
\multicolumn{1}{r}{$S_1$} & \multicolumn{1}{r}{$3$} & \multicolumn{1}{r}{$60$} & \multicolumn{1}{r}{$0$} & \multicolumn{1}{r}{$-1.63$} &\multicolumn{1}{r}{$-0.55$} & \multicolumn{1}{r}{$0.87$} & \multicolumn{1}{r}{$19.5$}\\
\multicolumn{1}{r}{$S_3$} & \multicolumn{1}{r}{$3$} & \multicolumn{1}{r}{$40$} & \multicolumn{1}{r}{$0$} & \multicolumn{1}{r}{$-1.83$} &\multicolumn{1}{r}{$-1.82$} & \multicolumn{1}{r}{$1.83$} & \multicolumn{1}{r}{$4.7$}\\
\bottomrule
\end{tabular}
\caption{Numerical data for the optimal arrangements computed in three-dimensions under strategies $S_{1}$ and $S_3$. All values computed by averaging the instant signal over the 10 latest learning episodes.}
\label{table:forced3drecap}
\end{center}	
\end{table}



\section{Conclusion}\label{section:disc} 

{Optimization of conjugate natural and forced heat transfer systems is achieved here training a fully connected network with a novel single-step PPO deep reinforcement algorithm, in which it gets only one attempt per learning episode at finding the optimal. The numerical reward fed to the network is computed with a finite elements CFD environment solving stabilized weak forms of the
coupled Navier--Stokes and heat equations} with a combination of variational multi-scale modeling, immerse volume method, and multi-component anisotropic mesh adaptation.

{Convergence is assessed by alleviating the natural convection induced enhancement of heat transfer in a two-dimensional, differentially heated square cavity controlled by 
piece-wise constant fluctuations of the sidewall temperature. The approach is also relevant to forced convection problems, as single-step PPO shows capable of improving the homogeneity of temperature across the surface of two and three-dimensional hot workpieces under impingement cooling. Several control strategies are considered, in which the position of multiple cold air injectors is optimized relative to a fixed workpiece position, each of which mimics a different levels of design constraint. The flexibility of the numerical framework also allows solving the inverse problem, i.e., optimizing the workpiece position relative to a fixed injector distribution, which is relevant in situations where the design cannot be changed. 
The approach is beneficial in two important respects: first, it is efficient, even though the parameter spaces are large and it may be costly to identify optimal control parameters from simple parametric searches. Second, and more significantly, it is capable of determining additional optimal configurations, as the results of the inverse problem under symmetrical actuation indicate that the workpiece is best positioned offset from the symmetry axis, which had not been anticipated. Such results clearly stress that single-step PPO (and DRL in general) can be effective to explore and discover new solutions from unforeseen parameter combinations.}

Fluid dynamicists have just begun to gauge the ability of DRL to design optimal control strategies. {The efforts for developing single-step PPO are ongoing and remain at an early stage, so we do not expect the approach to compete right away with more established methods, for instance Evolution strategies (ES), a popular class of algorithms imitating principles of organic evolution processes 
as rules for black-box optimum seeking. ES rely on a stochastic description of the variables to optimize, i.e., they consider probability density functions, not deterministic variables. Simply put, at each generation (or iteration) new candidate solutions are sampled isotropically by variation of the current parental individuals according to a multivariate normal distribution. Recombination and mutation transformations are applied (that amount respectively to changing the mean and adding a random, zero-mean perturbation), after which the individuals with the highest cost function are selected to become the parents in the next generation. Improved variants include the covariance matrix adaptation evolution strategy (CMA-ES), that speeds up convergence by updating its full covariance matrix (which amounts to learning a second-order model of the objective function).
In present form, single-step PPO can be thought as an evolutionary-like algorithm with simpler heuristics (i.e., without an evolutionary update strategy, as the optimal model parameters are learnt via gradient ascent), so it is our guess that the performance should be comparable to that of standard ES algorithms with isotropic covariance matrix.
Besides consolidating the acquired knowledge,} future research should thus aim at improving efficiency (by fine-tuning the hyper parameters, or using pre-trained deep learning models) and convergence (by coupling with a surrogate model trained on-the-fly, using non-normal probability density functions, or modifying the balance between exploration and exploitation, as PPO {prevents large updates of the policy to avoid the issue of performance collapse}). 
\red{For complex configurations representative of industrial applications, the implementation of properly designed numerical rewards (under partial state information) and noise reduction techniques is another issue that deserves consideration, as pointed out in~\cite{Fan2020}.}

Scope is another key ingredient to {keep} pushing forward the  state of the art. {The next step} is to tackle more complex test cases exhibiting {flow unsteadiness and turbulence, which the CFD environment} is perfectly suited to do via a combination of Reynolds-averaged Navier--Stokes modeling~\cite{sari18,guiza2020} and second-order, semi-implicit time discretization~\cite{meli19jcp}. We believe that this will highlight even more clearly the relevance of the methodology, as~\cite{ghraieb2020} speculates that DRL should be able to handle chaotic systems without suffering from the shortcomings and limitations of the adjoint method, and it is shown in~\cite{beintema2020controlling} to outperform a canonical linear proportional-derivative controller in controlling turbulent natural convection. The long-term objective would be to enrich the description of the test cases using multi-physics modeling (e.g., radiative heat transfer, phase transformation) {in order to} pave the way toward flexible, ready-to-use control of industrially relevant applications, such as thermal comfort for building design or manufacturing processes.

\section*{Acknowledgements} 
This work is supported by the Carnot M.I.N.E.S. Institute through the M.I.N.D.S. project.

\bibliographystyle{unsrt}
\bibliography{biblist}


\end{document}